\documentclass[superscriptaddress,twocolumn,10pt,prx,aps,showpacs,longbibliography,floatfix]{revtex4-2}

\usepackage[dvipsnames]{xcolor}
\usepackage{titletoc}

\definecolor{darkred}{rgb}{0.6,0.05,0.05}
\definecolor{darkgreen}{rgb}{0.05,0.6,0.05}
\definecolor{darkblue}{rgb}{0.05,0.05,0.6}
\usepackage[colorlinks]{hyperref}
\hypersetup{colorlinks,
linkcolor=darkred,
filecolor=darkgreen,
urlcolor=darkblue,
citecolor=darkgreen}

\usepackage{textcomp,mathcomp}
\usepackage{mathtools}
\usepackage{amsmath}
\usepackage{amssymb}
\usepackage[normalem]{ulem}
\usepackage{amsthm}
\usepackage{xfrac}
\usepackage{dsfont}
\usepackage{siunitx}
\usepackage{physics}
\usepackage{graphicx}
\usepackage{hyperref}
\usepackage[capitalise]{cleveref}
\usepackage{dcolumn}
\usepackage{bm}

\usepackage{multirow}
\usepackage{tabularx}
\usepackage{booktabs}

\usepackage{soul}

\definecolor{armygreen}{rgb}{0.29, 0.33, 0.13}
\definecolor{palatinatepurple}{rgb}{0.41, 0.16, 0.38}
\definecolor{sangria}{rgb}{0.57, 0.0, 0.04}

\newcommand{\LL}{\mathcal{L}}
\newcommand{\rhot}{\rho}
\newcommand{\sss}{\rho_{\rm ss}}

\newcommand{\eig}[1]{\hat{\rho}_{#1}}

\renewcommand{\L}{\mathcal{L}}

\makeatletter
\newcommand*\bigcdot{\mathpalette\bigcdot@{.5}}
\newcommand*\bigcdot@[2]{\mathbin{\vcenter{\hbox{\scalebox{#2}{$\m@th#1\bullet$}}}}}
\makeatother

\newcommand{\de}{{\rm d}}

\usepackage{xfp}
\newcommand\ExtendedDataFigure{
  \xdef\presupfigures{\arabic{figure}}
  \renewcommand\thefigure{\fpeval{\arabic{figure}-\presupfigures}}
  \renewcommand{\figurename}{Extended Data FIG.}
  \setcounter{page}{1}
    \makeatletter
    \renewcommand{\thepage}{Ext\arabic{page}}
}

\newcommand\SupplementaryMaterials{
  \xdef\presupfigures{\arabic{figure}}
  \xdef\presupsections{\arabic{section}}
  \xdef\presuptable{\arabic{table}}
  \xdef\presupequation{\arabic{equation}} 
  \renewcommand\thefigure{S\fpeval{\arabic{figure}-\presupfigures}}
  \renewcommand\thesection{S\fpeval{\arabic{section}-\presupsections}}
  \renewcommand\theequation{S\fpeval{\arabic{equation}-\presupequation}}  
    \renewcommand{\thetable}{S\fpeval{\arabic{table}-\presuptable}}
    \renewcommand{\figurename}{FIG.}
\setcounter{page}{1}
    \makeatletter
    \renewcommand{\thepage}{S\arabic{page}}    
}

\begin{document}

\author{Guillaume Beaulieu}
\thanks{These two authors contributed equally}
\affiliation{Hybrid Quantum Circuits Laboratory (HQC), Institute of Physics, \'{E}cole Polytechnique F\'{e}d\'{e}rale de Lausanne (EPFL), 1015 Lausanne, Switzerland}
\affiliation{Center for Quantum Science and Engineering, \\ \'{E}cole Polytechnique F\'{e}d\'{e}rale de Lausanne (EPFL), CH-1015 Lausanne, Switzerland}
\author{Fabrizio Minganti}
\thanks{These two authors contributed equally}
\affiliation{Center for Quantum Science and Engineering, \\ \'{E}cole Polytechnique F\'{e}d\'{e}rale de Lausanne (EPFL), CH-1015 Lausanne, Switzerland}
\affiliation{Laboratory of Theoretical Physics of Nanosystems (LTPN), Institute of Physics, \'{E}cole Polytechnique F\'{e}d\'{e}rale de Lausanne (EPFL), 1015 Lausanne, Switzerland}
\author{Simone Frasca}
\affiliation{Hybrid Quantum Circuits Laboratory (HQC), Institute of Physics, \'{E}cole Polytechnique F\'{e}d\'{e}rale de Lausanne (EPFL), 1015 Lausanne, Switzerland}
\affiliation{Center for Quantum Science and Engineering, \\ \'{E}cole Polytechnique F\'{e}d\'{e}rale de Lausanne (EPFL), CH-1015 Lausanne, Switzerland}
\author{Vincenzo Savona}
\affiliation{Center for Quantum Science and Engineering, \\ \'{E}cole Polytechnique F\'{e}d\'{e}rale de Lausanne (EPFL), CH-1015 Lausanne, Switzerland}
\affiliation{Laboratory of Theoretical Physics of Nanosystems (LTPN), Institute of Physics, \'{E}cole Polytechnique F\'{e}d\'{e}rale de Lausanne (EPFL), 1015 Lausanne, Switzerland}
\author{Simone Felicetti}
\affiliation{Institute for Complex Systems, National Research Council (ISC-CNR) and Physics Department, Sapienza University, P.le A. Moro 2, 00185 Rome, Italy}
\author{Roberto Di Candia}
\affiliation{Department of Information and Communications Engineering, Aalto University, Espoo 02150, Finland}
\affiliation{Dipartimento di Fisica, Universit\`a degli Studi di Pavia, Via Agostino Bassi 6, I-27100, Pavia, Italy}
\author{Pasquale Scarlino}
\email[E-mail: ]{pasquale.scarlino@epfl.ch}
\affiliation{Hybrid Quantum Circuits Laboratory (HQC), Institute of Physics, \'{E}cole Polytechnique F\'{e}d\'{e}rale de Lausanne (EPFL), 1015 Lausanne, Switzerland}
\affiliation{Center for Quantum Science and Engineering, \\ \'{E}cole Polytechnique F\'{e}d\'{e}rale de Lausanne (EPFL), CH-1015 Lausanne, Switzerland}

\title{Observation of first- and second-order dissipative phase transitions \\in a two-photon driven Kerr resonator}

\date{\today}

\begin{abstract}
In open quantum systems, first- and second-order dissipative phase transitions (DPTs) can emerge in the thermodynamic limit from the competition between unitary evolution, driving terms, and dissipation.
The order of a DPT is defined by the continuity properties of the steady state.
Until now, second-order DPTs have predominantly been investigated theoretically, while first-order DPTs have been observed in key experiments based on the theory of the single-photon driven Kerr resonator.
We present here the first comprehensive experimental and theoretical analysis of \textit{both} first and second-order DPTs in a \textit{two-photon} (i.e., parametrically) driven Kerr superconducting resonator.
Firstly, we characterize the steady state and its main features at the second- and first-order critical points: squeezing below vacuum and coexistence of two phases with different photon numbers, respectively. 
Then, by continuously monitoring the system along quantum trajectories, we study the non-equilibrium dynamics across the critical points.
We witness the hysteresis cycles associated with the first-order DPT and
the spontaneous symmetry breaking due to the second-order DPT.
Applying the spectral theory of the Liouvillian superoperator, we develop efficient procedures to quantify the critical slowing down associated with the timescales of these processes. 
When scaling towards the thermodynamic limit, these timescales span five orders of magnitude.
Our results corroborate the predictions derived using the Liouvillian theory of DPTs.
This work stands as a compelling example of engineering and controlling of criticality in superconducting circuits. It marks a significant advancement in the use of two-photon driven Kerr resonators for criticality-enhanced quantum information applications.
\end{abstract}

\maketitle

\newcommand{\nocontentsline}[3]{}
\newcommand{\tocless}[2]{\bgroup\let\addcontentsline=\nocontentsline#1{#2}\egroup}

\tocless\section{Main} \label{sec:introduction}

\begin{figure}[!htb]
        \centering
        \includegraphics[height = 6cm, width = \textwidth, keepaspectratio]{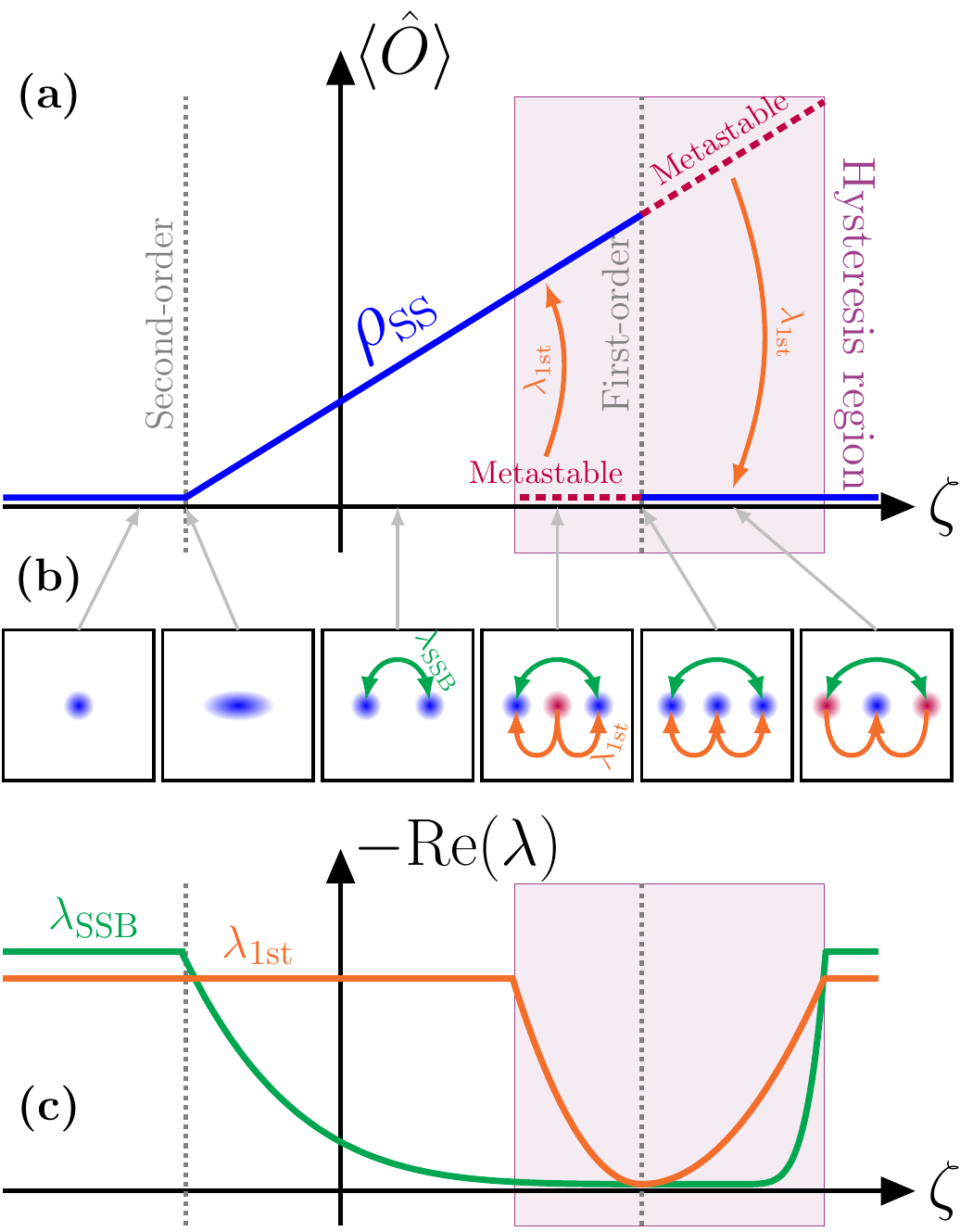}
        \includegraphics[height = 6cm, width = \textwidth, keepaspectratio]{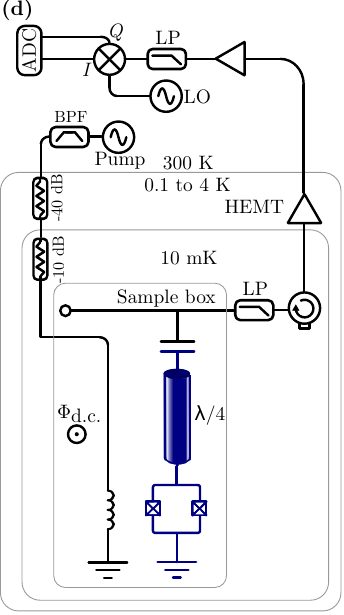}
        \caption{
        \textbf{Theory of dissipative phase transitions and schematic of the experimental set-up}.  
        (a) Illustration of dissipative phase transitions (DPTs) according to Ref.~\cite{MingantPRA18_Spectral}.
        Sweeping a control parameter $\zeta$, the expectation value of the order parameter $\langle \hat{O} 
        \rangle = \operatorname{Tr}[\sss(\zeta) \hat{O}]$ (blue curve) changes discontinuously (first-order DPT), or continuously with non-continuous derivative (second-order DPT).
        The purple dashed lines indicate the metastable states associated with hysteresis across the first-order DPT.
        (b) Phase-space-like representation of the system steady (blue) and metastable (purple) states across the DPTs.
        The arrows within each panel indicate the decay of an initial state towards the steady state. The green arrows represents the decay of a non-symmetric state at a rate $\lambda_{\rm SSB}$. 
        The orange arrows are associated with the metastable state of the first-order DPT, decaying at a rate $\lambda_{\rm 1st}$.
        (c) The Liouvillian gaps $\lambda_{\rm SSB}$ in green ($\lambda_{\rm 1st}$ in orange) associated with the second-order (first-order) DPT.
        (d) Schematic illustrating the device and the experimental setup. The device is a $\lambda/4$ coplanar waveguide resonator, capacitively coupled on one side to a feedline used only to collect the emitted signal via heterodyne detection (see Supplementary). On the other side, the cavity is terminated to ground via a SQUID.
        A magnetic field is applied through the SQUID, tuning both the resonance frequency and the Kerr nonlinearity. 
        A second waveguide, inductively coupled to the SQUID, is used to supply a coherent pump tone around twice the resonant frequency of the cavity ($\omega_p \simeq 2\omega_r$). 
        The pump results in a two-photon drive for the cavity (see Supplementary and e.g., Refs.~\cite{WilsonPRL10,Krantz_2013}).
        No other input signal is sent into the resonator.  
        }
         \label{Fig:scheme}
\end{figure}

Dissipative phase transitions (DPTs) are critical phenomena in which the steady state of the system -- or an observable associated with it (e.g., the order parameter) -- changes nonanalytically upon an infinitesimal change in a control parameter [see Fig.~\ref{Fig:scheme}(a)] \cite{MingantPRA18_Spectral,KesslerPRA12,CarmichaelPRX15,soriente2021distinctive}. DPTs extend the concepts of quantum and thermal phase transitions to systems out of their thermal equilibrium and placed in interaction with an environment \cite{CarmichaelPRX15,KesslerPRA12}. The investigation of DPTs is of paramount importance given their occurrence in various physical systems, spanning the fields of quantum optics \cite{FitzpatrickPRX17,FinkPRX17}, condensed matter \cite{RodriguezPRL17,FinkNatPhys18}, and quantum information and technology \cite{gravina2022critical,di2023critical,petrovnin2023microwave}. Therefore, the lack of established extremal principles to describe the steady states associated with DPTs (such as the minimization of thermodynamic potentials) calls for an effort to understand and characterize these critical phenomena.

DPTs can be either of first or second-order. 
First-order DPTs are characterized by a jump in the steady state and order parameter, together with phase coexistence, metastability, and hysteresis [see Figs.~\ref{Fig:scheme}(a-c)] \cite{MingantPRA18_Spectral,KesslerPRA12}.
First-order DPTs have been observed experimentally in several systems, including trapped ions~\cite{cai2021observation}, ultracold bosonic gasses \cite{Benary2022}, nonlinear photonic or polaritonic modes \cite{RodriguezPRL17,FinkNatPhys18,Zejian2022}, and circuit QED platforms \cite{Brookes2021,Chen2023,FinkPRX17,sett2022emergent}.

Second-order DPTs are characterized by symmetries and their spontaneous breaking, and display a continuous but non-differentiable steady state and order parameter as illustrated in Fig.~\ref{Fig:scheme}(a) \cite{MingantPRA18_Spectral}. 
As such, they present a jump in the {\it derivative} of the order parameter, which requires an exceptional degree of controllability of the system to be observed \cite{Ferri2021,LeuchPRL16}.
The peculiar characteristics of second-order DPTs are predicted to enhance efficient encoding of quantum information \cite{gravina2022critical} and bring advantageous metrological properties \cite{Garbe2020,Chu21,ilias22,Garbe22,Gietka22,di2023critical,ding2022enhanced}. 
These features further motivate the interest in an experimental characterization of the static and dynamical properties of second-order DPTs. 

Critical phenomena are commonly studied in many-body systems in the thermodynamic limit, where the number of constituents asymptotically diverges.
However, quantum phase transitions can also take place in finite-component systems, where the thermodynamic limit corresponds to a rescaling of the system parameters~\cite{hwang_quantum_2015,CasteelsPRA17-2,Peng19,felicetti2020universal}.
A preeminent role in the study of finite-component first-order DPTs has been played by nonlinear quantum-optical oscillators~\cite{Chen2023,Brookes2021,FinkNatPhys18,RodriguezPRL17}. 
An experimental analysis of the unexplored fundamental properties of first- and second-order DPTs requires to engineer drives and dissipative processes.
Superconducting circuits \cite{Blais2021} offer the 
necessary level of control to engineer these processes \cite{siddiqi2005direct,yamamoto2008flux,Ma2019}, while also allowing the parameter rescaling required to witness finite-component phase transitions.

In this article, we use a \textit{two-photon} driven superconducting Kerr resonator, and conduct a thorough experimental analysis of both its \textit{first- and second-order} DPTs.
As a first step, we scale the system towards the thermodynamic limit and analyze its steady state properties. 
We demonstrate the quantum nature of the system at the second-order DPT, showing squeezing below vacuum. 
Furthermore, we observe the coexistence of multiple metastable states in the vicinity of the first-order DPT, a feature that cannot be captured when neglecting the quantum effects of dissipation.
Then, we focus on the dynamical properties associated with both transitions by probing the system
dynamics through time resolved measurements. 
We analyze the data with novel theoretical tools, based on quantum trajectories and Liouvillian spectral theory, and extract the characteristic timescales.
From this analysis, we characterize the metastable states and quantify the critical slowing down of the two DPTs.

\begin{figure*}[!htb]
        \centering
        \includegraphics[width = 0.98 \textwidth]{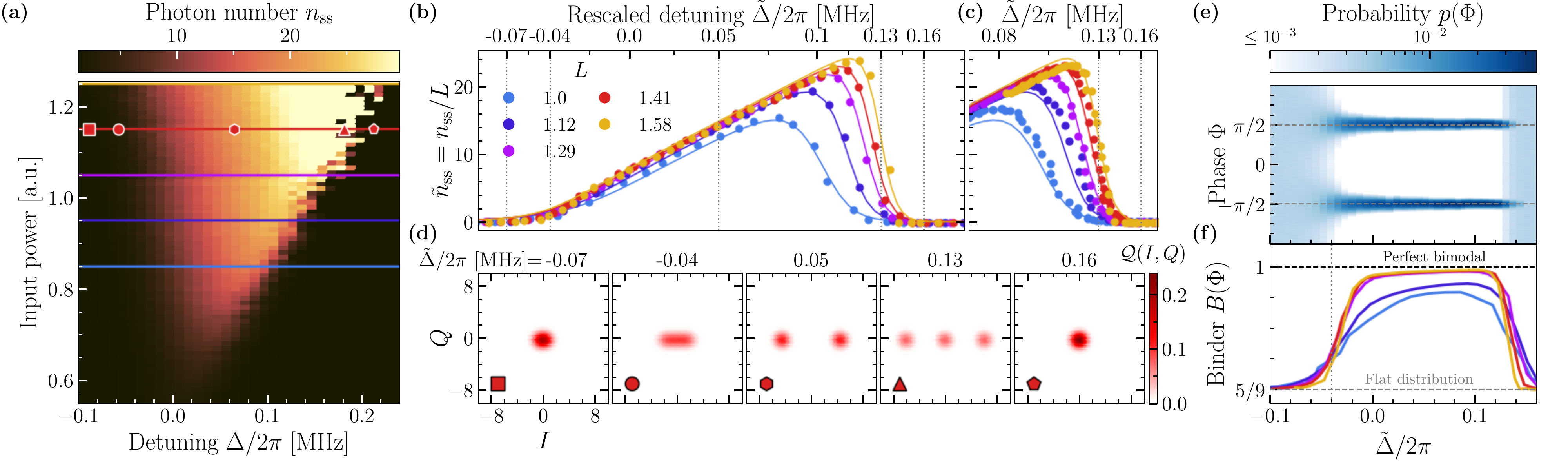}
        \caption{
        \textbf{Characterization of the steady state}.
        (a) Phase diagram showing the number of photons in the resonator as a function of the detuning $\Delta$ and input power, obtained by heterodyne detection of the emitted field. 
        The three phases are indicated by: (i) square marker (the vacuum at negative detuning); (ii) hexagon marker (the bright phase); (iii) pentagon marker (the vacuum at positive detuning).         
        The passage between these phases is accompanied by a second- [(i)$\to$(ii), circle marker] and first-order DPTs [(ii)$\to$(iii), triangle marker].
        (b) Rescaled number of photons $\tilde{n} = n/L$ as a function of the rescaled detuning $\tilde{\Delta} =\Delta/L $ and rescaled drive $G = \tilde{G} L$ for increasing scaling parameter $L$, with $\tilde{G} = 65.5$ KHz (see also text and Methods for details).
        Circles indicate the experimental data, and solid lines are obtained from the numerical simulation of Eq.~\eqref{Eq:lindblad}.
        The emergent discontinuities at negative and positive detuning with increasing $L$ signal the presence of a second- and first-order DPT in the thermodynamic limit, respectively.
        (c) Higher-resolution characterization of the abrupt change in $\tilde{n}$ across the first-order DPT.
        (d) Husimi-$Q$ function estimated through heterodyne detection. The markers correspond to those in panel (a), and the values of $\tilde{\Delta}$ corresponds to the vertical dotted gray lines in (b).
        (e) Histogram of the measured phase $\Phi$ along single trajectories for $L = 1.41$.
        (f) Bimodality coefficient (i.e., Binder cumulant) $B(\Phi)$, defined in the main text, calculated from the probability distribution of $\Phi$ for $L=1.41$.
        The vertical dotted line signals the value of the rescaled detuning where all curves cross.
        }
         \label{Fig:phase_diagram}
\end{figure*}

\vspace{15pt}
\tocless\subsection{Steady state properties and phase diagram}

The device, shown in Fig.~\ref{Fig:scheme}(d), is a superconducting cavity made nonlinear by terminating one end to ground via a superconducting quantum interference device (SQUID). A two-photon, i.e., parametric, drive is applied to the cavity by modulating the magnetic flux through the SQUID at nearly twice the resonance frequency of the cavity~\cite{WilsonPRL10,Krantz_2013,lin2014josephson}. The emitted signal is collected through a feedline coupled to the other end of the cavity, then filtered and amplified with a total gain $\mathcal{G}$ before being measured. Both signal quadratures ($\hat I$ and $\hat Q$) are acquired using time-resolved heterodyne detection (see Supplementary). This system is modeled by the Hamiltonian
\begin{equation}\label{Eq:Hamiltonian}
    \hat{H}/\hbar = \Delta \hat{a}^\dagger \hat{a} + \frac{U}{2} \hat{a}^\dagger \hat{a}^\dagger \hat{a} \hat{a} + \frac{G}{2} \left( \hat{a}^\dagger\hat{a}^\dagger + \hat{a}\hat{a} \right),
\end{equation}
where $\hat{a}$ is the photon annihilation operator, $\Delta= \omega_r - \frac{\omega_p}{2}$ is the pump-to-cavity detuning, and $G$ is the two-photon drive field amplitude. In this study, we use $\Delta$ as the control parameter across the transition [see $\zeta$ in Figs.~\ref{Fig:scheme}(a-c)]. Since the system interacts with the feedline, fluxline, and other uncontrolled bath degrees of freedom, its evolution is modeled via the Lindblad master equation
\begin{equation}\label{Eq:lindblad}
\begin{split}
    \frac{\partial \rhot}{\partial t}  = -\LL \rhot  & = - \frac{i}{\hbar} [\hat{H}, \rhot] + \kappa (n_{\rm th} +1) \mathcal{D}[\hat{a}] \rhot \\ &  + \kappa n_{\rm th} \mathcal{D}[\hat{a}^\dagger] \rhot + \kappa_{\phi} \mathcal{D}[\hat{a}^\dagger \hat{a}] \rhot+ \kappa_2 \mathcal{D}[\hat{a}^2] \rhot,
\end{split}    
\end{equation}
where $\LL$ is the Liouvillian superoperator, whose spectrum is key in characterizing DPTs \cite{KesslerPRA12,MingantPRA18_Spectral}. 
The dissipators are defined as $\mathcal{D}[\hat{A}]\rhot = \hat{A} \rhot \hat{A}^\dagger - \{\hat{A}^\dagger \hat{A}, \rhot\}/2$, and the rates $\kappa$, $\kappa_{\phi}$, and $\kappa_2$ are associated with the total photon loss, dephasing, and two-photon loss, respectively. 
Finally, $n_{\rm th}$ is the thermal photon number. Throughout the experiment, the resonator frequency is fixed at $\omega_r/2\pi = \SI{4.3497}{\giga\Hz}$, corresponding to a Kerr nonlinearity of $U/2\pi = \SI{7}{\kilo\Hz}$, and  $\kappa/2\pi=\SI{77}{\kilo\Hz}$. The other parameters of the experiment are theoretically estimated to be $\kappa_{\phi}/2\pi= 
 4.4$ kHz, $\kappa_2=78$ Hz, and $n_{th}=0.055$. 
As described in Methods, the value of $G$ is  measured, and then refined through a theoretical estimation. 
The methods used for determining these parameters are described in the Supplementary. 

\begin{figure}[h!]
    \centering
    \includegraphics[width=\textwidth]{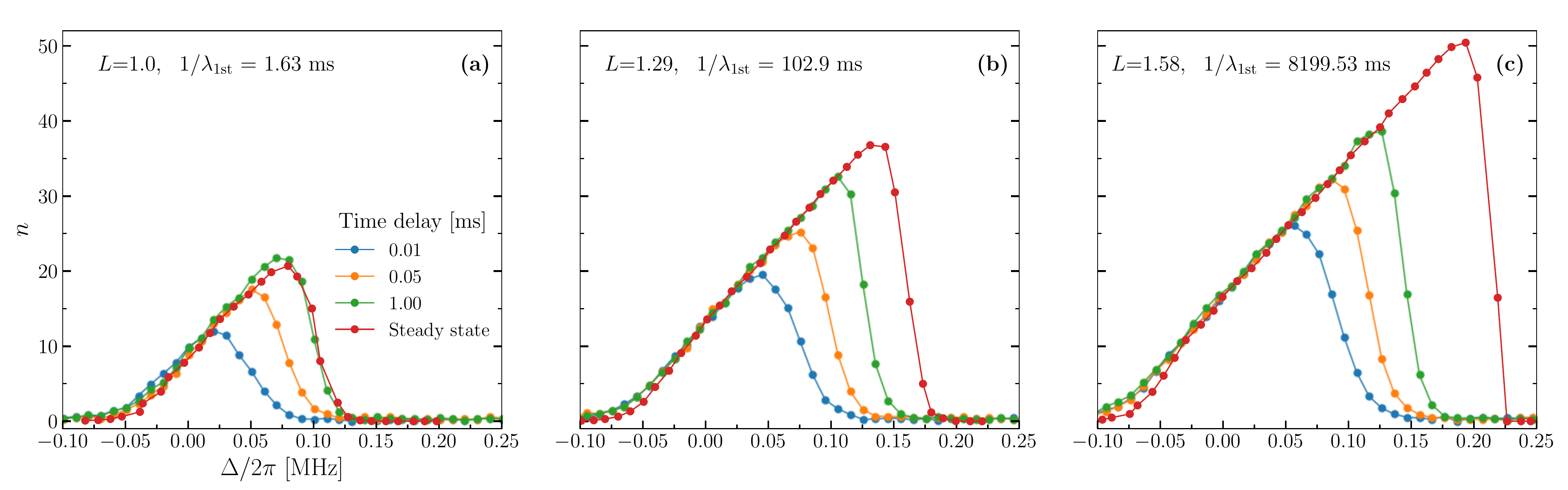}
    \caption{{\textbf{Photon number in the transient regime}}.
    To show the importance of waiting an appropriately long time in order to measure the steady state, we sweep down the detuning from $\Delta = 2$ MHZ (deep in the vacuum) for (a) $L=1$, (b) $L=1.29$, and (c) $L=1.58$. 
    For each point in the plot, we wait for the delay time indicated in the legend before decreasing the detuning (for further detail, see the explanation of the hysteresis curves in the main text).
    When such a delay time is significantly smaller than the inverse of the Liouvillian gap (i.e., the typical timescale to reach the steady state), the measured photon number is smaller $n_{\rm ss}$.
    This effect becomes more prominent when the Liouvillian gap becomes smaller, which happens for larger values of $L$.
    }
    \label{fig:Ext_data_1}
\end{figure}

We begin our study by characterizing the system steady state $\sss$, formally defined by $\partial_t \sss =0$. To this end, we initialize the system in the vacuum state, then switch on the two-photon drive $G$ at frequency $\omega_p$, and start acquiring the signal quadratures at frequency $\omega_p/2$ after a waiting time $\tau_{wait}$. 
Knowing the output gain $\mathcal{G}$ and total loss rate $\kappa$, the field quadratures $\hat{I}$ and $\hat{Q}$ of the cavity are then reconstructed (see Methods and Supplementary). 
The intracavity photon number is $n_{\rm ss}= \expval{\hat{a}^\dagger \hat{a}}_{\rm ss} = \expval{\hat{I}^2 +
\hat{Q}^2}$. In Fig.~\ref{Fig:phase_diagram}(a), $n_{\rm ss}$ is reported as a function of  $\Delta$ and input power, both tunable on demand. We stress that the required wait time $\tau_{wait}$ to reach the steady state can be orders-of-magnitude longer than the typical photon-lifetime $1/\kappa \sim \SI{2}{\micro\second}$ [see Extended Data Fig.~\ref{fig:Ext_data_1}], a clear indication of critical slowing down \cite{LandaPRL20}. 
From Fig.~\ref{Fig:phase_diagram}(a), we distinguish three regimes: (i) 
at large negative detuning, the system is in the vacuum state;
(ii) the system transitions from the vacuum state to a bright state without discontinuity. This happens at $\Delta \approx - G $ (see Methods);
(iii) at large positive detuning, $n_{\rm ss}$ falls abruptly from the high population phase to the vacuum \cite{WilsonPRL10}.

\begin{figure}
    \centering
    \includegraphics[width=\linewidth]{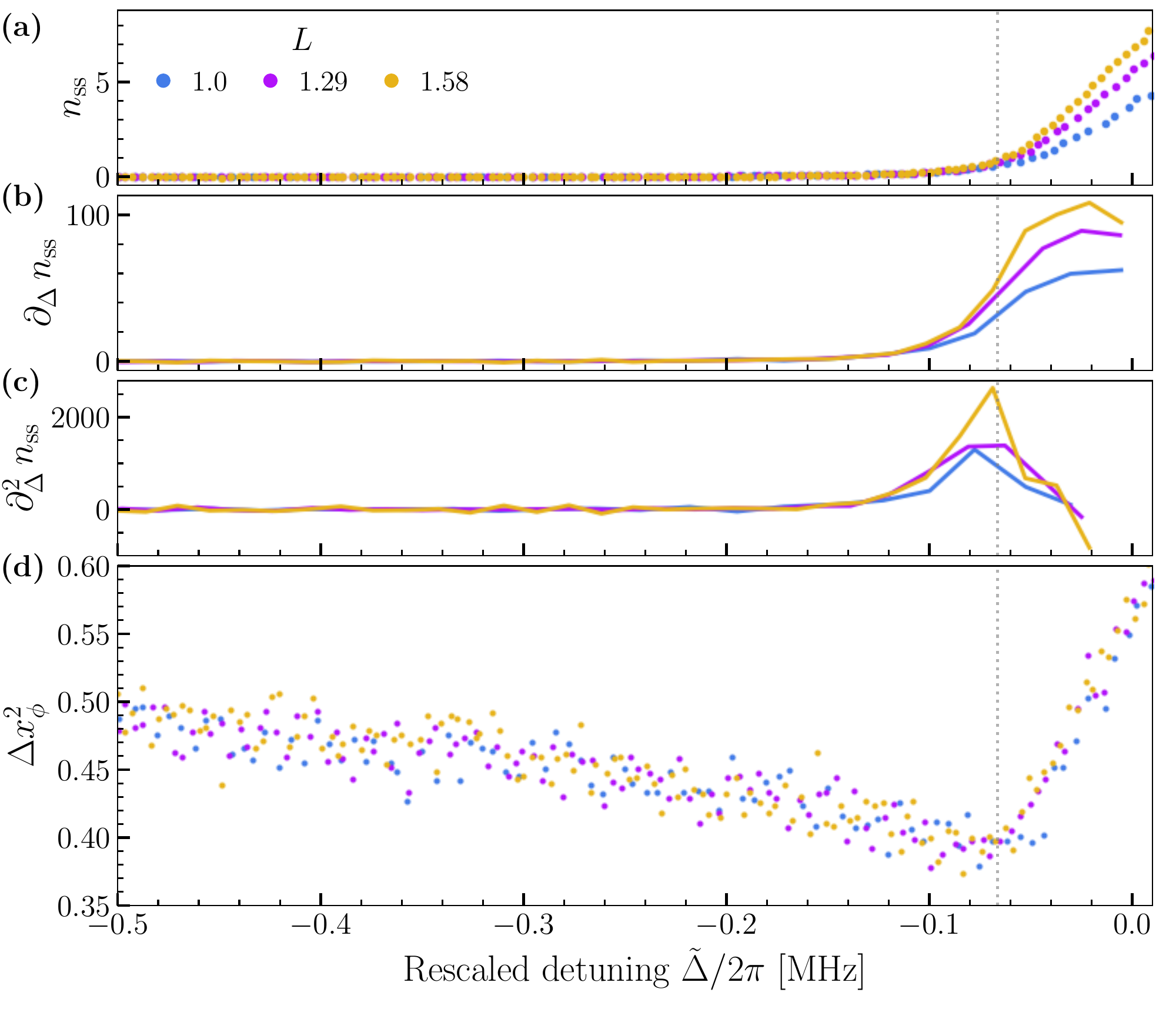}
    \caption{\textbf{Squeezing at the second-order DPT}.
    Photon number (a), its first (b), and second derivatives (c) calculated from the experimental data as a function of detuning.
    (d) The squeezing parameter $\Delta x^2_{\phi}$ evaluated across the second-order DPT.
    Notice that the minimum is in the vicinity of the critical point indicated by the maximum of the second derivative of the photon number in the panel (c). 
    The vertical dotted line indicates the expected minimum of the squeezing parameter obtained by numerical simulation of the steady state.}
    \label{fig:squeezing}
\end{figure}

To better characterize these regimes, we perform a rescaling of the parameters: $G = \tilde{G} L$ and $ \Delta = \tilde{\Delta} L$ (see Methods and Refs.~\cite{MingantPRA18_Spectral,CasteelsPRA16}). The rescaling parameter is defined such that $L=1$ corresponds to an estimated pump amplitude of $G=65.5$ kHz. 
In the experiment, the rescaling is achieved by increasing the two-photon drive amplitude and correspondingly spanning a larger region of detuning.
In Fig.~\ref{Fig:phase_diagram}(b), we compare the curves of the re-scaled steady state intracavity population $\tilde{n}_{\rm ss} = n_{\rm ss}/L$ for the same $\tilde{G}$ and range of $\tilde{\Delta}$, while increasing $L$.
The solid lines in the figure are the theoretical curves obtained by numerical simulation of the model in Eq.~\eqref{Eq:lindblad} and show an excellent agreement with the experimental data.
As $L$ increases, the emergence of a continuous but non-differentiable change in the photon number at negative detuning, and a discontinuous jump at positive detuning can be observed [see also Fig.~\ref{Fig:phase_diagram}(c)]. These are the fingerprints of second- and first-order DPTs, respectively, as also depicted in Fig.~\ref{Fig:scheme}(a).
The histograms of the measured $\hat{I}$ and $\hat{Q}$ quadratures -- i.e., the Husimi functions of the steady state convoluted by the noise of the amplifier -- are plotted Fig.~\ref{Fig:phase_diagram}(d) for the three regimes mentioned above and at the critical points. 
As the detuning increases across the second-order DPT, the vacuum becomes squeezed [see also Fig.~\ref{fig:squeezing}(d)] and then separates into two coherent-like states with opposite phase. At the first-order critical point, two coherent-like states with large photon number coexist with a vacuum-like state.

We introduce here a new procedure to characterize criticality, rooted in the theory of quantum trajectories and DPTs (see Supplementary) \cite{RotaNJP18}.
The critical nature of the system is also evident when considering the ``conjugate'' parameter of the photon number, i.e., the system's phase $\Phi(t) = \arg(I(t) + i Q(t))$. 
The probability distribution $p(\Phi)$ is reconstructed from the histogram of $I(t) = \langle \hat{I}(t)\rangle$ and $Q(t) = \langle \hat{Q}(t)\rangle$. 
Figure~\ref{Fig:phase_diagram}(e) shows histograms of $\Phi$ as a function of the rescaled detuning for a fixed value $L$. While $\Phi$ is uniformly distributed in the vacuum phase, it displays two narrow peaks in the bright phase, corresponding to the coherent-like states in Fig.~\ref{Fig:phase_diagram}(d).
We report the bimodality coefficient (binder cumulant) $B(\Phi) \equiv m_2^2/ m_4$ in Fig.~\ref{Fig:phase_diagram}(f), where the $j^{th}$ moments are $m_j = \int_{-\pi}^{\pi} p(\Phi) \Phi^j$ with $ p(\Phi)$ being the probability distribution of $\Phi$.
The transition between $B(\Phi)=5/9$ (flat distribution) and $B(\Phi) \simeq 1$ (bimodality) is smooth at the second-order and abrupt at the first-order critical points, thus reinforcing the evidence for DPTs.

\begin{figure*}
    \centering
    \includegraphics[width=\textwidth]{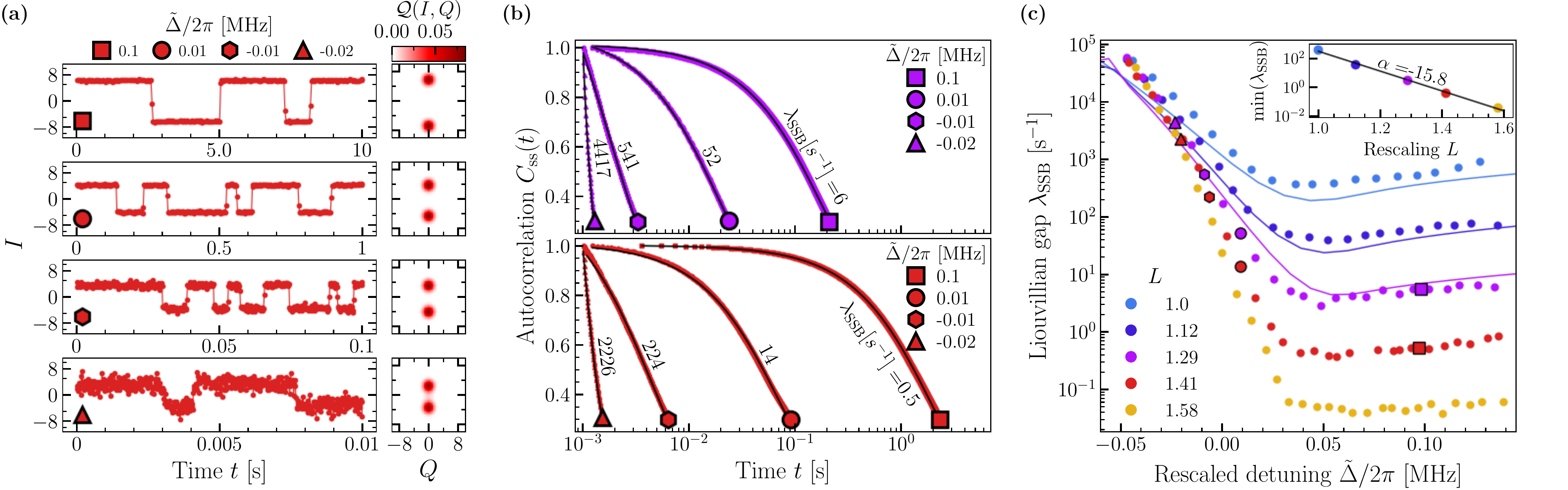}
    \caption{\textbf{Analysis of the second-order DPT.} (a) 
    A segment of the measured quantum trajectories. 
    As a function of time, we plot $I(t)$ for $L=1.41$ at various rescaled detunings $\tilde{\Delta}= \Delta/L$, indicated by the marker in each panel.
    Random jumps between two opposite values of the quadrature occur as time passes. 
    These correspond to the switches between $\rho_{\rm SSB}^{+}$ and $\rho_{\rm SSB}^{-}$ described in the main text.
    Using the entire collected signal, we recover a bimodal Husimi function shown on the right. 
    (b) The autocorrelation function $C_{\rm ss}(t)$ (see Eq.~\eqref{Eq:Correlators} and Methods), obtained from single trajectories as those shown in panel (a). 
    The markers at the end of the curves represent the values of $\tilde{\Delta}$, and the colors indicate the scaling parameter ($L=1.29$: purple, $L = 1.41$: 
    red). 
    The Liouvillian gap can be extracted from fitting these curves using Eq.~\eqref{Eq:Correlators}. The fit are represented by the black lines.
    (c) The fitted Liouvillian gap $\lambda_{\rm SSB}$ as a function of $\tilde{\Delta}$ for different scaling parameters $L$, such that $G = \tilde{G} L$ with $\tilde{G} =65.5$ kHz. 
    Points are the experimental data, while the solid lines describe the theoretical prediction obtained by diagonalizing the Liouvillian in Eq.~\eqref{Eq:lindblad}.
    The inset shows the minimum of $\lambda_{\rm SSB}$  as a function of the rescaling parameter $L$.
    The black line shows the fit of the function $\lambda_{\rm SSB} \propto \exp(\alpha L)$ to the data.
    }
     \label{Fig:second_order}
\end{figure*}

\vspace{10pt}
\tocless\subsubsection{Quantum nature of the transitions}

Quantum fluctuations play a fundamental role at both transitions. 
In fact, at the critical point of the second-order DPT, the steady state is squeezed below vacuum.
We define the squeezing parameter as the minimal variance $ \Delta x_{\phi}^2 \equiv \langle \hat x^2_\phi\rangle  - \langle \hat x_\phi\rangle^2$  of the quadrature $\hat x_\phi = (\hat ae^{-i\phi}+\hat a^\dag e^{i\phi})/\sqrt{2}$, spanning all possible $\phi$.
Figure~\ref{fig:squeezing}(d) shows the squeezing parameter as a function of the detuning (see Methods for details on its estimation). 
At large negative detuning
$ \Delta x_{\phi}^2= 1/2$ because the steady state is the vacuum. 
The minimum of the squeezing parameter is below $1/2$ (squeezing below vacuum). 
The position of this minimum closely aligns with the second-order critical point, i.e., the maximum of the second derivative of the photon number, as shown in Figs.~\ref{fig:squeezing}(a-c).
This analysis supports the claim that quantum fluctuations play an important role at the second-order DPT.

\begin{figure}[h!]
    \centering
    \includegraphics[width=\textwidth]{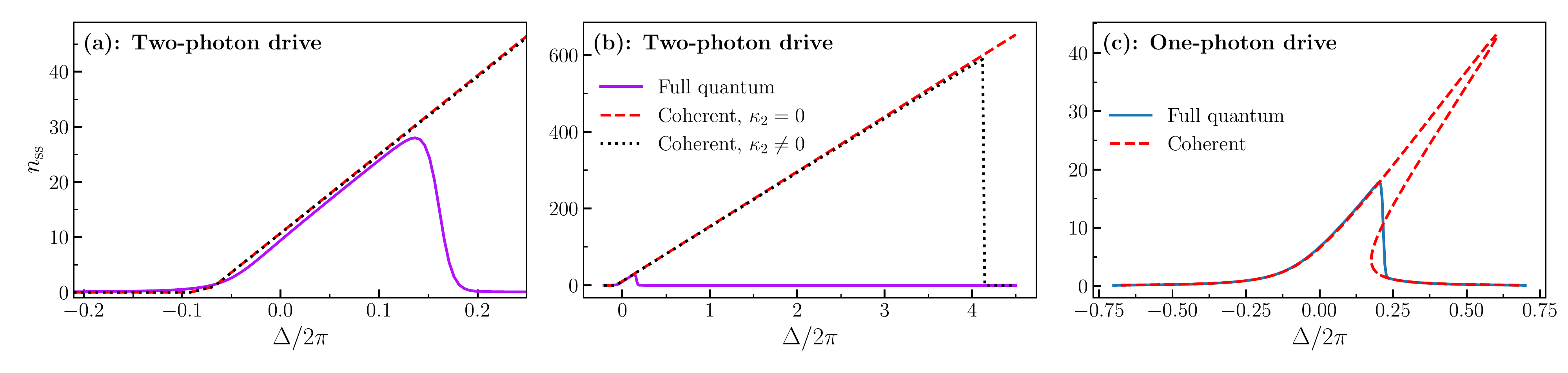}
    \caption{\textbf{First order transition in the one- and two-photon driven Kerr resonator.}
    (a-b) For $L=1.29$, we numerically compute the steady-state photon number $n_{ss}$ using Eq.(~\ref{Eq:lindblad}) and compare it to that obtained through a coherent-state approximation. In the latter case, we consider $\kappa_2 =0$ (red dashed line) and  $\kappa_2 \neq 0$ (black dotted line). While none of these approximations correctly capture the transition point, it is only for $\kappa_2 \neq 0$ that the bright phase becomes unstable at large detuning.
    In this regard, we argue that the two-photon decay plays a fundamental role in determining the metastability region.
    (c) We show here for comparison the one-photon driven Kerr resonator, with Hamiltonian $\hat{H} = - \Delta \hat{a}^\dagger \hat{a} + U/2 \hat{a}^{\dagger\, 2} \hat{a}^2 + F (\hat{a}^\dagger + \hat{a})$, with dissipator $\kappa \mathcal{D}[\hat{a}]$.
    Also in this case, the semiclassical approximation cannot capture the point of the transition.
    However, it can correctly predict the presence of the transition and the finite size of the metastability region.
    For (a,b) parameters as in the main text. For (c): $F/2\pi = 0.25$ MHz, $U/2\pi = 0.014$ MHz, and $\kappa/2\pi = 0.077$ MHz.   }
    \label{fig:Ext_data_2}
\end{figure}

In the case of the first-order DPT, the effect of quantum fluctuations is fundamental to correctly determine the transition point and the region of metastability.
In the one-photon driven Kerr resonator, first-order DPTs  have been observed across multiple platforms \cite{RodriguezPRL17,FinkNatPhys18,Brookes2021,Chen2023,Benary2022}. 
From a theoretical viewpoint, in the one-photon driven resonator the presence of metastability, and thus criticality, can be argued using a semi-classical model, i.e., assuming a coherent state, and just one-photon loss.
This is not the case for the two-photon driven Kerr resonator.
A coherent-state approximation for the equations of motion obtained by considering one-photon losses alone cannot predict the presence of this DPT. 
The region of metastability requires two-photon decays to be correctly captured by a coherent-state approximation.
Criticality can \textit{only} be theoretically obtained within a full quantum picture, as shown in Extended Data Fig.~\ref{fig:Ext_data_2}

\vspace{15pt}
\tocless\subsection{Dynamical properties}

\tocless\subsubsection{Second-order}

Having characterized the steady state critical properties, we now focus on the dynamical properties. 
A distinctive feature of second-order DPTs is spontaneous symmetry breaking (SSB) [see Figs.~\ref{Fig:scheme}(a-c)].
The Eq.~\eqref{Eq:lindblad} is invariant under the transformation $\hat{a} \to -\hat{a}$. 
This weak $Z_2$ symmetry \cite{AlbertPRA14,LeePRL13,JinPRX16} imposes constraints on steady state of the system (see Methods).
Namely, when collecting the signal, for each measured quadrature $(I, Q)$, it must be equally probable to measure $(-I, -Q)$.
As such, the presence of a $Z_2$ symmetry enforces $\langle \hat{I} \rangle_{\rm ss} =\langle \hat{Q} \rangle_{\rm ss} =0$.
SSB is defined as the presence of states $\rho_{\rm SSB}^\pm$ that, despite being stationary, do not respect the previous condition
\cite{MingantPRA18_Spectral}.
These states can only emerge in the thermodynamic limit $L\to \infty$, or for classical analogues where the number of excitations can be taken to be infinite \cite{dykman1980fluctuations,leuch2016parametric}. 
At finite values of $L$, however, the emergence of SSB is signaled by critical slowing down: $\rho_{\rm SSB}^\pm$ are not stationary, but they decay towards $\sss$ at a rate $\lambda_{\rm SSB} \ll 1/\kappa$ \cite{MingantPRA18_Spectral}, as sketched in Fig.~\ref{Fig:scheme}(c).
For the two-photon driven Kerr resonator model, $\rho_{\rm SSB}^{\pm}\simeq \ketbra{\pm \alpha}$ and $\sss = (\rho_{\rm SSB}^{+}+ \rho_{\rm SSB}^{-})/2$, where $\ket{\alpha}$ is a coherent state \cite{BartoloPRA16}.  
Theoretically, this rate corresponds to one of the Liouvillian eigenvalues (see Methods and Supplementary).

\begin{figure}[h!]
    \centering
    \includegraphics[width=\textwidth]{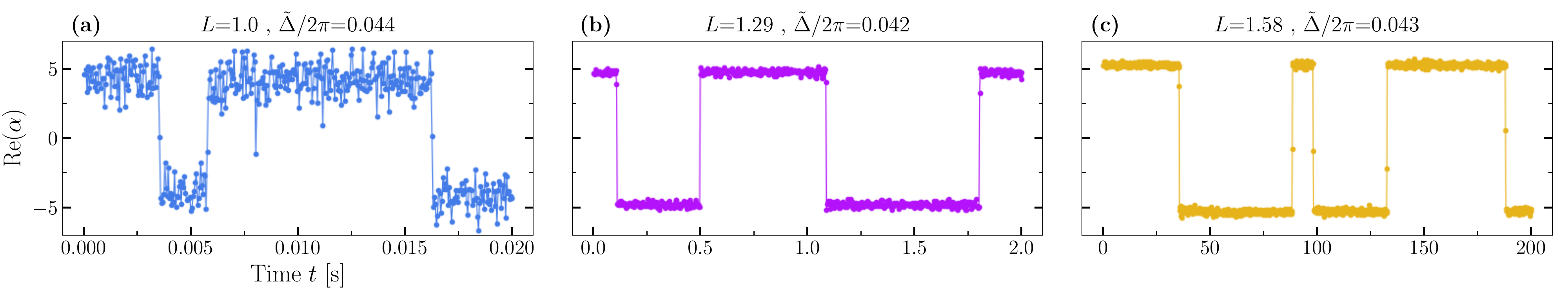}
    \caption{\textbf{Sample of the measured heterodyned signal for quantum trajectories with different \textit{L}.}
    As a function of time, we plot the real part of the cavity field (acquired through the $\hat{I}$ quadrature) for approximately the same rescaled detuning.
    The colors indicate the different values of $L$ as in Fig.~\ref{Fig:second_order} in the main text. Notice that the three panels present different $x$-scales.
    Random jumps between two opposite values of the quadrature occur as time passes and the rate of jumps is orders of magnitude slower than the photon-loss rate $\kappa$. 
    }
    \label{fig:Ext_data_3}
\end{figure}

The continuous measurements along single quantum trajectories shown in Fig.~\ref{Fig:second_order}(a) display jumps between the states $\rho_{\rm SSB}^\pm$.
Notice how the observed rate of phase jumps is significantly larger than the typical photon lifetime $1/\kappa \sim \SI{2}{\micro\second}$ and further decreases with increasing value of $L$ (see the Extended Data Fig.~\ref{fig:Ext_data_3}). 

\begin{figure}[h!]
    \centering
    \includegraphics[width=0.5\textwidth]{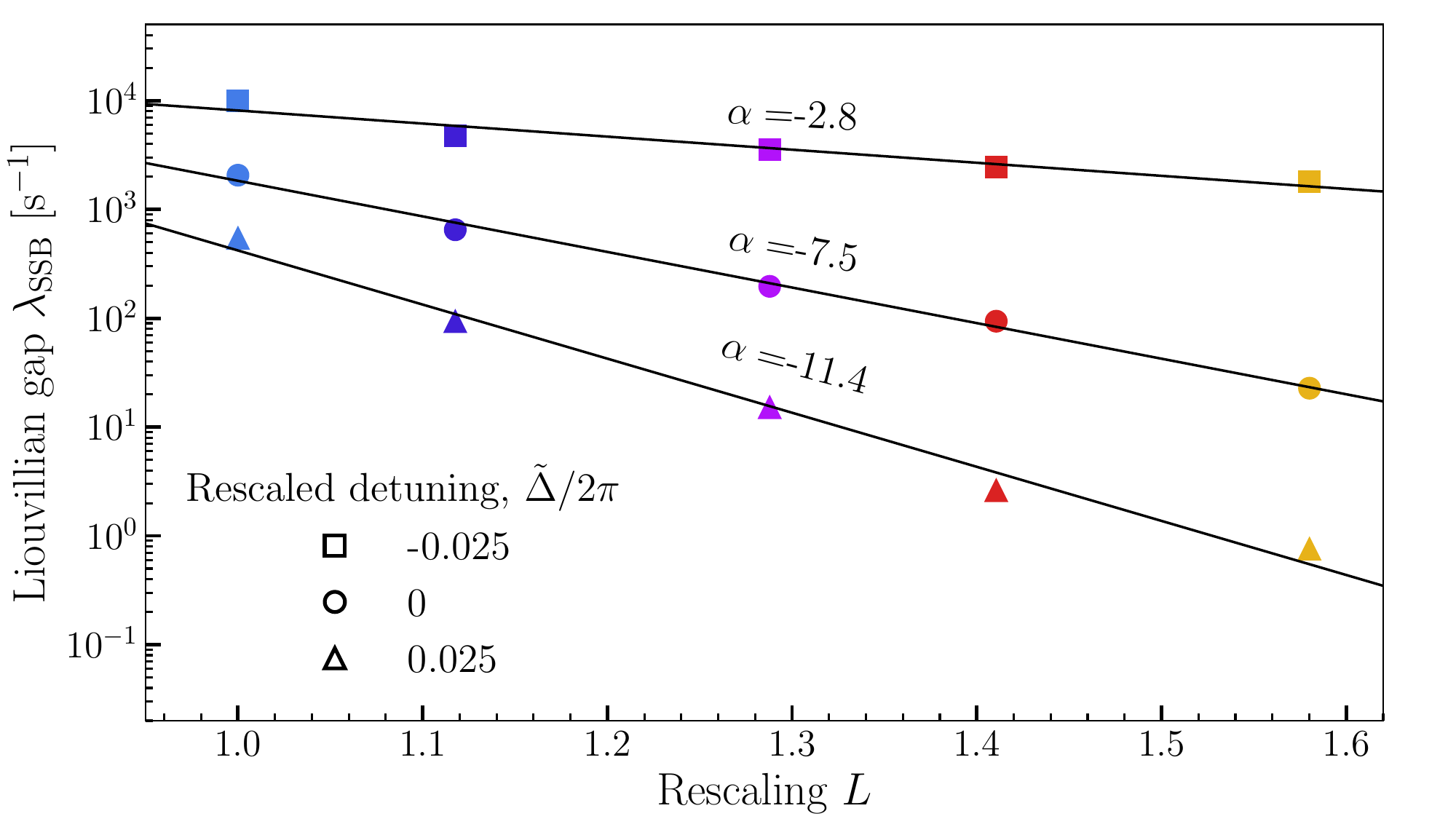}
    \caption{\textbf{Scaling of the SSB Liouvillian gap at different detuning.}
    As a function of $L$, we plot the Liouvillian gap $\lambda_{\rm SSB}$ shown in Fig.~\ref{Fig:second_order}(c) for different values of the rescaled detuning $\tilde{\Delta}$.
    The scaling rate strongly depends on the choice of $\tilde{\Delta}$, supporting the conclusions on the use of detuning as a resource for quantum computing discussed in \cite{gravina2022critical,RuizPRA23,venkatraman2023driven}.}
    \label{fig:Ext_data_4}
\end{figure}

In order to quantify the critical slowing down, we have derived a method to extract $\lambda_{\rm SSB}$ from the steady state auto-correlation function.
As proven in the Methods and Supplementary, in the limit in which critical slowing down takes place, one has
\begin{equation}\label{Eq:Correlators}
\begin{split}
    C_{\rm ss}(t) &= \lim_{\tau, \, T \to \infty} \frac{1}{T}\int_{\tau}^{\tau+T} \frac{I(\tau') I(t+ \tau')}{I^2(\tau')} d \tau' \\
    &\simeq \exp{-\lambda_{\rm SSB} t}
\end{split}    
\end{equation}
where $I(\tau)$ is the measured quadrature at time $\tau$ along a single quantum trajectory such as those shown in Fig.~\ref{Fig:second_order}(a). 
In the experiment, given the discrete nature of the signal, $C_{\rm ss}(t)$ is calculated by averaging over multiple times $\tau$ the product of $I(\tau)$  and $I(t+\tau)$.   
We plot the autocorrelation functions and their fit according to Eq.~\eqref{Eq:Correlators} in Fig.~\ref{Fig:second_order}(b).
From this, we finally obtain $\lambda_{\rm SSB}$, shown in Fig.~\ref{Fig:second_order}(c), as a function of the rescaled detuning $\tilde{\Delta}$ and for various $L$.
Remarkably, in our measurements, $\lambda_{\rm SSB}$ spans five orders of magnitude.
The numerical simulations for $\lambda_{\rm SSB}$  closely resemble the experimental data.
It is worth emphasizing that the Liouvillian eigenvalues associated with the DPTs strongly depend on the model parameters. 
This is shown in the Extended Data Fig.~\ref{fig:Ext_data_theory_wrong}.
Therefore, the validity of the model in Eq.~\eqref{Eq:lindblad} and of the chosen parameters is confirmed. By fitting the minimum of $\lambda_{\rm SSB}$ and plotting it as a function of $L$ [see inset of Fig.~\ref{Fig:second_order}(c)], we clearly see an exponential behavior, characteristic of finite-component phase transitions, indicating the presence of a true SSB in the thermodynamic limit $L\to \infty$.
Finally, notice that $\lambda_{\rm SSB}$ is associated to a bit-flip error rate in Kerr and dissipative cat qubits \cite{grimm_stabilization_2020}.  
As our results demonstrate, $\lambda_{\rm SSB}$ can be reduced by changing the detuning.
Moreover, we see that $\lambda_{\rm SSB}(\Delta, L) \propto e^{\alpha(\Delta) L}$, where $\alpha(\Delta)$ strongly depends on $\Delta$, as also shown in Refs.~\cite{gravina2022critical,venkatraman2023driven, RuizPRA23}.
This is also highlighted in greater details in the Extended Data Fig.~\ref{fig:Ext_data_4}. These observations demonstrate how criticality can be exploited for quantum information processing \cite{PRLLieu20}.

\begin{figure*}[!htb]
        \centering
        \includegraphics[width=\textwidth]{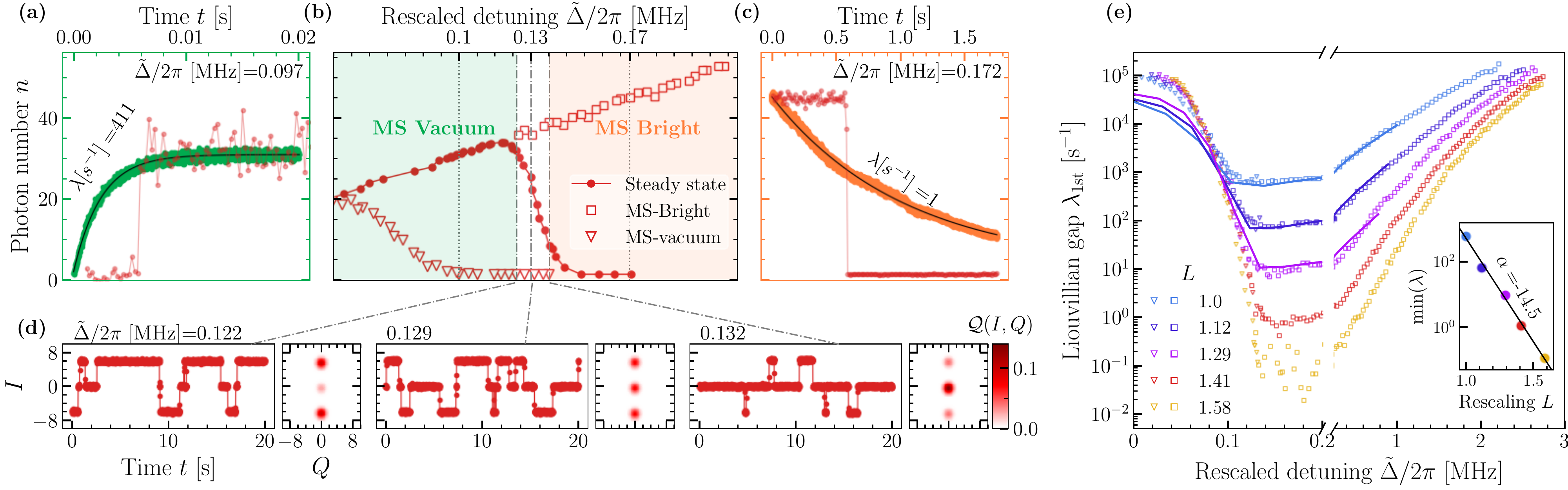}
        \caption{\textbf{Analysis of the first-order DPT.} (a-d) For $L=1.41$, metastability around the critical detuning $\Delta_c/2\pi\approx \SI{0.13}{\mega\Hz}$ where the first-order transition takes place. $\Delta_c$ corresponds to the detuning for which the first-derivative of $n_{\rm ss}$ with respect to detuning is maximal.
        (b) The photon number $n$ both in the steady state (circles) and in the metastable regimes (squares and triangles) as a function of detuning. 
        The photon number in the metastable regimes have been obtained by initializing the system 
        at $\Delta< \Delta_c$ ($\Delta>\Delta_c$)
        in the vacuum (in the high-population) phase and waiting for a time $1/\kappa$. 
        (a) For $\Delta< \Delta_c$, the system is initialized in the vacuum, and it evolves towards the bright phase. The red curve is the measured photon number along a single trajectory, while the green curve is the average over $1000$ trajectories, and is fitted by Eq.~\eqref{Eq:photon_number_time} (black line).
        (c) As in (a), but for $\Delta> \Delta_c$, where the system is initialized in the bright phase.
        (d) Phase coexistence takes place in the proximity of the critical point $\Delta \simeq \Delta_c$.
        Single trajectory display random jumps between the vacuum and the bright phase.
        From left to right, $\Delta$ increases and the relative weights of the two phases change, as it can be observed in the Husimi functions.
        (e) Liouvillian gap $\lambda_{\rm 1st}$ extrapolated using Eq.~\eqref{Eq:photon_number_time} from data similar to those in panels (a-c).
        Markers indicate the experimental data, obtained by fitting the decay from either the vacuum or the bright phase towards the steady state, while the solid lines are the results of the numerical diagonalization of the Liouvillian in Eq.~\eqref{Eq:lindblad}. The inset shows the minimum of $\lambda_{\rm 1st}$  as a function of the rescaling parameter $L$.
    The black line shows the fit of the function $\lambda_{\rm 1st} \propto \exp(\alpha L)$ to the data.
        }
         \label{Fig:first_order}
\end{figure*}

\begin{figure}[!h]
        \centering
        \includegraphics[width=\columnwidth]{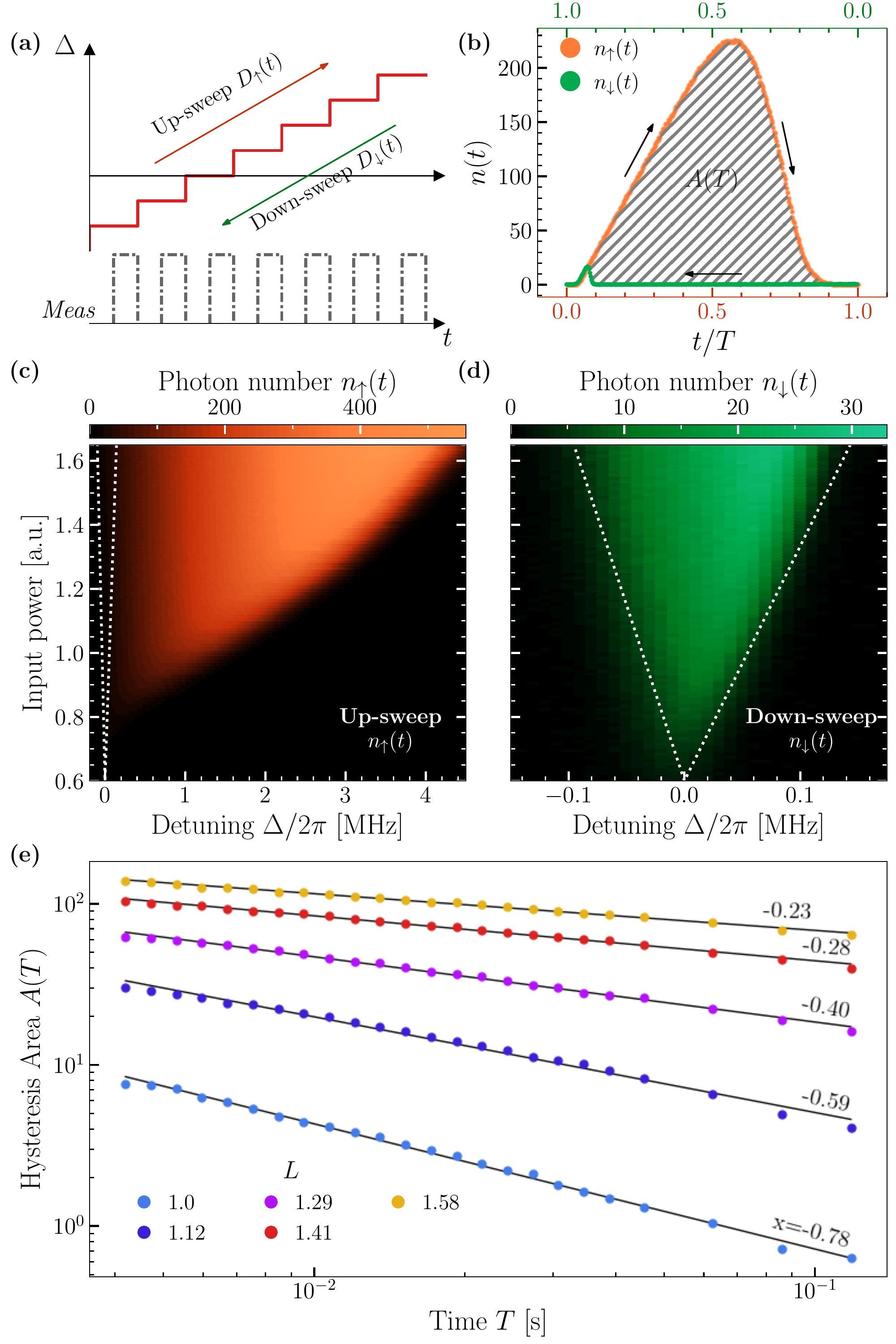}
        \caption{\textbf{Analysis of the hysteresis due to the first-order DPT.}
        (a) Schematic of the measurement protocol to obtain the hysteresis area. The up-sweep is $\Delta_{\uparrow}(t) = \Delta_{\rm min} + D \, t$, for $D = (\Delta_{\rm max} - \Delta_{\rm min})/T$. Similarly, the down-sweep is $\Delta_{\downarrow}(t) = \Delta_{\rm max} - D \, t$. 
        Details of the measurement can be found in the Supplementary.
        (b) The area of hysteresis defined in Eq.~\eqref{eq:hysteresis} for $T=3.5$ ms, $D/2\pi = \SI{1000}{\mega \Hz \per \second}$  and $L=1.41$. 
        (c) Phase diagram of the photon number for an up-sweep with $D/2\pi = \SI{1000}{\mega \Hz \per \second}$.  
        (d) As in (c), but for a down-sweep. In both (c) and (d), the white dotted line indicates the same portion of the phase diagram.       
        (e) As a function of $T$, the hysteresis area for various $L$. The black lines have been obtained by fitting the data with the power-law $A(T)\propto T^{x}$. 
        $\Delta_{\rm max}/2\pi = \SI{4}{\mega \Hz}$ and $\Delta_{\rm min}/2\pi=\SI{-0.21}{\mega \Hz}$.
        }
         \label{Fig:Hysteresis}
\end{figure}

\vspace{15pt}
\tocless\subsubsection{First-order transition}

We now focus on the dynamical properties of the first-order DPT. Similarly to the second-order DPT, criticality can only occur in the thermodynamic limit.
In the case of finite $L$, however, the emergence of DPT results again in critical slowing down, associated this time to a rate $\lambda_{\rm 1st}$.
In particular, the photon number at a given time $t$ follows \cite{MingantPRA18_Spectral}
\begin{equation} \label{Eq:photon_number_time}
    n(t) \simeq n_{\rm ss} + \delta n  \,e^{-\lambda_{\rm 1st} t},
\end{equation}
where $\delta n$ depends on the initial state. 
Following this definition, we identify three regimes, summarized in Fig.~\ref{Fig:first_order}(b), in the proximity of the critical point $\Delta_c$ of the  first-order DPT: 
(i) $\Delta<\Delta_c$ shown in Fig.~\ref{Fig:first_order}(a);
(ii) $\Delta\simeq\Delta_c$ in Fig.~\ref{Fig:first_order}(d); and
(iii) $\Delta>\Delta_c$ in Fig.~\ref{Fig:first_order}(c).

In (i), single quantum trajectories remain in the vacuum for a long time before randomly jumping to the bright phase (red curve).
Once the bright phase is reached, the system never jumps back to the vacuum.
Averaging over many trajectories (the green curve) results in $n(t)$ following  Eq.~\eqref{Eq:photon_number_time}.
We conclude that the steady state $\sss$ is the bright phase, while the vacuum is metastable with lifetime $1/\lambda_{\rm 1st}$. 
In (ii), single trajectories show that the state jumps between the bright and the vacuum phase. The relative time they spend in each of these phases determines the composition of $\sss$.
This is also evident from the Husimi functions that reflect the phase coexistence between the vacuum and the bright phase.
This region of coexistence shrinks as the thermodynamic limit is approached (not shown).
Finally, in (iii) the quantum trajectories display a jump between the bright phase and vacuum (red curve). Averaging over many trajectories (orange curve) results in an exponential decay from $\delta_n \simeq n(t=0)$ to the vacuum following Eq.~\eqref{Eq:photon_number_time}. In this regime, $\sss$ is the vacuum, and $\lambda_{\rm 1st}$ describes the decay of the bright metastable phase. 

\begin{figure}[h!]
    \centering
    \includegraphics[width=\textwidth]{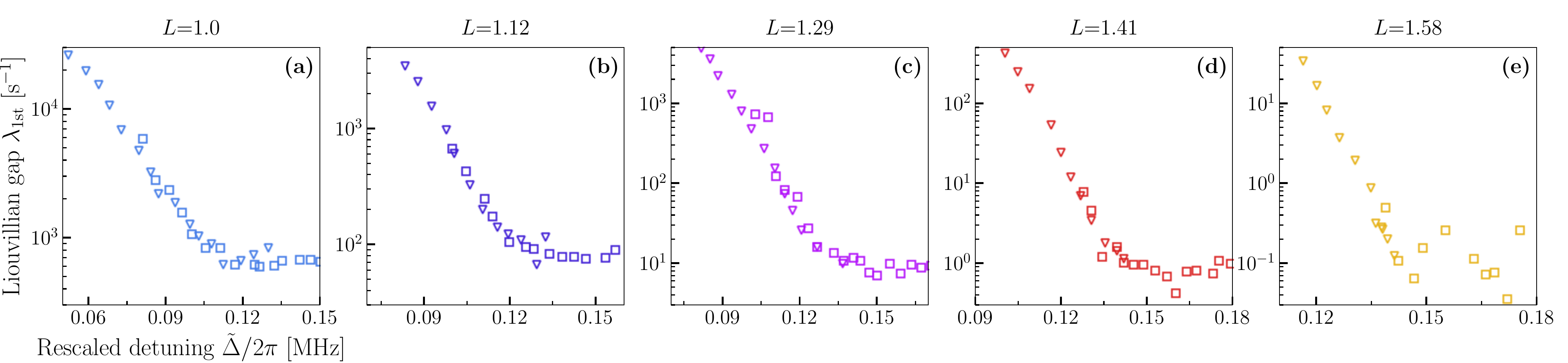}
    \caption{\textbf{Zoom on the Liouvillian gap at the first-order DPT.}
    For the same $L$ considered in the main text, and with the same color code, zoom on the minimum of Liouvillian gap shown in Fig.~\ref{Fig:first_order}.
    Both methods of extrapolating the Liouvillian gap (starting from vacuum, triangle markers;  starting from the bright phase, crosses) coincide at the minimum.    }
    \label{fig:Ext_data_5}
\end{figure}

We plot the Liouvillian gap $\lambda_{\rm 1st}$ in Fig.~\ref{Fig:first_order}(e), demonstrating the emergence of critical slowing down associated with the first-order DPT as we approach the thermodynamic limit.
Notice that both data extrapolated in the regions (1) and (3) match at the critical point, confirming the theoretical prediction of Ref.~\cite{MingantPRA18_Spectral} (see also the zoom on the region in the Extended Data Fig.~\ref{fig:Ext_data_5}).
Furthermore, as shown in the inset, we also observe an exponential dependence for the minimum of $\lambda_{\rm 1st}$ with respect to the scaling parameter $L$.

\begin{figure}[h!]
    \centering
    \includegraphics[width=\textwidth]{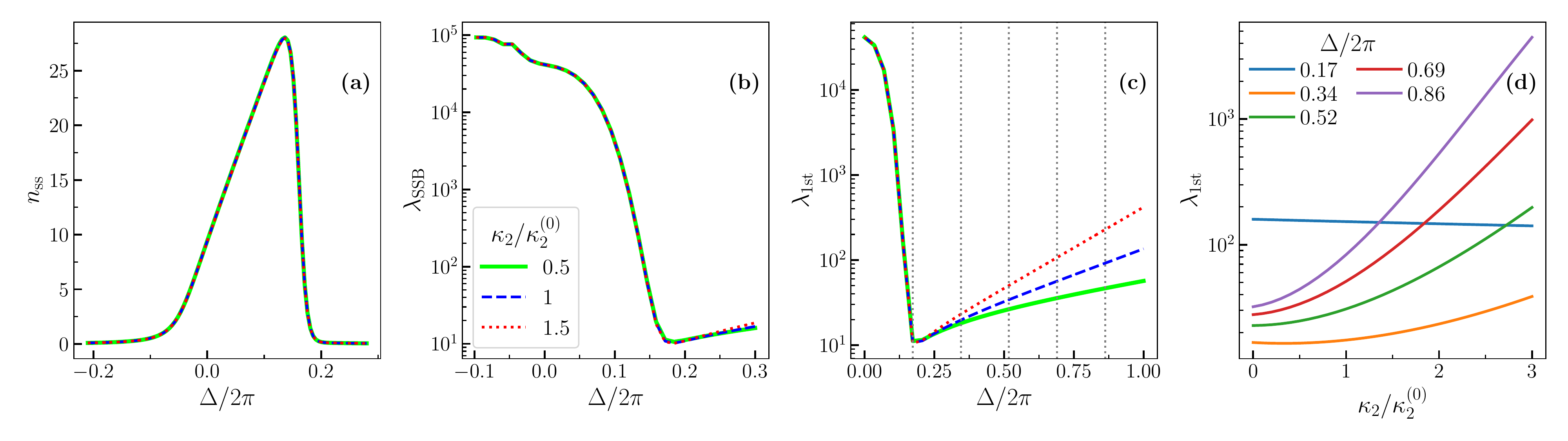}
    \caption{\textbf{Effect of two-photon dissipation.}
    For $L=1.29$, we numerically simulate the effect of changing $\kappa_2$ with respect to the original value $\kappa_2^{(0)}=78$ Hz used in all other figures.
    While (a) the photon number and (b) $\lambda_{\rm SSB}$ are only marginally affected by a change in $\kappa_2$, (c) $\lambda_{\rm 1st}$ is particularly sensitive to its value, especially at large detuning where the bright phase is metastable.
    (d) This dependence becomes more pronounced as the detuning is increased. The chosen detunings in (d) correspond to the vertical dashed lines in (c).
    We conclude that, in Kerr-dominated regimes where $\kappa_2 \ll U$, measuring the Liouvillian gap could be used as an efficient way to assess the value of $\kappa_2$.}
    \label{fig:Ext_data_6}
\end{figure}

As we previously discussed, two-photon dissipation plays a fundamental role in the correct theoretical description of the observed first-order DPT.
 While in the previous simulations shown in Figs.~\ref{Fig:phase_diagram} and \ref{Fig:second_order}, $\kappa_2$ played only a marginal role, it now determines the dependence of $\lambda_{\rm 1st}$ with respect to $\Delta$.
This is shown in greater detail in the Extended Data Fig.~\ref{fig:Ext_data_6}.
Given the sensitivity of $\lambda_{\rm 1st}$ to very small changes in the value of $\kappa_2$, measuring $\lambda_{\rm 1st}$ is a promising tool for determining  $\kappa_2$ in Kerr-cat based quantum devices \cite{grimm_stabilization_2020}.

As the critical region is characterized by metastable states, 
whose lifetime is of the order $1/\lambda_{\rm 1st}$, an hysteretic behavior in $\Delta$ is expected. 
As sketched in Fig.~\ref{Fig:Hysteresis}(a),
the detuning is ramped between $\Delta_{\rm min}$ and $\Delta_{\rm max}$, both outside the hysteresis range, according to $\Delta_{\uparrow}(t) = \Delta_{\rm min} + D \, t$, with $D = (\Delta_{\rm max} - \Delta_{\rm min})/T$ and $\Delta_{\downarrow}(t) = \Delta_{\rm max} - D \, t$, where $T$ is the sweep time.
Hysteresis is immediately visible when comparing Figs.~\ref{Fig:Hysteresis}(c) and (d).
For a quantitative description of the effects of hysteresis, we calculate the loop area [see Fig.~\ref{Fig:Hysteresis}(b)] defined as
\begin{equation}\label{eq:hysteresis}
    A(T) = \frac{\int_0^{T}\, dt \, \left[ n_{\uparrow}(t) - n_{\downarrow}(t) \right] }{T}.
\end{equation}
where $n_{\uparrow}(t)$ [$n_{\downarrow}(t)$] is the average intracavity population at time $t$ along a sweep. 
As shown in Figs.~\ref{Fig:Hysteresis}(e), by fitting the data by a power law, we find that $A(T) \propto T^{x}$.
The loop area shrinks as $T$ is increased. 
In addition, the area expands when increasing the parameter $L$. 
This proves that the hysteresis is indeed linked to $\lambda_{\rm 1st}$. Our analysis confirms the theoretical prediction \cite{CasteelsPRA16} and other experimental verifications \cite{RodriguezPRL17}.

\vspace{15pt}
\tocless\subsection{Conclusions }

We have established the occurrence of both first- and second-order dissipative phase transitions in a single superconducting Kerr resonator under parametric drive. 
This was demonstrated by conducting a comprehensive study, encompassing both static and dynamic properties of these finite-component DPTs, as we rescaled the system parameters towards the thermodynamic limit.
The scaling has been implemented by increasing the drive amplitude and correspondingly spanning a larger range of detuning.
We measured the timescales characterizing the critical slowing down of both DPTs, and developed an efficient method using autocorrelation measurements to extract these timescales.
We framed and interpreted our results within the formalism of the Liouvillian theory, showing an excellent agreement between the experiment and numerical simulations.
Our analysis unambiguously demonstrates the quantum nature of these critical phenomena, showing that quantum fluctuations and quantum dissipative processes are the main drive of the observed transitions. 
This control of the critical dynamics of a finite-component solid-state device paves the way to the technological applications of critical phenomena.
In particular, it serves as proof of concept towards the use of criticality and cat states for noise-biased bosonic codes~\cite{gravina2022critical}, and it lays the foundation for the realization of dissipative critical quantum sensors~\cite{di2023critical}.

\vspace{15pt}
\begin{center}
{ \textbf{ACKNOWLEDGEMENTS}}
\end{center}

The authors thank Alberto Biella, L\'eo P. Peyruchat, Marco Scigliuzzo, and Gianluca Rastelli for the stimulating discussions, Alberto Mercurio for the support in developing the codes for the numerical simulations, and Davide Sbroggio for helping in the fabrication process.
P.S. acknowledges support from the Swiss National Science Foundation through Projects No. 206021\_205335 and from the SERI through grant 101042765\_SEFRI MB22.00081.
V.S. acknowledge support by the Swiss National Science Foundation through Projects No. 200020\_185015 and 200020\_215172, 
P.S. and V.S. acknowledge support from the EPFL Science Seed Fund 2021 and 
Swiss National Science Foundation project
UeM019-16 - 215928.
R.D. acknowledges support from the Academy of Finland, grants no. 353832 and 349199.
\vspace{10pt}

\noindent \textbf{Contributions}\\

These authors contributed equally: Guillaume Beaulieu and Fabrizio Minganti.

F.M., R.D., S.Fe., and P.S. devised the research project. G.B., S.Fr., and P.S. designed the experiment. G.B. and S.Fr. fabricated the device. G.B. performed the measurements. F.M. numerically simulated the model. G.B. and F.M. analyzed the data. R.D. and S.Fe. derived the theory for the measurement and moment reconstruction. P.S. and V.S. supervised the experimental and theoretical parts of the project, respectively. All authors contributed to the writing of the paper.

\tocless\section{Methods}

\tocless{\subsection*}{Fabrication and setup}

The device is made of a \SI{150}{\nano\meter} thick aluminium layer deposited by e-beam evaporation on a \SI{525}{\micro\meter} thick silicon substrate. The coplanar waveguides are fabricated by photolithography followed by wet etching. The \SI{6.42}{\milli\meter} long CPW resonator is grounded through two Al/AlOx/Al Josephson junctions of area \SI{0.56}{\micro\square\meter}, forming a SQUID.
 The junctions were fabricated by e-beam lithography and deposited using double-evaporation technique inside a Plassys MEB550SL system. The participation ratio $\gamma$  of the SQUID nonlinear Josephson inductance over the bare cavity inductance is $\gamma= \SI{3.13e-2}{}$. The finished device is bonded using Al wire to a custom printed circuit board, which is screwed to a copper mount anchored at the mixing chamber stage of a dilution refrigator with base temperature of \SI{10}{\milli\kelvin}.  Two high-permeability magnetic shields protect the sample against external magnetic fields.

During the measurement, a NbTi coil placed underneath the sample provides a constant DC flux bias of $F=\phi_{ext}/\phi_0=\pi/6$. Under this static field, the resonance frequency and Kerr nonlinearity are of $\omega_r/2\pi=\SI{4.3497}{\giga\Hz}$ and  $U/2\pi=\SI{7}{\kilo\Hz}$. The internal and external photon loss rate, originating from the coupling to the feedline and other spurious baths are respectively of $\kappa_{ext}/2\pi=\SI{60}{\kilo\Hz}$ and $\kappa_{int}/2\pi=\SI{17}{\kilo\Hz}$.
All of these parameters are extracted by fitting the measured scattering coefficients using input-output relations.

A detailed description of the device, the fabrication process, the experimental setup, and the pulse sequence used in the measurements can be found in the Supplementary.

\vspace{15pt}
\tocless{\subsection*}{Parameters estimation}

In addition to the measured parameters, we need to quantify the pump amplitude $G$, the dissipative rates $\kappa_2$ and $\kappa_{\phi}$, and the number of thermal photons $n_{\rm th}$ to model our system.
$\kappa_2$ is a two-photon dissipation rate, arising through the same processes that convert the incoming pump tone into the two-photon drive \cite{Carmichael_BOOK_2}.
$\kappa_{\phi}$ is the dephasing rate mainly due to the nonlinearity of the resonator.
To estimate these parameters, we explore the parameter space though a simulated annealing algorithm, and then we search for the parameters that better fit the experimental data for photon number, $\lambda_{\rm 1st}$, and $\lambda_{\rm SSB}$.
Details on this procedure can be found in the Supplementary.

When estimating $G$, its initial guess has been obtained by measuring the steady state photon number $n_{ss}$ as a function of the detuning $\Delta$. In the mean-field approximation $n_{\rm ss}$ is given by
\begin{equation}
    n_{\rm ss} = \frac{\Delta-\sqrt{\left | G \right |^2-(\kappa+\kappa_{\phi})^2}}{\sqrt{U^2 + \kappa_2^2}}.
\end{equation}
We experimentally and theoretically find that such an approximation is valid far from the transition points.
In a regime where $\kappa_{\phi}\ll \kappa$, $G$ can thus be easily estimated by extrapolating $n_{\rm ss}(\Delta)$ to the x-intercept ($n_{\rm ss}(\Delta)=0$), even without knowing the Kerr nonlinearity. 
We find that the value of $G$ obtained via annealing simulation is within a few $\%$ of the initial guess.

\vspace{15pt}
\tocless{\subsection*}{Acquisition of the signal}

In the experiment, we measure $I_m$ and $Q_m$, demodulated at half of the pump frequency, of the emitted field using time-resolved heterodyne detection. Each acquired point is the integrated signal over a time interval $\tau_{int}$ varying from 2 to \SI{50}{\micro\second}, depending on the necessary time resolution. Given the timescale of the process and the desired accuracy, a single quantum trajectory is constructed by concatenating $10^{3}$ to $10^7$ quadrature measurements obtained sequentially. From the measured $I_m$ and $Q_m$, the quadratures of the intracavity field ($I$,$Q$) are obtained by removing the effect of the amplification chain and its associated noise. These correspond to the measure of $\hat{I} \equiv (\hat{a}^\dagger +\hat{a})/{2} $ and $\hat{Q} \equiv i (\hat{a}^\dagger -\hat{a})/{2}$.

Having obtained the $I$ and $Q$ data, we can reconstruct the physics of other observable from the higher-order moments.
From a theoretical perspective, this procedure is equivalent to constructing the probability $p(I,Q)$ from the measured quadratures, and then computing the  moments of this distribution.
For instance, we estimate a generic $\langle \hat{I}^m\rangle = \iint \, d {I} \, d {Q} \, p(I,Q) \, I^m$.
Notice that, as the measurement acquisition time is finite, the estimation of $p(I,Q)$ may be inaccurate in regimes characterized by rapid fluctuations.

\vspace{15pt}
\tocless{\subsection*}{Thermodynamic limit}

The thermodynamic limit for the two-photon driven Kerr resonator has been discussed in details in Ref.~\cite{BartoloPRA16}. 
Considers a lattice of  $L$ coupled two-photon resonators, described by the Hamiltonian $\hat{H} = \sum_j \hat{H}_j +\sum_{i, j}  \hat{H}_{i,j} $, with $\hat{H}_j  = \Delta \hat{a}^\dagger_j \hat{a}_j + U \hat{a}^\dagger_j \hat{a}^\dagger_j \hat{a}_j \hat{a}_j /2 + G  (\hat{a}^\dagger_j\hat{a}^\dagger_j + \hat{a}_j\hat{a}_j )/ 2$ and
$\hat{H}_{i,j} = J \hat{a}_i^\dagger \hat{a}_j +{\rm h.c.} $. Re-writing the Hamiltonian using the Fourier modes, keeping only the mode $\hat{a}_{0} = \hat{a}_{k=0}$, and fixing the Kerr nonlinearity $U$ as the unity of the model, results in
\begin{equation}
\begin{split}
    \hat{H}_{k=0} = & (\Delta -2 J) \,L \, \hat{a}^\dagger_0 \hat{a}_0 + \frac{U}{2} \hat{a}^\dagger_0 \hat{a}^\dagger_0 \hat{a}_0 \hat{a}_0 \\ &+ \frac{G L}{2}  \left(\hat{a}^\dagger_0\hat{a}^\dagger_0 + \hat{a}_0\hat{a}_0 \right).
\end{split}    
\end{equation}
This leads, up to a shift in the detuning, to a rescaling of the single resonator Hamiltonian.
Similarly, the photon loss term scales as $\kappa \to L \kappa$, while $\kappa_{\phi}$ and $\kappa_2$ remain unchanged.
Scaling the parameter $L$ in the single resonator thus mimic the scaling of the uniform $k=0$ mode of a lattice of $L$ resonators towards the thermodynamic limit.

In the experiment, we re-scale $\Delta$ and $G$, but not $\kappa$. 
As the data demonstrate, $\kappa$ plays only a marginal role in determining the proprieties of the second-order DPT and of the bright phase.
Indeed, $\Delta \gg \kappa$ at the second-order critical point. 
However, $\kappa$ plays a more significant role in determining the critical point for the first-order DPT \cite{gravina2022critical}.

\vspace{15pt}    
\tocless{\subsection*}{Symmetry and Liouvillian eigenvalues}
The equation of motion remains unchanged upon the transformation $\hat{a} \to -\hat{a}$, thereby establishing the model's invariance under the $Z_2$ symmetry.
The presence of this weak  $Z_2$ symmetry can be formalized through the action of the parity operator $\hat{\Pi} = \exp{i \pi \hat{a}^\dagger \hat{a}}$. 
Indeed, the steady state is such that $\sss = \hat{\Pi} \sss \hat{\Pi}$ \cite{AlbertPRA14}.
In a phase-space representation, this condition translates to $\sss$ being symmetric upon a point reflection with respect to the origin, as one clearly sees in Fig.~\ref{Fig:second_order}(a) where
$\sss \simeq (\ketbra{\alpha} + \ketbra{-\alpha})/2$.

As a consequence of the symmetry, the Liouvillian, represented as a matrix \cite{MingantPRA18_Spectral}, has a block-diagonal structure, with two independent blocks, $\LL_1$ and $\LL_2$.
By diagonalizing the Liouvillian, we obtain its eigenvalues $\lambda_j$ and eigenoperators $\rho_j$. In particular, the dynamics of any state can be recast as
\begin{equation}\label{Eq:eigendecomposition_rho}
\rho(t) = \sss + \sum_j c_j e^{-\lambda_j t} \rho_j,
\end{equation}
where the coefficients $c_j$ depend only on the initial state of the system.

Within this picture, we can directly assign a precise meaning to all the states and rates discussed in the paper.
When we diagonalize $\LL_1$, we find the eigenvalue $\lambda_0 =0$ associated with the steady state $\sss \equiv \rho_0$.
We then see that, in the critical region, a second eigenvalue $\lambda_{\rm 1st}$ approaches zero. 
Using Eq.~\eqref{Eq:eigendecomposition_rho}, one can demonstrate that
\begin{equation}
n(t) = n_{\rm ss} + \sum_j e^{-\lambda_j t} \delta n_j 
\xrightarrow[]{t \gg 1/\kappa} n_{\rm ss} + \delta n \, e^{-\lambda_{\rm 1st} t},
\end{equation}
where $\delta n = c_{\rm 1st} \operatorname{Tr}(\rho_{\rm 1st} \hat{a}^\dagger \hat{a})$, $\rho_{\rm 1st}$ being the operator associated with $\lambda_{\rm 1st} $.
This formula is Eq.~\eqref{Eq:photon_number_time}, used to extrapolate the Liouvillian eigenvalue associated with the critical slowing down due to the first-order DPT.

The physics of SSB is theoretically described by $\LL_2$.
We call $\lambda_{\rm SSB}$ the eigenvalue of $\LL_2$ whose real part is closest to zero. 
An unambiguous signature of critical slowing down and SSB is then given by the observation $\lambda_{\rm SSB} \to 0 $ as $L$ is increased.
$\lambda_{\rm SSB}$ describes the rate at which
$\rho_{\rm SSB}\simeq \ketbra{\pm \alpha}$ evolves towards $\sss$.
Experimentally, it corresponds to the rate at which the system jumps between the states of opposite phase.

\vspace{15pt}
\tocless{\subsection*}{Autocorrelation function}

A convenient way to extract $\lambda_{\rm SSB}$ in the bright phase [see Fig.~\ref{Fig:phase_diagram}(b)] is 
to measure the system along a single quantum trajectory $\ket{\psi_n(t)}$, representing a single realization of the heterodyne measurement. 
The autocorrelation is defined as
\begin{equation}
    C_n(\tau, t) = \left|\frac{
    \expval{\psi_n(\tau)|\hat{a}|\psi_n(\tau)} \expval{\psi_n(t+\tau)|\hat{a}|\psi_n(t+\tau)}}{
    \expval{a}_{\rm ss}^2
    } \right|.
\end{equation}
At each time $\tau$, the system is in one of the states $\ket{\pm \alpha}$, and the rate at which it jumps to the opposite state is $\lambda_{\rm SSB}$.
Averaging over several trajectories (ideally, infinitely many) taken at a long-enough time $t \gg 1/\kappa$ and for $\tau \gg 1/\kappa$, the steady state correlation function is then defined as
\begin{equation}\label{Eq: autocorellation}
    C_{\rm ss}(t) = \sum_{n=1}^{N} \frac{C_n(\tau \gg 1/\kappa, t)}{N} \simeq \exp{-\lambda_{\rm SSB} t}.
\end{equation}
The last equality is proved in the supplementary, and it is rooted in a quantum trajectories interpretation of the Liouvillian dynamics, and of its symmetries.
Given the ergodic nature of the system discussed in the Supplementary,  $C_{\rm ss}(t)$ can also be computed along a very long trajectory $\ket{\psi_{1}(t)}$. In this case, 
\begin{equation}
    C_{\rm ss}(t) =\frac{1 } {T}\int_{0}^{T \gg 1/\kappa} d t' C_{1 \rm traj}(\tau + t' , t)
\end{equation}
This procedure provides the advantage of isolating the timescale of SSB. As such, this technique is a straightforward way to measure the Liouvillian eigenvalue.

\clearpage

\onecolumngrid
\ExtendedDataFigure

\onecolumngrid

\begin{center}
\textbf{\Large Extended Data Figures for:\\
\bigskip
Observation of first- and second-order dissipative phase transitions in a two-photon driven Kerr
resonator}
\end{center}

\begin{figure}[h!]
    \centering
    \includegraphics[width=\textwidth]{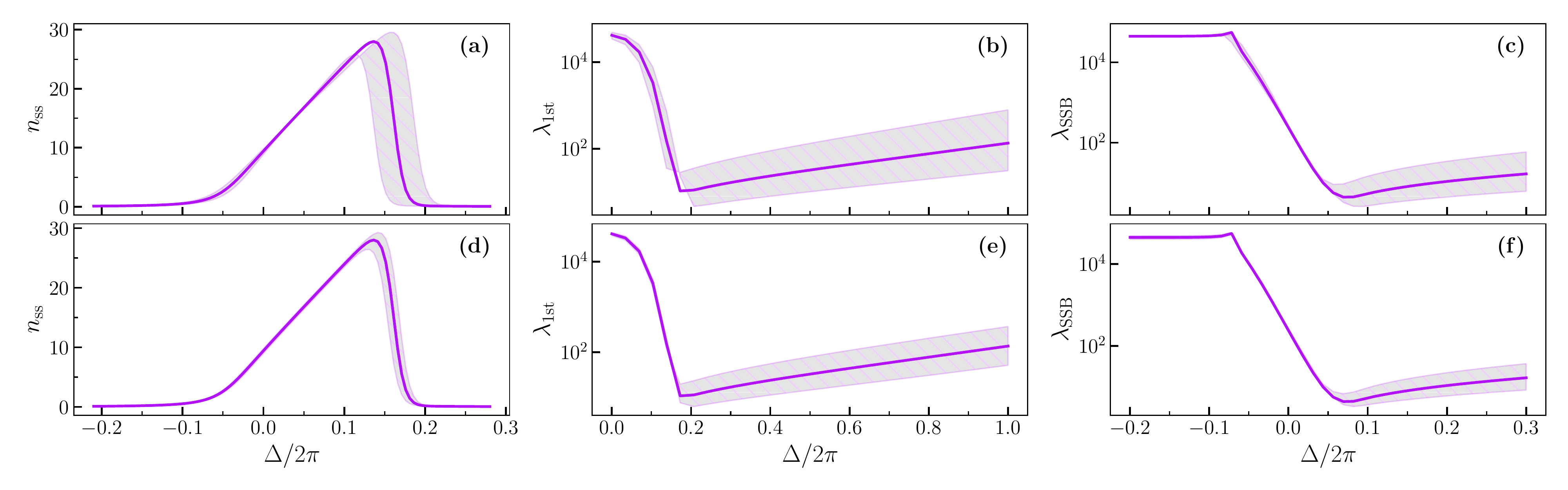}
    \caption{\textbf{Numerical results upon a small change of the parameters.}
    In this figure we use Eq.~\eqref{Eq:lindblad} to derive (a,d) The photon number; (b,e) The Liouvillian eigenvalue $\lambda_{\rm 1st}$; and (c,f) The Liouvillian eigenvalue $\lambda_{\rm SSB}$. 
    The solid line indicates the results obtained for parameters  used Figs.~\ref{Fig:phase_diagram},~\ref{Fig:second_order}, and \ref{Fig:first_order} and $L=1.29$.
    The shaded area encloses the numerical results: (a-c) for $U$ and  $G$  are both increased and decreased by 10\% relative to the values used to compute the solid line; (d-f) same for $\kappa$.
    Far from the transition, these variations do not significantly impact the data. However, they lead to noticeable differences in the critical region, particularly in the case of the Liouvillian eigenvalues.
    }
    \label{fig:Ext_data_theory_wrong}
\end{figure}

\clearpage

\onecolumngrid

\SupplementaryMaterials

\begin{center}
\textbf{\Large Supplemental Material for:\\
\bigskip
Observation of first- and second-order dissipative phase transitions in a two-photon driven Kerr
resonator}
\end{center}

\tableofcontents

\section{Setup and device}

\subsection{Device design}

\begin{figure*}[!htb]
        \centering
        \includegraphics[width=0.8\textwidth]{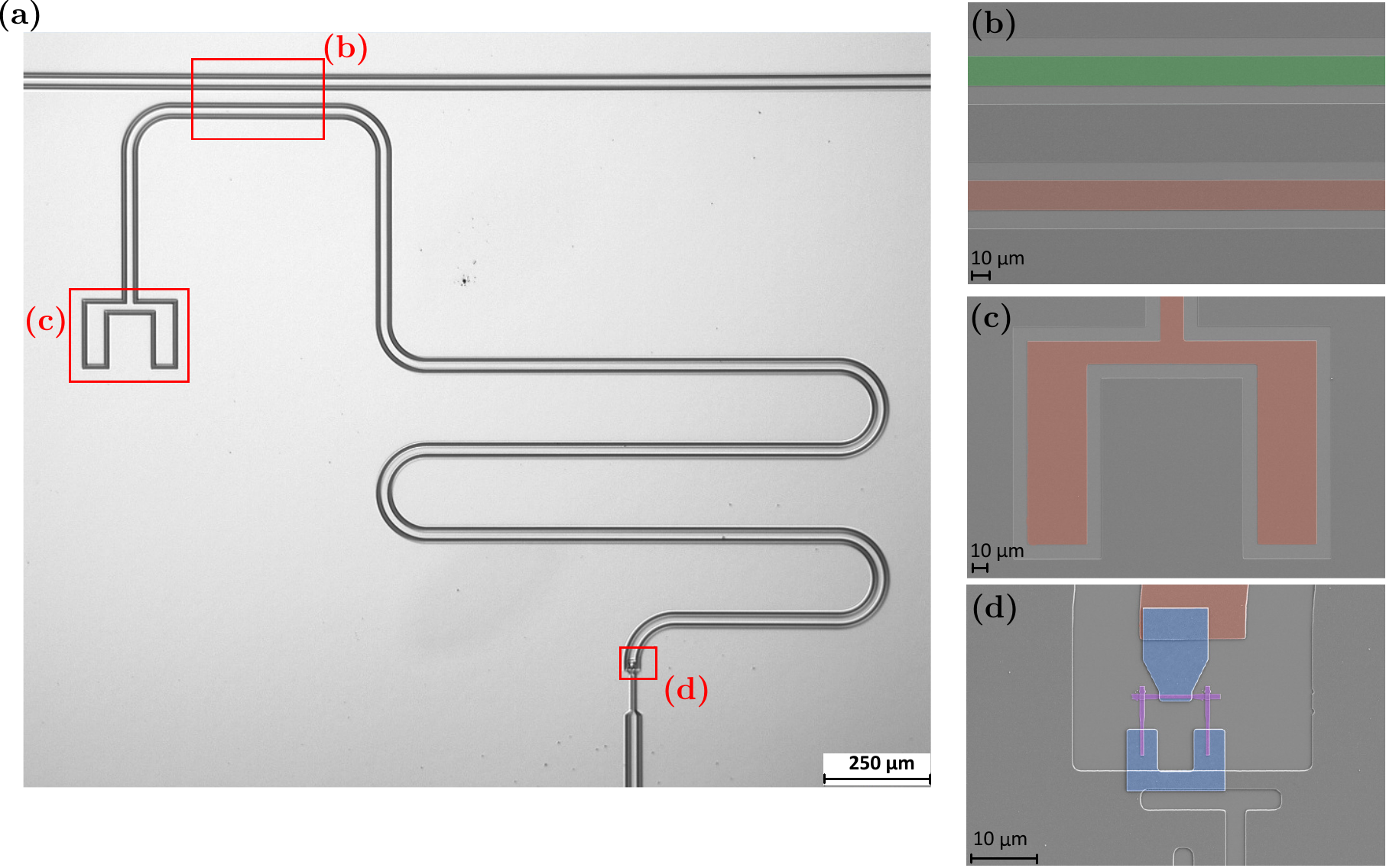}
        \caption{\textbf{Optical and SEM images of the device} (a) Micrograph of the device: a $\lambda/4$ resonator terminated by a SQUID on one end and capacitively coupled to a feedline on the other. The three regions indicated by the red squares correspond to the locations where SEM images were taken. (b) Capacitive coupling between the feedline (green) and the resonator (red). The spacing between the centers of the two coplanar wave guides is \SI{70}{\micro\meter}. (c) Open end of the resonator (red), designed for coupling to other elements in future experiments. (c) SQUID formed by two Josephson junctions (purple) galvanically connected to the resonator (red) by a patch (blue). The upper part of the fluxline is visible beneath the lower patch. 
        }
         \label{Fig:device}
\end{figure*}

Figure \ref{Fig:device}(a) shows an optical micrograph of the sample used in the experiment. The main component of the circuit is \SI{6.18}{\milli\meter} long $\lambda/4$ resonator, made flux-tunable by terminating one end to ground via a DC superconducting quantum interference device (SQUID) \cite{dykman2012fluctuating}. The SQUID consists of two identical tunnel junctions designed to have an area of (0.75 x 0.75)\SI{}{\micro\square\meter}. To drive the resonator parametrically, an L-shaped flux line is inductively coupled to the SQUID [see Fig.~\ref{Fig:device}(d)]. As shown Fig.~\ref{Fig:device}(b), the other end of the resonator is capacitively coupled to a feedline, which in our measurements is used solely to collect the emitted signal. After conducting the measurements, the feedline is repurposed for sending a single photon-drive to extract the device parameters, as discussed in Supplementary Sect.~\ref{section: supplementary experimental setup}. All waveguides are coplanar with a \SI{16.47}{\micro\meter} wide centerline separated from the ground plane by a \SI{10}{\micro\meter} gap, resulting in a characteristic impedance of approximately 50 $\Omega$.

\subsection{Device fabrication}

The waveguides are made of a \SI{150}{\nano\meter} thick aluminium layer deposited by e-beam evaporation (rate of \SI[per-mode = symbol]{0.2}{\nano\meter\per\second} under a vacuum of $\sim$\SI{1.1e-6}{} Torr) onto a \SI{525}{\micro\meter} thick silicon substrate. Prior to Al deposition, the substrate is thoroughly cleaned of organic residues using a piranha solution, and the native oxide is fully etched using 1\% hydrofluoric acid (HF). During the first patterning step, alignment marks are defined by a photolithography and lift-off process. These marks are deposited by e-beam evaporation of a \SI{5}{\nano\meter} thick layer of Ti followed by a 55 nm thick layer of Pt. Next, the waveguides are defined via photolithography and wet etching (2min30s in TechniEtch Alu80 at \SI{27}{\degreeCelsius}). The Al/AlOx/Al Josephson junctions forming the SQUID are fabricated through the following steps: first, a bilayer resist (\SI{500}{\nano\meter} of MMA EL9 and \SI{450}{\nano\meter} of PMMA 495k A8 developed in a 3:1 MiBK solution for 2min) is exposed using e-beam lithography, then Al is e-beam evaporated using the double-angle technique under ultra-high vacuum (UHV) inside a Plassys MEB550SL3, and finally, a lift-off procedure is performed. The top and bottom Al layers of the junctions are deposited at a rate of \SI[per-mode = symbol]{0.5}{\nano\meter\per\second} to achieve respective thicknesses of \SI{50}{\nano\meter} and \SI{120}{\nano\meter}. 
The junction barrier is grown during a static oxidation step carried out under a pressure of 0.15 Torr in pure O$_2$ atmosphere for 10 minutes. To connect the SQUID to the resonator and to the ground plane, a \SI{200}{\nano\meter} thick Al patch is deposited by e-beam evaporation (deposition rate of \SI[per-mode = symbol]{0.5}{\nano\meter\per\second}) and patterned by e-beam lithography and lift-off [see Fig.~\ref{Fig:device}(d)]. To ensure a good electrical contact between the patch and the resonator/junctions, in situ Ar ion plasma milling is used to remove the native Al oxide. Finally, the substrate is diced into chips of size 4 x \SI{7}{\milli\meter} using a nickled bonded diamond blade. The chip is then bonded with Al wire to a custom printed circuit board, which is screwed to a copper mount. To prevent slotline modes, bridge bounds connecting the ground plane across the chip are also added.

\subsection{Experimental setup}\label{section: supplementary experimental setup}

\begin{figure*}[h]
        \centering
        \includegraphics[width=\textwidth]{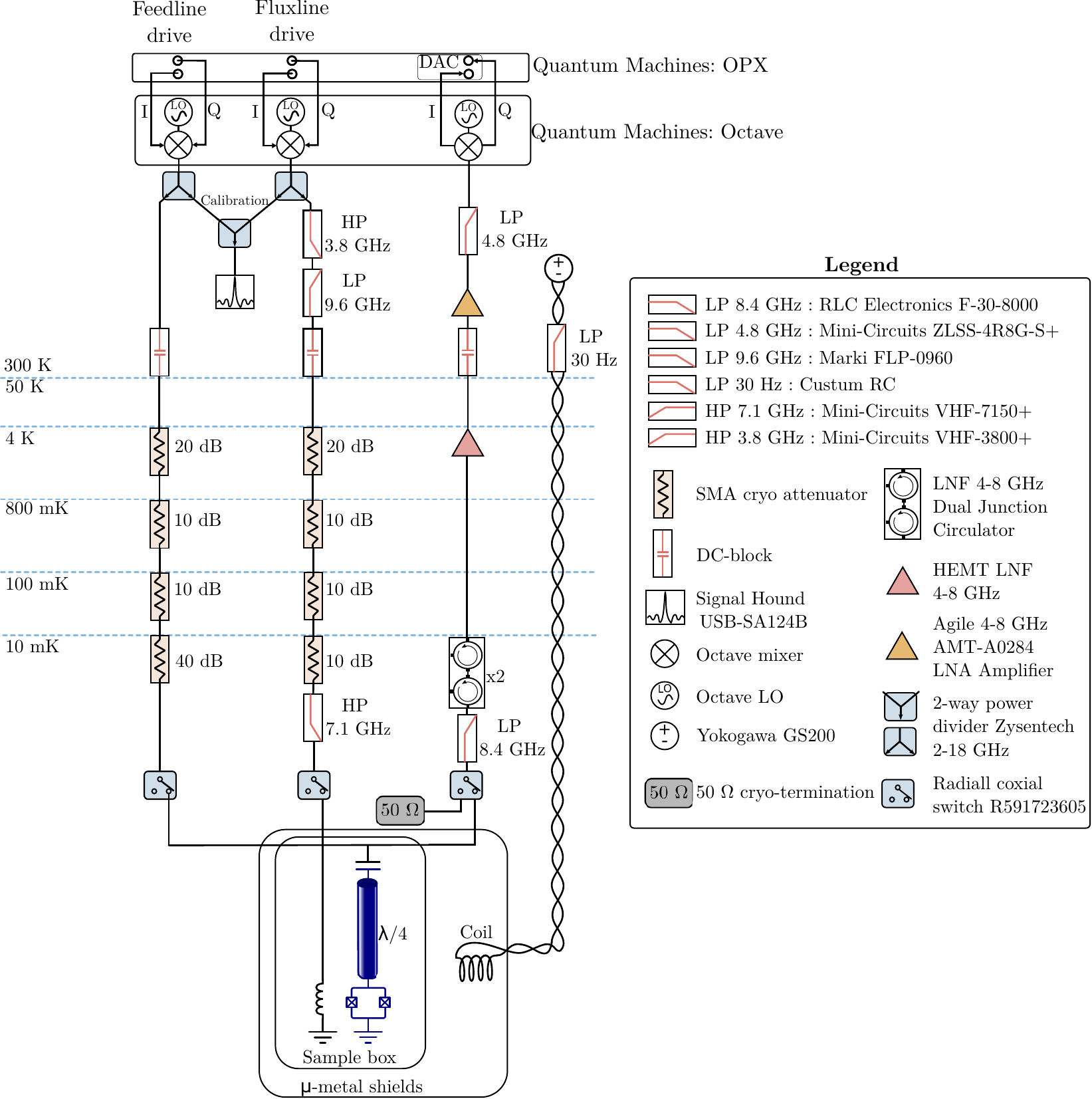}
        \caption{\textbf{Schematic of the experimental setup.} A two photon-drive is generated by sending a signal to the flux line, which modulates the magnetic flux in the SQUID at frequency $\omega_p \sim 2\omega_r$. The driving field in the fluxline is generated by mixing (upconversion) the intermediate frequency signals $I$ and $Q$ from the OPX+ with one of the octave's local oscillator (set at \SI{8.6}{\giga\Hz}). The parametric excitation of the cavity results in the emission of a signal at $\omega_p/2$, which is subsqeuntly amplified, and filtered to remove any component at $\omega_p$. The output signal is mixed (downconversion) with another local oscillator of the octave (set at \SI{4.3}{\giga\Hz}) to obtain the two quadratures $I$ and $Q$ at an intermediate frequency. These intermediate frequency signal are demodulated and integrated within the OPX+ over a time interval $\tau_{int}$, resulting in a single pair of $I$ and $Q$ expressed in volts. }
         \label{Fig:experimental set up}
\end{figure*}

The packaged sample is mounted in a high-purity copper enclosure which is thermally anchored at the mixing chamber stage of a BlueFors dilution refrigerator with a base temperature of \SI{10}{\milli\kelvin}.  To generate a DC flux bias on the sample, a coil made of NbTi wire is screwed underneath the sample holder. Two high permeability metal cans provide shielding against external magnetic fields. A schematic of the cryogenic and room temperature measurement setup is shown in Fig.~\ref{Fig:experimental set up}. An OPX+ and Octave modules are used to generate the single-photon (at $\omega \sim \omega_r)$ and the two-photons drives (at $\omega_p \sim 2\omega_r)$ respectively directed to the feedline and the pump line. These signals are produced by modulating the octave’s local oscillators with the I/Q signals originating from the OPX+. Throughout the experiment, the local oscillator used for the single-photon drive remains off to prevent any leakage field in the feedline. It is turned on only before/after the measurements are completed to extract the device parameters. After exiting the Octave, the two drives are split using a 2-way power divider ZSPD-20180-2S. Half of the signal enters the fridge, while the other half is directed to a spectrum analyzer (Signal Hound USB-SA124B) to monitor the drive amplitude and compensate for any drift. The two input lines have 20 dB, 10 dB, and 10 dB attenuators respectively positioned at the 4 K, 800 mK, and 100 mK stages.  At the base plate, the input of the feedline has a 40 dB attenuator to limit the Kerr shift, whereas the pump line has a 10 dB attenuator. Multiple filters are incorporated along the pump line to eliminate higher or lower harmonics of the driving field. The output signal, collected via the feedline, passes through two circulators (LNF 4-8 GHz Dual Junction Circulator) and travels in a NbTi low-loss superconducting line before being amplified by a 4-8 GHZ LNF High-Electron-Mobility Transistor (HEMT)  amplifier at the 4K plate. The output signal is further amplified at room temperature using a low noise amplifier (Agile AMT-A0284) before being demodulated in the Octave and digitized in the OPX+. Additionally, two filters are also placed along the output path to eliminate any signal coming from the flux line at $\omega_p$. The switches positioned at the base plate are used to connect to different devices inside the fridge. Additionally, the switch on the output line is also connected to a \SI{50}{\ohm} cryogenic termination which is used to perform a Planck spectroscopy experiment (see Supplementary Sec.~\ref{sect: supplementary calibration}). To set a static flux bias, a DC source (Yokogawa GS200) is connected to the coil attached under the sample. 

\subsection{Characterization of the device parameters}

\begin{figure*}[h]
        \centering
        \includegraphics[width=\textwidth]{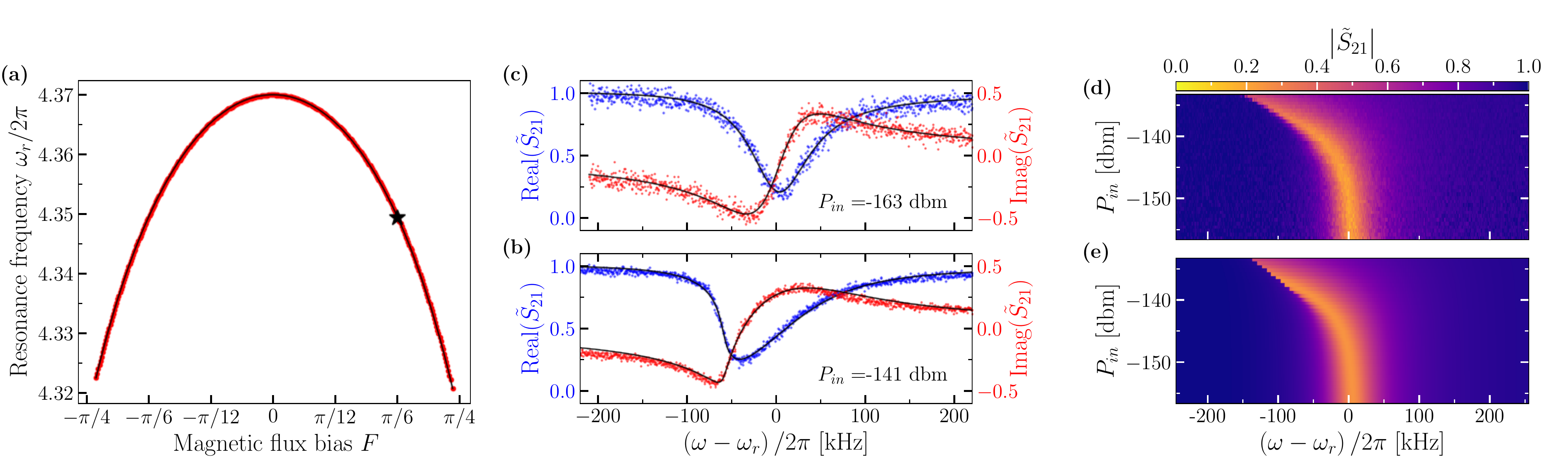}
        \caption{ \textbf{Measurement of the device parameters.} (a) Red markers show the measured resonance frequency $\omega_r$ for varying magnetic flux bias $F$. The solid black line indicates the fit to Eq.~(\ref{eq:flux_response_approx}). The black star marks the operating point $F\approx \pi/6$ used throughout the experiment. (b)-(c) Real (blue) and imaginary (red) part of the measured rescaled scattering coefficient  $\tilde{S}_{21}= S_{21}/a e^{j\alpha}e^{-j\omega \tau}$ at low (b) and high (c) input power. Notice that for high input power, the simple Lorentzian response observed at low power becomes distorted due to the nonlinearity of the resonator. The black lines indicate the fit to Eq.~(\ref{Eq: final_fit}). (d) Absolute value of the measured rescaled scattering coefficient for increasing input power, and (d) the corresponding 2D fit to Eq.~(\ref{Eq: final_fit}).} 
        \label{Fig:parameter_char}
\end{figure*}

The first step in the characterization of the device is measuring the scattering coefficient $S_{21}$ at low power while varying the magnetic flux bias. The scattering response is measured using a Vector Network Analyzer (Rohde \& Schwarz, ZNA26 series) connected between the feedline and output line. The magnetic flux is varied by adjusting the current sent from the Yokogawa GS200 source to the NbTi coil underneath the sample. For each magnetic flux value, the resonance frequency $\omega_r$ is determined by fitting the scattering response using Eq.~\eqref{Eq: final_fit} (see Supplementary Sec.~\ref{section: supplementary scattering response } for the derivation of $S_{21}$). The result of this measurement is shown in Fig.~\ref{Fig:parameter_char}(a). Using Eq.~\eqref{eq:flux_response_approx}, the flux response of the resonance frequency is fitted to determine the ratio of the SQUID to the cavity inductance (participation ratio)
$\gamma \approx \SI{3.1e-2}{}$
and bare cavity resonance frequency (without SQUID)
$\omega_{\lambda/4} \approx \SI{4.5068}{\giga\Hz}$.

 All the measurements discussed in the main text were performed at a magnetic flux bias of $F\approx\pi/6$, corresponding to the black star in Fig.~\ref{Fig:parameter_char}(a). Figure~\ref{Fig:parameter_char}(b) shows the real (blue markers) and imaginary (red markers) part of the corrected scattering response $\tilde{S}_{21}= S_{21}/a e^{j\alpha}e^{-j\omega \tau}$ at this operating point for an input power of -163 dbm. The resonance frequency, as well as the external and internal coupling parameters, are determined by fitting the data with Eq.~(\ref{eq:flux_response_approx}). Repeating the measurement 12 times, we obtain the values: $\omega_{r} = \SI{4.3497}{\giga\Hz}$, $\kappa_{ext} = \SI{60}{\kilo\Hz}$ and,  $\kappa_{int} = \SI{17}{\kilo\Hz}$. The value of $\omega_{r}$ presented here is obtained from measurements taken within a few minutes of interval. However, the resonance frequency is characterized by small fluctuations over long period of time. These fluctuations pose no problem to the experiment, as the relevant quantity to characterize the resonator is not $\omega_{r}$, but rather the detuning between the pump and cavity. As briefly discussed in Sec.~\ref{supp section : measurement protocol}, the detuning can be straightforwardly corrected before each experiment.

The Kerr nonlinearity $U$ can be directly calculated from the measured participation ratio and bare resonance frequency using Eq.~(\ref{eq:Kerr}). Neglecting the SQUID capacitane ($C_J=0$), a value of $U=\SI{7}{\kilo\Hz}$ is obtained. The derivation of this equation and the validity of the approximation are detailed in Sec.~\ref{Sect:circuit parameters}. 

A second method to estimate the Kerr nonlinearity is fitting the scattering response $S_{21}$ at higher power, where the nonlinearity influences the scattering coefficient (see Sec.~\ref{section: supplementary scattering response }). The absolute value of the measured scattering coefficient as a function of the input power at the device $P_{in}$ is shown in Fig.~\ref{Fig:parameter_char}(d). Performing a 2D fit using Eq.~(\ref{Eq: final_fit}), a value $U=\SI{6}{\kilo\Hz}$ is obtained, thus confirming our initial estimate and the accuracy of our calibration. The fit is shown for the all the input powers in Fig.~\ref{Fig:parameter_char}(e), and in Fig.~\ref{Fig:parameter_char}(c) for a specific power of $P_{in}=-141$ dBm. The values experimentally extracted for the circuit parameters are summarized in Table~\ref{table:parameter}.

\begin{table}[h!]
\centering
\caption{\textbf{Experimental values of the circuit parameters}}
\begin{tabular}{>{\centering\arraybackslash} p{2.75cm}>{\centering\arraybackslash} p{2.75cm}>{\centering\arraybackslash} p{2.75cm} >{\centering\arraybackslash} p{2.75cm} >{\centering\arraybackslash}p{2.75cm} >{\centering\arraybackslash} p{2.75cm}}
\toprule
$\omega_r/2\pi$ [GHz] & $\omega_{\lambda/4}/2\pi$ [GHz] & $\mathbf{\gamma}$ & $U/2\pi$ [kHz] & $\kappa_{ext}/2\pi$ [kHz]  & $\kappa_{int}/2\pi$ [kHz]\\
\midrule
 4.3497 & 4.5068 & \SI{3.13e-2}{} & 
 7 & 60 & 17 \\

\bottomrule
\end{tabular}
\label{table:parameter}
\end{table}

\subsection{Calibration of Input Attenuation, Amplifier Noise, and Gain}\label{sect: supplementary calibration}

The two-photon pump populates the cavity.
The signal emitted from the cavity, due to the emission into the waveguide at a rate $\kappa_{\rm ext}$,  passes through filters, amplifiers (cryo and room temperature) and cables before being collected. The  power gain $\mathcal{G}$ of the output line  takes into account all of these components, and relates the field measured at room temperature $\hat c(t)$ to the output field of the cavity at cryogenic temperature $\hat b_{out}^{(r)}(t)$ through the relation 
\begin{align}\label{transf}
\hat c(t) = \sqrt{\mathcal{G}}~\hat b_{out}^{(r)}(t)+\sqrt{\mathcal{G}-1}~\hat h^\dag(t).
\end{align}
where the mode $\hat h(t)$ is a white noise (i.e., $\langle\hat h(t)\hat h^\dag(t')\rangle=(n+1)\delta(t-t')$) that comes from a combination of cable losses, amplifier noise, and IQ mixer noise, and usually follows Gaussian statistics. For a given temporal filter $w(t)$ with normalization $\int dt~|w(t)|^2=1$, one can consider the mode $\hat c=\int dt\,w(t)\hat c(t)=\int d\omega\,\tilde{w}(f)\hat c(f)$, where $\tilde{w}(f)$ 
 and $\hat c(f)$ are  the Fourier transform of $w(t)$ and $\hat c(t)$. The input-output relations for the filtered fields are
 \begin{align}
\hat c= \sqrt{\mathcal{G}}~\hat b_{out}^{(r)}+\sqrt{\mathcal{G}-1}~\hat h^\dag,
 \end{align}
where $\hat b_{out}^{(r)}=\int dt\,w(t)\hat b_{out}^{(r)}(t)$ and $\hat h=\int dt\,w(t)\hat h(t)$.

We perform a Planck spectroscopy experiment to calibrate $\mathcal{G}$ and the power $n$ of the mode $\hat h$, i.e., $n=\langle \hat h^\dag \hat h\rangle$~\cite{PhysRevLett.109.250502}. Using the switch positioned a the MXC stage of the cryostat, a $\SI{50}{\ohm}$ cryogenic termination is connected to the output line (see Sec.~\ref{section: supplementary experimental setup}). This termination, thermalized at the MXC temperature $T$, acts as a black body emitter with average number of photons at frequency $f$ given by $\bar{n}_{T}(f)=1/\left \{ \exp[hf/(k_BT)]-1 \right \}$.
Therefore, the gain of the line can be calculated by comparing the theoretically known emitted power of thermal radiation from the attenuator in a bandwidth $B$, i.e., $P_{T}=\bar{n}_{T}(f) hfB$, to the measured power at room temperature $P_m$ within the same bandwidth (i.e., we integrate with a filter $\tilde{w}(f)=1_{f\in[f_0- B/2,f_0+B/2]}$).

\begin{figure*}[h]
        \centering
        \includegraphics[width=0.5\textwidth]{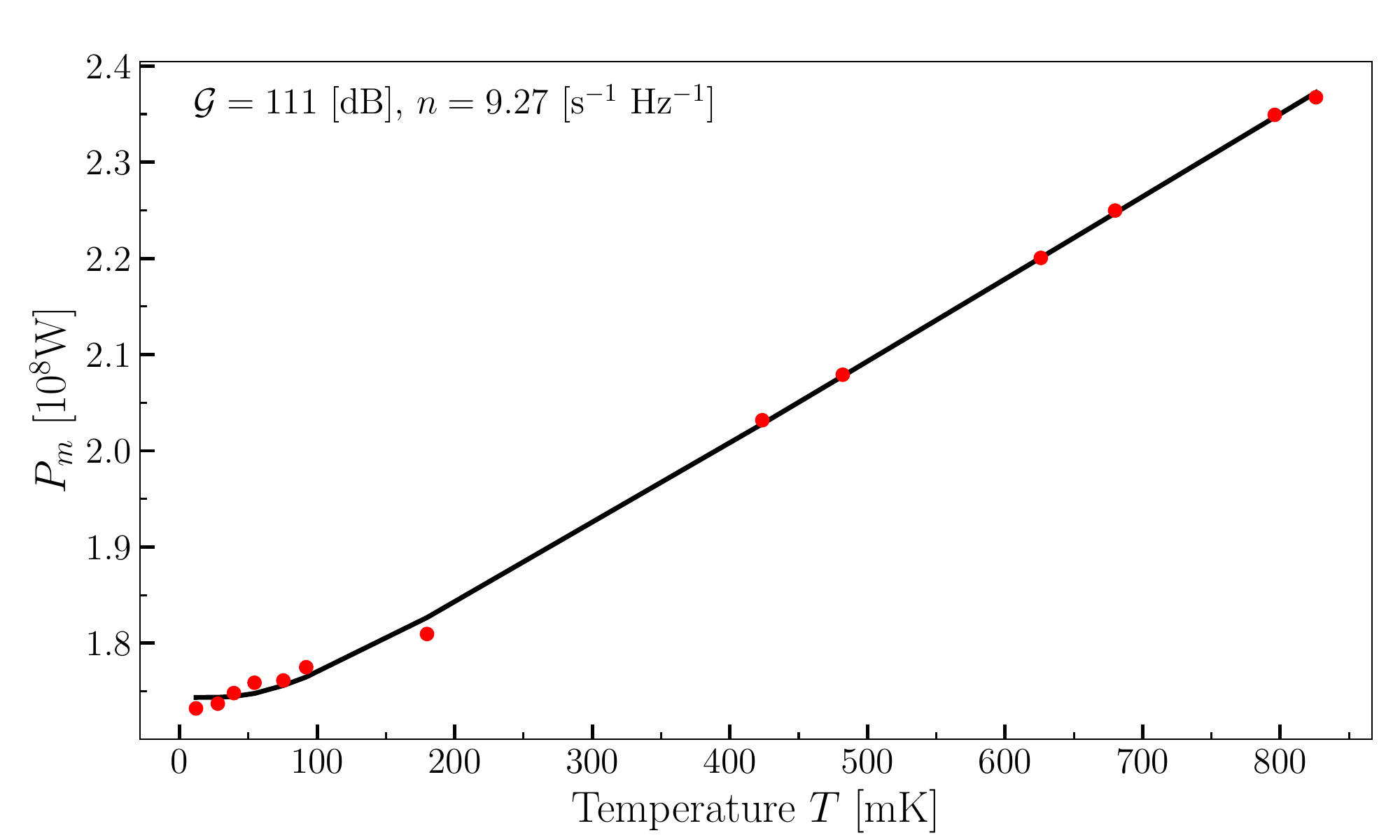}
        \caption{ \textbf{Gain calibration by Planck spectroscopy  measurement}. (a) Blue markers indicate the measured output power at a frequency $f=\SI{4.3}{\giga\Hz}$ in a bandwidth $B=\SI{5}{\kilo\Hz}$ as a function of the temperature. The solid black line is the fit with Eq.~\ref{Eq: gain_calibration}.}. 
         \label{Fig:calibration}
\end{figure*}

To raise the temperature of the attenuator, the turbo pump of the cryostat is turned off, reducing the flow of the $^{3}$He/$^{4}$He mixture, and consequently, the cooling power. Simultaneously, heat is applied through a heater at the MXC stage. The emitted power is given by 
\begin{equation}\label{Eq: gain_calibration}
     P_{m}= \frac{\langle\hat{I}_m^2 +
\hat{Q}_m^2 \rangle }{Z_0} = B \mathcal{G} h f \left [ \frac{1}{2} \coth \left( \frac{h f}{2 k_B T} \right) + n\right ],
\end{equation}
where $k_B=\SI[per-mode = symbol]{1.38e-23}{\joule\per\kelvin}$ is the Boltzmann constant, $h=\SI{6.63e-34}{\joule\second}$ is the Planck constant. Fig.~\ref{Fig:calibration} shows the measured output power as a function of $T$. Here, $\hat I_m=\int_{f_0-B/2}^{f_0+B/2} df~\hat I_m(f)$ and $\hat Q_m=\int_{f_0-B/2}^{f_0+B/2} df~\hat Q_m(f)$, where $\hat I_m(f)$ and $\hat Q_m(f)$ are the Fourier transform of $\hat I_m(t)$ and $\hat Q_m(t)$, which are the quadratures of $\hat c(t)$ up to a dimensional normalization factor. The output power was measured at frequency $f=\SI{4.3}{\giga\Hz}$ (near the resonance frequency) in a bandwidth of $B=\SI{5}{\kilo\Hz}$. To ensure proper thermalization of the termination, we repeated the output power measurements at 10-minute intervals for each temperature setting until successive readings converged.

From the fit with Eq.~\ref{Eq: gain_calibration}, we obtain a power gain of $\mathcal{G}=\SI{111}{\decibel}$ and a mean photon noise of $n=\SI{9.27}{\per \second \per \Hz}$ corresponding to a noise temperature of $\SI{1.91}{\kelvin}$.

Additionally, knowing the gain factor $\mathcal{G}$ of the output line allows for a straightforward characterization of the input attenuation of the feedline. The sample is used as a pass-through by applying a magnetic flux to shift its resonance frequency away from the measurement frequency. A known power $P_d$ is then sent into the input line. This power is related to measured output signal power $ P_{m}$ by the following relation relation 
 \begin{equation}
     P_{m}=\mathcal{G}\left[ AP_d + B h f n\right ].
 \end{equation}

We obtain an input attenuation $A=\SI{-85}{\decibel}$.

\subsection{Moments reconstruction and squeezing parameter} \label{sect : state reconstruction}
For reconstructing the moments of $\hat b_{out}^{(r)}(t)$, we use the so-called reference-state method~\cite{PhysRevA.86.032106, di2014dual}. This consists in reconstructing the moments of the noise mode $\hat h(t)$ using the vacuum as a reference state. Then, we use this characterization to retrieve the moments of the resonator output $\hat b_{out}^{(r)}(t)$. The method assumes a pre-knowledge of the measurement-line gain $\mathcal{G}$, which can be characterized with the method presented in the previous subsection.

We define the complex envelope operator
\begin{align}\label{eq: output field gain}
\hat S \equiv \frac{1}{\sqrt{Z_0 h f}}\frac{\hat I_m+i\hat Q_m}{\sqrt{\mathcal{G}}} =\hat b_{out}^{(r)}+\hat h^\dag,
\end{align}
The real and imaginary parts of the complex envelope operator $\hat S$ represent the measured $\hat{I}_m$ and $\hat{Q}_{m}$ quadratures up to a normalization factor. 
One can easily derive the relation~\cite{PhysRevA.86.032106,di2014dual}
\begin{align}\label{Eqset}
\langle (\hat S^\dag)^n\hat S^m \rangle_{\rho_{\hat b_{out}^{(r)}}} = \sum_{i=0}^{n}\sum_{j=0}^m {n\choose i}{m \choose j} \langle (\hat b_{out}^{(r)\,\dag})^i\hat b_{out}^{(r)\,j}\rangle\langle \hat h^{n-i}(\hat h^\dag)^{m-j}\rangle.
\end{align}
Once the anti-normal ordered moments of the noise $\langle \hat h^{n}(\hat h^\dag)^{m}\rangle$ is known, the set of linear equations in \eqref{Eqset} can be solved for $\langle (\hat b_{out}^{(r)\,\dag})^n\hat b_{out}^{(r)\,m}\rangle$. If $\hat b_{out}^{(r)}$ is in a vacuum state, Eq.~\eqref{Eqset} reduces to
\begin{align}\label{noise}
\langle (\hat S^\dag)^n\hat S^m \rangle_{|0\rangle\langle0|}= \langle \hat h^{n}(\hat h^\dag)^{m}\rangle,
\end{align}
since $\langle (\hat b_{out}^{(r)\,\dag})^n\hat b_{out}^{(r)\,m}\rangle_{|0\rangle\langle 0|}=0$ for $n,m\not=0$. From Eq.~\eqref{noise} we can witness non-Gaussianity of the noise mode by looking at the cumulants~\cite{PhysRevLett.109.250502}. We finally invert Eq.~\eqref{Eqset} for a generic input state. Notice that due to the increasing amount of terms involved for increasing $n$ and $m$, higher moments reconstruction will have more statistical noise. However, the experimental samples are enough to reconstruct faithfully the moments up to $n+m=2$. Let us write down the reconstruction formulas up to $n+m=2$, as they are useful to characterize the squeezing:
\begin{align}
\langle \hat b_{out}^{(r)}\rangle&= \langle\hat S\rangle-\langle \hat h^\dag\rangle \label{inv1}\\
\langle \hat b_{out}^{(r)\,2}\rangle &= \langle \hat S^2\rangle - \langle (\hat h^\dag)^2\rangle-2\langle \hat b_{out}^{(r)}\rangle\langle \hat h^\dag\rangle \label{inv2}\\
\langle \hat b_{out}^{(r)\,\dag} \hat b_{out}^{(r)}\rangle &=\langle \hat S^\dag \hat S\rangle-\langle\hat h \hat h^\dag\rangle-\langle\hat b_{out}^{(r)}\rangle\langle \hat h \rangle -\langle \hat b_{out}^{(r)\, \dag}\rangle\langle \hat h^\dag \rangle. \label{inv3}
\end{align}
Again, non-Gaussianity of the mode $\hat b_{out}^{(r)}$ can be witnessed by computing higher moments, since Gaussian distributions are defined by the first-moment vector and the covariance matrix. So far we have discussed how to retrieve the moments of a filtered field $\hat b_{out}^{(r)}$. If one is interested in continuous monitoring of the moments of $\hat b_{out}^{(r)}(t)$, one can use a filter picked in $t$, i.e., consider $\frac{1}{\sqrt{\Delta T}}\int_{t-\Delta T/2}^{t+\Delta T/2} d\tau~\hat b_{out}^{(r)}(\tau)\simeq \hat b_{out}^{(r)}(t)\sqrt{\Delta T}$.

With the output line calibrated, the intracavity field $\hat a(t)$ can be inferred with the input-output relations $\hat b_{out}^{(r)}(t)=\sqrt{\frac{\kappa_{ext}}{2}}~\hat a(t) + \hat{b}_{in}^{(r)}(t)$ (Eq.~\ref{Eq: input-ouput} in Supplementary Sec.~\ref{section: supplementary scattering response }). Consequently, substituting this relation in Eq.~\ref{eq: output field gain} in the absence of any input field in the feedline (i.e., setting $\hat{b}_{in}^{(r)}(t)$ to the vacuum), one can derive all the ordered moments of the integrated cavity mode $\hat a=\int dt~w(t)\hat a(t)$, by inverting
\begin{align}
\langle \hat b_{out}^{(r)\,\dag k} \hat b_{out}^{(r)\,l}\rangle  =\left(\frac{\kappa_{ext}}{2}\right)^{\frac{k+l}{2}}\langle\hat a^{\dag k}\hat a^l\rangle.
\end{align}

Let us define the quadrature of $\hat a$ as $\hat x_\phi = \frac{1}{\sqrt{2}}(\hat ae^{-i\phi}+\hat a^\dag e^{i\phi})$. One can use Eqs.~\eqref{inv1}-\eqref{inv3} to reconstruct the variance $\langle \Delta x_{\phi}^2\rangle = \langle \hat x^2_\phi\rangle  - \langle \hat x_\phi\rangle^2$. Notice that the quadrature variance of the vacuum state is set to $1/2$.

\renewcommand{\paragraph}[1]{
\vspace{5pt}\noindent \textbf{#1}}

\section{Measurement protocols}
\label{supp section : measurement protocol}

\begin{figure*}[t]
        \centering
        \includegraphics[width=0.5\textwidth]{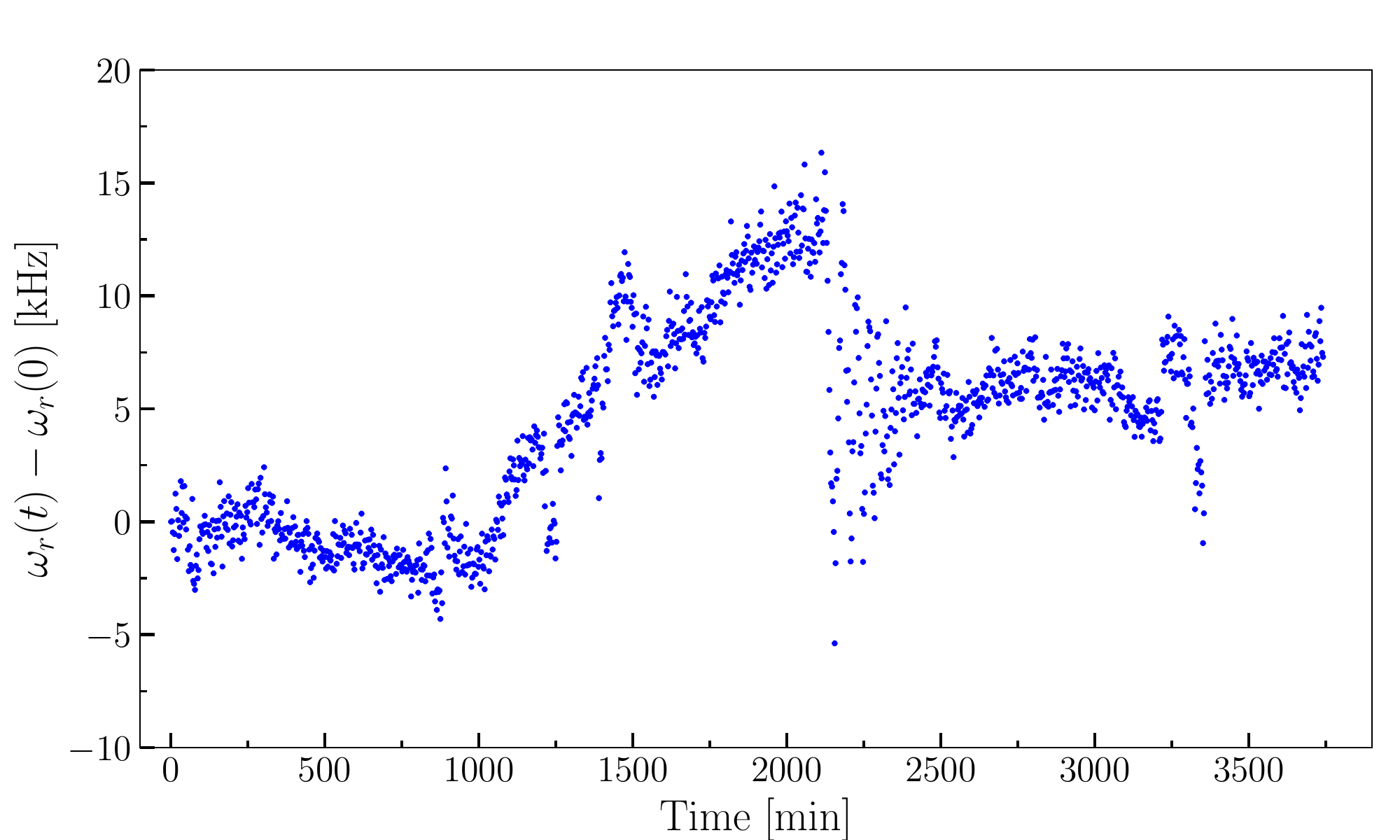}
        \caption{ \textbf{Resonance frequency drift.} Each marker, acquired every 2 minutes over $\sim$ 60 hours, corresponds to the resonance frequency extracted from the scattering response $S_{21}$. }
         \label{Fig:resonance fluctuation}
\end{figure*}

\paragraph{Calibration of the cavity frequency:} Prior to each measurement, the amplitude of the flux line signal is measured using the Signal Hound spectrum analyzer, and any amplitude drift is corrected. Furthermore, the resonance frequency is measured for every value of detuning during the experiment. This is necessary because the cavity resonance frequency slightly drifts over time, as a consequence of environmental noise (magnetic noise, vibrations in the lab, etc.). Figure~\ref{Fig:resonance fluctuation} shows multiple measurements of the resonance frequency taken at intervals of 2 minutes over a period of $\sim$ 60 hours. The standard deviation over the full time interval is $\sim \SI{4}{\kilo\Hz}$. This drift corresponds to a noticeable change in detuning and, if not accounted for, could lead to inconsistent results. Fortunately, since the relevant quantity in the experiment is the detuning, and not the cavity frequency alone, a drift in $\omega_r$ can be compensated by adjusting the pump frequency. To achieve this, we first evaluate the resonance frequency by measuring the scattering response $S_{21}$ using a weak probe signal sent into the feedline, and fitting the result with Eq.~(\ref{Eq: final_fit}). Then, the pump frequency is adjusted to obtain an accurate value of detuning. It is also important to note that the drift has negligible effect on the value of the Kerr nonlinearity [see Eq.~(\ref{eq:Kerr})].

\paragraph{Acquisition of a quantum trajectory}. After the calibration of the pump amplitude and detuning, each measurement involves three main steps: (I) initializing the system in the desired state, (II) activating the two-photon pump $G$ at frequency $\omega_p$, and (III) acquiring the quantum trajectory through heterodyne measurement. A single quantum trajectory is constructed from $N$ quadrature measurements acquired sequentially, with a certain time delay $\tau_{delay}$ between each acquisition, while the two-photon pump is still on. Each quadrature measurement corresponds to the demodulated signal quadratures ($I_m(t)$,$Q_m(t)$) at frequency $\omega=\omega_p/2$, integrated over a time interval $\tau_{int}$ at the ADC of the OPX+. The quadratures ($I(t)$,$Q(t)$) of the intracavity field are then obtained by removing the effect of the amplification chain and its associated noise from the measured data (see Sect.~\ref{sect: supplementary calibration} for more details). The details of each measurement, including the specific parameters used, are discussed
below.

\begin{figure*}[h]
        \centering
        \includegraphics[width=0.8\textwidth]{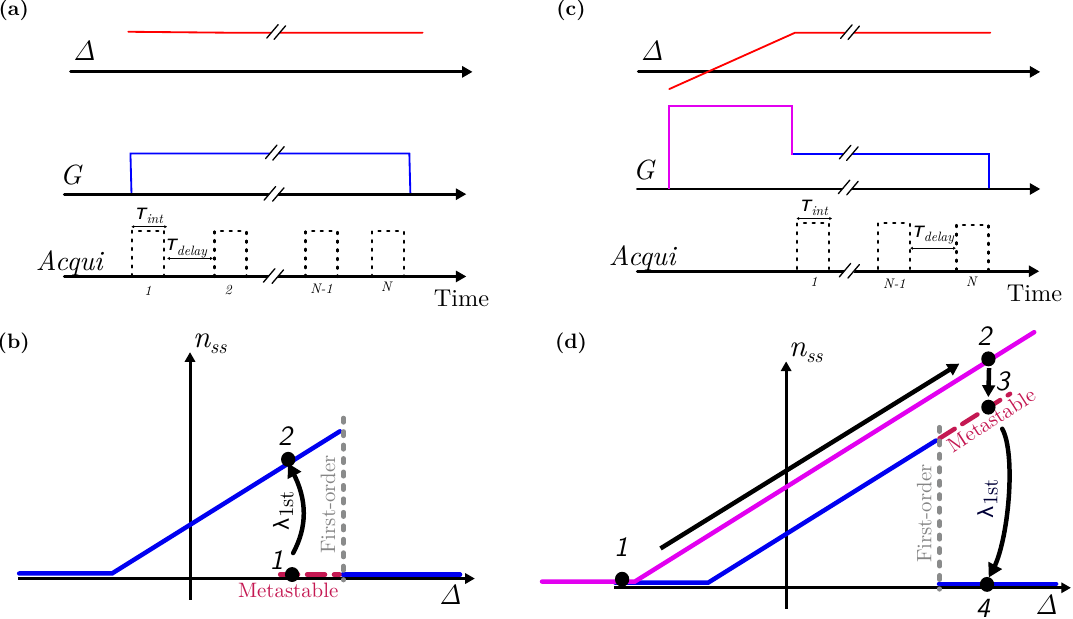}
        \caption{ \textbf{Pulse sequence for the measurement of $\lambda_{1st}$.} (a) The pulse sequence for measuring he vacuum metastability consists in activating the two-photon pump $G$ (blue solid line) at a fixed detuning $\Delta$ (red solid line) and start acquiring the quadratures (dotted black line). The quadratures are acquired at time intervals $\tau_{delay}$ with each acquisition corresponding to the signal integrated over a time interval $\tau_{int}$. A single trajectory consists of $N$ acquisitions. (b) Schematic illustrating the measurement of vacuum metastability. The system is initialized in the vacuum state (black dot 1) by setting  $G=0$. After a certain time $\sim 1/\lambda_{1st}$, the system jumps to the steady-state bright phase (black dot 2). (c) The pulse sequence for measuring the bright phase metastability involves: (i) turning on a strong two-photon drive $G$ (purple solid line) at large negative detuning $\Delta$, (ii) increasing the detuning until the desired value is reached (red solid line), and (iii) abruptly lowering the value of $G$ to reach the metastable bright phase (blue solid line). (d) Schematic illustrating the measurement of bright phase metastability. The bright phase initialization process begins by setting the system in the vacuum state at strong pump $G$ (purple curve) and large netgative detuning (black dot 1). The detuning is then progressively increased until the desired value is reached (black dot 2), before abruptly lowering the pump to be in a metastable bright phase (black dot 3). After a certain time $\sim 1/\lambda_{1st}$, the system jumps to the steady-state vacuum (black dot 4).  }
         \label{Fig:pulse sequence}
\end{figure*}

\paragraph{Dynamics at the first order DPT, i.e., measurement of $\lambda_{1st}$:} To characterize the metastability of the vacuum, the system is initialized in the vacuum state, i.e., we set $G=0$ which corresponds to no drive applied to the system. Then, the two-photon drive $G$ is switched on at frequency $\omega_p$. The corresponding the detuning is $\Delta = \omega_r - \omega_p/2$ (the latter having been adjusted for any frequncy drift). 
A quantum trajectory is then acquired following the procedure described above. After measuring the trajectory, the two-photon drive is turned off and a waiting time $\tau_{wait}$ is required for the system to return to the vacuum state before the next measurement can be performed. For a given pump frequency $\omega_p$ (i.e., for a given detuning), the same measurement protocol is repeated $M$ times giving $M$ different trajectories. The average over these trajectories is then fitted by Eq.~(\ref{Eq:photon_number_time}) to extract $\lambda_{1st}$. A schematic illustrating the measurement of the metastable vacuum is shown Fig.~\ref{Fig:pulse sequence}(a)-(b). \\

A similar protocol is followed to characterize the metastability of the bright phase, with the only difference being the initialization process. To initialize the system in the bright phase and observe the decay towards the vacuum, we apply the following protocol [see Fig.~\ref{Fig:pulse sequence}(d)] :
\begin{itemize}
    \item 
    The system is initialized in the vacuum state by choosing a pump frequency $\omega_p$, such that the detuning is large and negative $\Delta\approx \SI{-0.3}{\mega\Hz}$.
    \item Then, a strong two-photon drive of $G=135$ kHz is turned on. For comparison, the maximal drive considered in the figures of the main text  is $G=103.5$ kHz.
    \item With the drive on, the detuning is continuously swept at a rate of \SI[per-mode = symbol]{1}{\mega\Hz\per\milli\second} until the desired value of detuning is reached. Under these conditions, the system is in the bright state with a large number of photons.
    \item At this point, the two-photon drive is abruptly reduced to the desired value. The system then rapidly evolves to the desired bright metastable state.
    
\end{itemize}
This procedure ensures the system is in the bright phase at a given detuning and pump amplitude. 
The pulse sequence for this measurement is illustrated in Fig.~\ref{Fig:pulse sequence}(c) and the evolution of the system's state is also schematically represented in Fig.~\ref{Fig:pulse sequence}(d).

Since $\lambda_{\rm 1st}$ varies over several orders of magnitude depending on $\Delta$ and $G$, the measurement duration for each trajectory - determined $\tau_{int}$ and $N$ - is calibrated based on an estimate of $\lambda_{\rm 1st}$ given by an initial sampling of 100 trajectories. For $\lambda_{\rm 1st}< \SI{100}{\per\second}$, the integration time is set to $\tau_{int}=\SI{10}{\micro\second}$, while for $\lambda_{1st}>\SI{100}{\per\second}$, $\tau_{int}=\SI{50}{\micro\second}$. All measurements are done with a time delay $\tau_{delay}=0$. The value of $N$ ranged from $N=1000$ to $N=150000$, corresponding to a measurement time per trajectory varying from \SI{0.01}{\second} to \SI{7.5}{\second}. The choice of $N$ is made such that the measurement time is at least twice $1/\lambda_{\rm 1st}$. However, at $L=1.58$, for the values of detuning where the system is the slowest, the measurement time per trajectory is capped to \SI{7.5}{\second} in order to keep reasonable measurement times. The number of repetition is varied from $M=2000$ to $M=500$ depending on the measurement time per trajectory.

\paragraph{Dynamics at the second order, i.e., measurement of $\lambda_{SSB}$:} Measuring $\lambda_{SSB}$ involves recording a single very long quantum trajectory containing multiple jumps between the two coherent states $\ket{\alpha}$ and $\ket{-\alpha}$. Starting with the system in the vacuum state, the two-photon pump is switched on at a frequency $\omega_p$ and after a waiting time $\tau_{wait}$, a quantum trajectory is acquired following the procedure described above. The auto-correlation function of the trajectory is then calculated and fitted to Eq.~(\ref{Eq: autocorellation}) to obtain $\lambda_{SSB}$. Since $\lambda_{SSB}$ varies of several orders of magnitude depending on $\Delta$ and $G$, we first measure a few jump events to obtain an estimate of the duration - determined by $\tau_{int}$ and $N$ and $\tau_{delay}$ - required to measure multiple jumps. For $\lambda_{SSB}< \SI{100}{\per\second}$, the integration time is set to $\tau_{int}=\SI{10}{\micro\second}$, while for $\lambda_{SSB}>\SI{100}{\per\second}$, $\tau_{int}=\SI{50}{\micro\second}$. The time delay $\tau_{delay}$ varied from zero to \SI{1900}{\micro\second} and the number of samples $N$ ranged from \SI{2.5e6}{} to \SI{5e6}{}, resulting in measurement times for a single trajectory ranging from \SI{50}{\second} to \SI{20}{\minute}. The waiting time $\tau_{wait}$ was kept constant for all measurements at \SI{1}{\second}. For values of detuning and pump amplitude where $1/\lambda_{1st}>\SI{1}{\second}$, the initial samples in the trajectory, acquired before the system reached the steady state, were simply discarded.

\paragraph{Phase diagram, measurement of $n_{ss}$ and $\Phi$:} The steady state properties can be directly inferred from the data used to calculate $\lambda_{SSB}$. The steady state photon number $n_{ss}$ is obtained from averaging the quadrature squared over a single very long trajectory $n_{ss}=\expval{\hat{I}^2 +\hat{Q}^2}$. Note that near the second-order phase transition $1/\lambda_{SSB}$ becomes larger than $\tau_{int}$. As a consequence, jumps occurring during the signal integration lead to a smaller value of $n_{ss}$. The phase of the system $\Phi$ corresponds to the angle of the complex number $I+iQ$ for one pair of quadratures. From the $N$ pairs of quadratures in a single trajectory, an histogram can be constructed representing the phase distribution during the measurement. Fig.~\ref{Fig:phase_diagram}(e) is obtained by plotting such histograms as a function of the detuning $\Delta$.

\paragraph{Squeezing measurement:} The measurement of the squeezing follows a similar procedure to that of $\lambda_{SSB}$. The key difference is that it requires faster measurements to avoid jumps occurring within the integration time of the quadrature. Additionally, a large number of samples are necessary to accurately reconstruct the state (see Sec.~\ref{sect : state reconstruction} for the state reconstruction process). The quantum trajectories used to calculate the squeezing are all acquired with the same parameters: an integration time $\tau_{int}=\SI{2}{\micro\second}$, no delay $\tau_{delay}=\SI{0}{}$, a waiting time of $\tau_{wait}=\SI{0.5}{\second}$ to reach the steady-state, and a total of $N=\SI{10e6}{}$ samples. 

\paragraph{Hysteresis measurement:} To characterize the hysteretic behavior of the system, the detuning is ramped (by changing the pump frequency $\omega_p$) while keeping the two-photon pump amplitude $G$ constant. At each step of the detuning sweep, the quadratures are measured once, followed by a certain delay time $\tau_{delay}$ before moving to the next detuning value. Consequently, the detuning rate $D$ is determined by the integration time $\tau_{int}$, the delay time $\tau_{delay}$, and the change in detuning in each pump frequency increment. The complete sweep is repeated $M$ times, and the final result is obtained by averaging over all repetitions. The hysteresis maps shown in Figs.~\ref{Fig:Hysteresis}(a,b) are the average of 4000 repetitions done with $\tau_{int}=\SI{10}{\micro\second}$, $\tau_{delay}=\SI{16}{\nano\second}$ and $D/2\pi=\SI[per-mode = symbol]{1000}{\mega\Hz\per\second}$. In Fig.~\ref{Fig:Hysteresis}(c), a similar measurement is conducted to observe how the hysteresis area varies with the sweep rate. The sweep rate is changed from $D/2\pi=\SI[per-mode = symbol]{1000}{\mega\Hz\per\second}$ to $D=\SI[per-mode = symbol]{35}{\mega\Hz\per\second}$ by varying $\tau_{delay}$ from $\SI{16}{\nano\second}$ to $\SI{0.28}{\milli\second}$. The integration time remains constant at $\tau_{int}=\SI{10}{\micro\second}$ and the area is calculated from the average of $M=4000$ repetitions.

\section{Modeling of the system}
\subsection{Hamiltonian parameters}\label{Sect:circuit parameters}

\begin{figure*}[!h]
        \centering
        \includegraphics[width=0.7\textwidth]{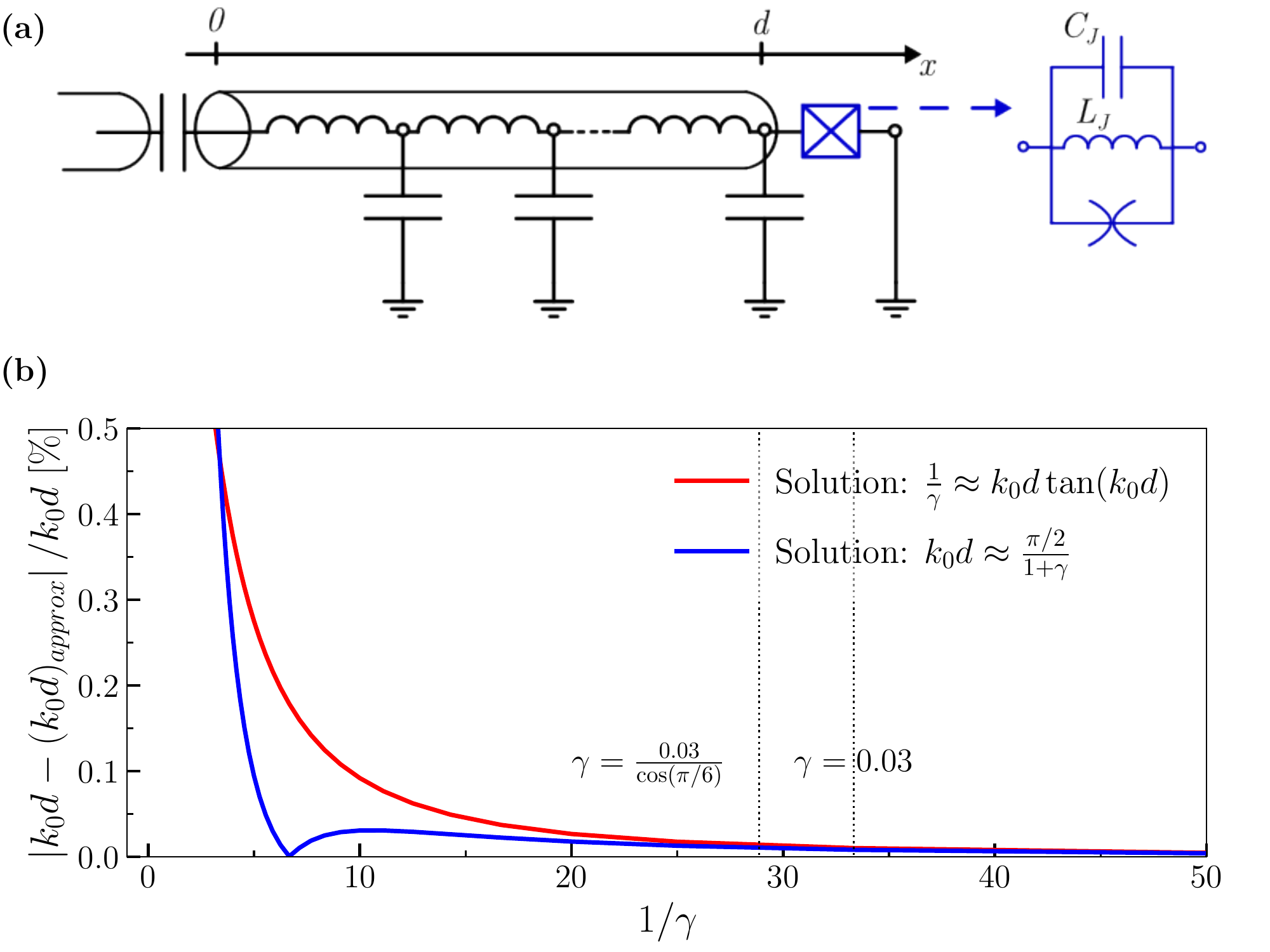}
        \caption{ \textbf{ 
        Distributed-element model of a $\lambda/4$ resonator terminated by a SQUID.} (a) The transmission line is modeled by a discrete chain of identical LC oscillators with inductance and capacitance per unit length of $l$ and $c$ respectively. The cavity has an open end at $x=0$ and is grounded through a SQUID at $x=d$. (b) Relative error of the fundamental mode solution $k_0$ as a function of the inductance ratio. The red line is the solution of Eq.~(\ref{Eq:eigenmode_general}) neglecting the capacitance ratio, i.e.  $C_J/C_{cav}=0$, while the blue line is the first order approximation of Eq.~(\ref{eq:first_order_approx}). The relative error is calculated with respect to the solution obtained with Eq.~(\ref{Eq:eigenmode_general}) for $C_J/C_{cav}=0.05$. The dotted line at $\gamma=0.03$ corresponds to the participation ratio of our device without any applied external flux, while $\gamma=0.03/\cos(\pi/6)$ is the participation ratio during the measurement. }
         \label{Fig:distributed_model}
\end{figure*}

The device is modeled by the following Hamiltonian
\begin{equation}\label{Eq:Hamiltonian_sup}
    \hat{H}/\hbar = \omega_r \hat{a}^\dagger \hat{a} + \frac{U}{2} \hat{a}^\dagger \hat{a}^\dagger \hat{a} \hat{a} + \frac{G}{2} \left( \hat{a}^\dagger\hat{a}^\dagger e^{-i\omega_p t} + \hat{a}\hat{a} e^{i\omega_p t} \right),
\end{equation}
with $\omega_r$ the resonance frequency, $U$ the Kerr nonlinearity and $G$ the two-photon drive amplitude. In the following, it will be explicitly demonstrated that this quantum mechanical model corresponds to a transmission line terminated by a SQUID as shown in Fig.~\ref{Fig:distributed_model}(a). Through this calculation, equations for the system parameters $\omega_r$ and $U$ will be derived. These equations provide valuable insights to design the device.

A transmission line of length $d$, with one end open and the other one grounded, forms a quarter wavelength resonator with eigenmode wavevectors $k_n =(\pi/d)(n+1/2)$. The corresponding eigenmode frequencies are $\omega_n=(\pi/d \sqrt{lc})(n+1/2)$, where $l$ and $c$ are respectively the inductance and capacitance per unit length \cite{MWallquist_strip}. The total inductance and capacitance of the cavity are simply expressed as $L_{cav}=ld$ and $C_{cav}=cd$. Adding a SQUID at the end of the transmission line ($x=d$) modifies the boundary condition, which results in an increase of the effective wavelength and, consequently, a deviation from the eigenmodes of a quarter wavelength resonator. To calculate this deviation, we follow the derivation of Refs.~\cite{Eichler_2014_josephson,MWallquist_strip,Wustmann_2013}. Throughout the following, emphasis will be placed on the fundamental mode, given its use in the experiment. 

The total Lagrangian of the system [see Fig.~\ref{Fig:distributed_model}(a)] is 
\begin{equation}
\label{Eq: lagrangian}
    \mathcal{L}=\underbrace{\int_{0}^d \left ( \frac{c}{2}\dot{\phi}(x,t)^{2}-\frac{1}{2l}{\phi}'(x,t)^{2} \right ) dx}_{\rm Cavity} +\underbrace{ \frac{C_{J}}{2}\dot{\phi}(d,t)^2+E_J \cos\left (\frac{\phi(d,t)}{\phi_{0}} \right ) }_{\rm SQUID},
\end{equation}
where $C_J$ is the SQUID capacitance, $E_J$ is the SQUID Josephson energy and $\phi_0=\hbar/2e$ is the reduced flux quantum. Note that the two Josephson junctions forming the SQUID are considered to be identical with Josephson energy $E_{J,s}/2$. Under this condition, the SQUID can be modeled as a single junction with a tunable Josephson energy $E_J=E_{J,s} \left | \cos(F) \right |$, where $F$ is given by the external magnetic flux enclosed by the SQUID loop. This also allows to represent the SQUID as a simple inductor of value $L_J=\phi_0^2/E_J$. The first term of Eq.~(\ref{Eq: lagrangian}), defines the equation of motion of cavity field 
\begin{equation}
    \ddot{\phi}(x,t)-v^{2} {\phi}''(x,t)=0,
\end{equation}
where $v=1/\sqrt{lc}$ is the phase velocity. In addition to this wave equation, the cavity field is defined by two boundary conditions. Applying Kirchhoff current law at $x=d$ results in the first boundary condition
\begin{equation}\label{Eq: boundary squid}
    C_{J}\ddot{\phi}(d,t)+\frac{\phi_0}{L_j} \sin\left (\frac{\phi(d,t)}{\phi_{0}}  \right )+\frac{{\phi}'(d,t)}{l}=0.
\end{equation}
At the open end of the cavity $x=0$, the absence of current flow gives the second boundary condition
\begin{equation}
    \frac{{\phi}(0,t)'}{l}= 0.
\end{equation}
The latter condition fixes the form of the allowed eigenmodes, denoted by $m$, as 
\begin{equation}\label{Eq:eigenmode_general}
    \phi_m(x,t) = \phi_m(t) \cos(k_m x).
\end{equation}
Assuming a small amplitude for $\phi(d,t)$, Eq.~(\ref{Eq: boundary squid}) is linearized [$\sin\left ( \phi(d,t)/\phi_0 \right ) \approx \phi(d,t)/\phi_0$]. Substituting Eq.~(\ref{Eq:eigenmode_general}) into this linearized boundary condition leads to 
\begin{equation}\label{Eq:eigenmode}
    \frac{L_{cav}}{L_j}-\frac{C_J(k_m d)^2}{C_{cav}}=k_m d \tan(k_md).
\end{equation}
The solutions $k_m$ to this transcendental equation determine the spatial degree of freedom of the eigenmodes of the system in the linear regime. The device used in the experiment was designed such that the capacitance and inductance ratio are respectively roughly $C_J/C_{cav} \approx 0.05$ and $\frac{L_J}{L_{cav}} \approx 0.03$. As will be shown below, the inductance participation ratio $L_J/L_{cav}$, denoted  by $\gamma$, is a key design parameter. Neglecting SQUID capacitance ($C_J=0$) and expanding Eq.~(\ref{Eq:eigenmode}) to the first order gives the following approximate solution for the fundamental mode 
\begin{equation}\label{eq:first_order_approx}
    k_0d \approx \frac{\pi/2}{1+\gamma}. 
\end{equation}

 Expressed in terms of the resonance frequency $\omega_m=k_m v$, this can be rewritten

\begin{equation}\label{eq:flux_response_approx}
    \omega_0 \approx \frac{\omega_{\lambda/4}}{1+\gamma},
\end{equation}
where $\omega_{\lambda/4}$ is the resonance frequency of the fundamental mode of the bare cavity (for $L_J=0$)~\cite{Krantz_2013}. In Figure~\ref{Fig:distributed_model}(b), the relative error of the fundamental mode solution $k_0$ is shown as a function of the inductance ratio. This comparison is made both when neglecting the SQUID capacitance (red curve) and when using the first-order solution of Eq.~(\ref{eq:flux_response_approx}) (blue curve). The results clearly demonstrate, that within the parameter regime of the device, the capacitive term can be neglected, and that Eq.~(\ref{eq:flux_response_approx}) is an excellent approximation.

Knowing the spatial solution of the eigenmodes, circuit quantization is performed by substituting $\phi_m(x,t)$ into the Lagrangian Eq.~(\ref{Eq: lagrangian}). Interactions between the eigenmodes are neglected.
\begin{equation}
    \mathcal{L}=\int_0^d\left (\frac{c}{2}\dot{\phi}_m(t)^2\cos^2(k_mx) - \frac{k_m^2}{2l}\phi_m(t)^2 \sin^2(k_mx)  \right )dx 
    + \frac{C_J}{2}\dot{\phi}_m(t)^2 \cos^2(k_md) + E_J \cos \left ( \frac{\phi_m(t)\cos(k_md)}{\phi_0} \right ).
\end{equation}

Expanding the nonlinear potential to the second order: $E_J \cos\left (\frac{\phi_m(d,t)}{\phi_{0}} \right ) \sim - \frac{E_J}{2\phi_0^2}\phi_m(d,t)^2 + \frac{E_J}{24\phi_0^2}\phi_m(d,t)^4 $, gives
\begin{equation}\label{Eq: full lagrangian}
\begin{split}
\mathcal{L}=\int_0^d\left (\frac{c}{2}\dot{\phi}_m(t)^2\cos^2(k_mx) - \frac{k_m^2}{2l}\phi_m(t)^2 \sin^2(k_mx)  \right )dx 
    &+ \frac{C_J}{2}\dot{\phi}_m(t)^2 \cos^2(k_md) \\
&-\frac{E_J\phi_m(t)^2}{2\phi_0^2}\cos^2(k_md) + \frac{E_J \phi_m(t)^4}{24\phi_0^4} \cos^4(k_md).
\end{split}
\end{equation}

An effective LC oscillator can be defined  from the linear part of Eq.~(\ref{Eq: full lagrangian}). The oscillator as an effective capacitance $C_m$ and inductance $L_m$ defined as 
\begin{equation}\label{Eq: effective capacitance}
   C_m = c\int_0^d \cos^2(k_mx) dx + C_J \cos^2(k_md)=\frac{C_{cav}}{2} M_m,
\end{equation}
\begin{equation}\label{Eq: effective inductance}
     L_m^{-1} = \int_0^d\frac{k_m^2}{l} \sin^2(k_mx) dx +  \frac{E_J}{\{\phi_0^2}\cos^2(k_md)
   = \frac{(k_md)^2}{2L_{cav}} M_m,
\end{equation}
where, 
\begin{equation}
    M_m = \left [ 1+ \frac{\sin(2 k_m d)}{2k_md} + \frac{2C_J}{C_{cav}}\cos^2(k_md) \right ].
\end{equation}

Note that the equality of Eq.~(\ref{Eq: effective inductance}) is found by substituting $1/L_J$ by Eq.~(\ref{Eq:eigenmode}). Using the above definitions, the Lagrangian of Eq.~(\ref{Eq: full lagrangian}) simplifies to
\begin{equation}
    \mathcal{L}= \frac{C_m}{2} \dot{\phi_m}(t)^2  -\frac{1}{2 L_m} \phi_m(t)^2 + \frac{E_j \phi_m(t)^4}{24\phi_0^4} \cos^4(k_md). 
\end{equation}

A Legendre transformation of the Lagrangian (neglecting any mode interactions) results in the following Hamiltonian  
\begin{equation}
      \mathcal{H}= \frac{C_m}{2} q_m(t)^2  +\frac{1}{2 L_m} \phi_m(t)^2 - \frac{E_j \phi_m(t)^4}{24\phi_0^4} \cos^4(k_md), 
\end{equation}

where $q_m(t)=C_m \dot{\phi}_m(t)$ is the conjugate variable of $\phi_m(t)$. In the quantum regime $\phi_m(t)$ and $q_m(t)$ are operators satisfying the commutation relation $\left [ \hat{\phi}_m, \hat{q_m} \right ] = i\hbar$. These operators can be rewritten in terms of normal mode annihilation $\hat{a}$ and creation operators $\hat{a}^\dag$ 
\begin{equation}
\begin{split}
      \hat{\phi}_m &= \phi_{zpf,m} ( \hat{a}_m +\hat{a}_m^\dag) \\
      \hat{q}_m &= -iq_{zpf,m} ( \hat{a}_m - \hat{a}_m^\dag), 
\end{split}
\end{equation}
where $q_{zpf,m}=\sqrt{\hbar \omega_m C_m /2}$ and $\phi_{zpf,m}=\sqrt{\hbar/2\omega_m C_m}$. Under the rotating wave approximation and considering only the fundamental mode $m=0$, the Hamiltonian of the system can be expressed as

\begin{equation}
    \hat{H}/\hbar = \omega_r \hat{a}^\dagger \hat{a} + \frac{U}{2} \hat{a}^\dagger \hat{a}^\dagger \hat{a} \hat{a},
\end{equation}

where there resonance frequency is $\omega_r = \omega_0+U$ and Kerr nonlinearity is 
\begin{equation}\label{eq:Kerr}
      U = -\frac{E_J}{2\hbar} \left ( \frac{\phi_{zpf,0}}{\phi_0} \right )^4 \cos^4(k_0d)  = -\frac{\hbar \omega_0^2 L_{cav}}{2\gamma \phi_0^2}  \left [ \frac{ \cos^2(k_0d)}{(k_0d)^2 M_0} \right ] ^2. 
\end{equation}

Note that, since $\omega_0\gg U$, we can neglect the small photon number dependent frequency shifts due to the nonlinear term such that $\omega_r=\omega_0$.

\subsection{Open-system parameters}
\label{section: supplementary scattering response }

To model the entire system, the Hamiltonian of the $\lambda/4$ Kerr resonator has to be modified to incorporate the interaction with the environment, i.e., the surrounding bosonic baths. The system is coupled to three distinct baths: the feedline, which is separated into right- and left-propagating modes, and the intrinsic bath. 
As we are interested in fitting the scattering coefficients (depending on the intracavity photon number), only the dominant effects of photon loss are considered.
Following the approach of~\cite{Qi_scattering}, the total Hamiltonian can be described as 
\begin{equation}
\begin{split}
     &\hat{H}/\hbar = \underbrace{\omega_r \hat{a}^{\dag} \hat{a} + \frac{U}{2}\hat{a}^{\dag}\hat{a}^{\dag}\hat{a}\hat{a}}_{\text{Cavity}}+\underbrace{\int \, d \omega \,  \omega \,  \hat{b}_{\omega}^{(int)\,  \dag}\hat{b}_{\omega}^{(int)} + i g_{int}\left ( \hat{b}_{\omega}^{(int)\,  \dag} \hat{a}- \hat{b}_{\omega}^{(int)} \hat{a}^{\dag}\right )}_{ \text{Intrisic bath + coupling}}\\
     &+ \underbrace{\int \, d \omega \,  \omega \,  \hat{b}_{\omega}^{(r)\,  \dag}\hat{b}_{\omega}^{(r)} + i g_{r}\left ( \hat{b}_{\omega}^{(r)\,  \dag} \hat{a}- \hat{b}_{\omega}^{(r)} \hat{a}^{\dag}\right )}_{\text{Right propagating modes + coupling}} +
     \underbrace{\int \, d \omega \,  \omega \,  \hat{b}_{\omega}^{(l)\,  \dag}\hat{b}_{\omega}^{(l)} + i g_{l}\left ( \hat{b}_{\omega}^{(l)\,  \dag} \hat{a}- \hat{b}_{\omega}^{(l)} \hat{a}^{\dag}\right )}_{\text{Left propagating modes + coupling}}  ,
\end{split}
\end{equation}
where $\hat{b}_{int/l/r}$ are the harmonic oscillator modes associated with the internal, right- and left-propagating baths and $g_{int/r/l}$ represents the coupling strength between these modes and the resonator field $\hat{a}$.
We assumed these couplings to be independent on the frequency of bath modes.
Following standard input-output theory~\cite{Qi_scattering,introduction_quantum_noise}, the time evolution of $\hat{a}$ is defined by the quantum Langevin equation 
\begin{equation}
\label{Eq:time_langevin}
\begin{split}
    \dot{a}(t) &=-i\omega_r\hat{a}(t)-iU\hat{a}^{\dag}(t)\hat{a}^{2}(t)-\frac{\kappa_{int}+\kappa_{l}+\kappa_{r}}{2}\hat{a}(t) -\sqrt{\kappa_l}\hat{b}^{(l)}_{in}(t)-\sqrt{\kappa_r}\hat{b}^{(r)}_{in}(t) -\sqrt{\kappa_{int}}\hat{b}^{(int)}_{in}(t),
\end{split}
\end{equation}
where $\kappa_{int/l/r}\rightarrow \sqrt{ \frac{g_{int/l/r}}{2\pi}}$ and the input fields are defined as $\hat{b}_{in}^{(l/r)}(t)=\frac{1}{\sqrt{2\pi}}\int_{-\infty}^{\infty} d\omega e^{-i\omega t} \hat{b}_{\omega}^{(l/r)}(0)$. 
The input-output relations for the fields propagating in the feedline are   
\begin{equation}
\label{Eq: input-ouput}
\begin{split}
    \hat{b}_{out}^{(l)}(t) & =\hat{b}_{in}^{(l)}(t)+\sqrt{\kappa_l}\hat{a}(t),\\
    \hat{b}_{out}^{(r)}(t) & =\hat{b}_{in}^{(r)}(t)+\sqrt{\kappa_r}\hat{a}(t),\\
    \hat{b}_{out}^{(int)}(t) & =\hat{b}_{in}^{(int)}(t)+\sqrt{\kappa_{int}}\hat{a}(t).
\end{split}
\end{equation}

Assuming an equal coupling between the intra-resonator mode and both the left- and right-propagating mode ($\kappa_l=\kappa_r$),  we define the total external coupling $\kappa_{ext}$ as $\kappa_{ext} = 2\kappa_r$. By  Fourier transforming Eq.~\eqref{Eq:time_langevin} and substituting for $\kappa_{ext}$, we obtain 
\begin{equation}
\label{Eq:Fourier_langevin}
\begin{split}
    &i(\omega_r-\omega) \hat{a}(\omega)+iU\hat{a}^{\dag}(\omega)\hat{a}^{2}(\omega)+\frac{\kappa_{int}+\kappa_{ext}}{2} \hat{a}(\omega) = -\sqrt{\frac{\kappa_{ext}}{2}} \left (  \hat{b}^{(r)}_{in}(\omega)+\hat{b}^{(l)}_{in}(\omega)\right ) -\sqrt{\kappa_{int}}\hat{b}^{(int)}_{in}(\omega) . 
\end{split}
\end{equation}

To characterize the parameters of the system, only the right propagating input field is sent to the cavity, i.e., $\hat{b}_{in}^{(l)}(t)=0$ and $\hat{b}_{in}^{(int)}(t)=0$ are the vacuum mode. Furthermore, assuming that the intraresonator and input fields are coherent states, respectively defined as $\hat{a}\left |  \alpha \right \rangle = \alpha \left |  \alpha \right \rangle$ and $\hat{b}_{in}^{(r)}\left |  \beta_{in}^{(r)} \right \rangle = \beta_{in}^{(r)} \left |  \beta_{in}^{(r)} \right \rangle$, we can rewrite Eq.~(\ref{Eq:Fourier_langevin}) as
\begin{equation}
\label{Eq:Fourier_langevin_coherent}
\begin{split}
    &i(\omega_r-\omega) \alpha +iU \left | \alpha \right |^2 \alpha +\frac{\kappa_{int}+\kappa_{ext}}{2} \alpha = -\sqrt{\frac{\kappa_{ext}}{2}} \beta_{in}^{(r)}.
\end{split}
\end{equation}
This approximation, known as the semiclassical approximation, is justified either in the limit in which $U \left | \alpha \right |^2  \ll \kappa_{ext} $, and it is known to be predictive for the Kerr resonator far from the critical points \cite{BartoloPRA16,CasteelsPRA16}. 

Following Refs.~\cite{four_wave_mixing,Eichler_2014_josephson}, we multiply Eq.~(\ref{Eq:Fourier_langevin_coherent}) by its complex conjugate to derive the average photon number $\left | \alpha \right |^2$ in the resonator

\begin{equation}
        \left (\delta^2+\frac{1}{4}  \right )n-2\delta\xi n^{2}+\xi^2 n^{3}=\frac{1}{2},
\end{equation}
where the scale invariant quantities $\delta$, $\xi$ and $n$ are defined as 
\begin{equation}
\label{Eq: detla}
    \delta\equiv \frac{\omega-\omega_r}{\kappa_{in}+\kappa_{ext}},
\end{equation}

\begin{equation}
\label{Eq: xi}
\xi\equiv \frac{\left | \beta_{in}^{(r)} \right |^2\gamma_{ext}U}{\left ( \gamma_{ext}+\gamma_{int} \right )^3},
\end{equation}

\begin{equation}
\label{Eq: photon_number}
    n\equiv \frac{\left | \alpha \right |^{2}}{\left | \beta_{in}^{(r)} \right |^2}   \frac{\left ( \kappa_{ext}+\kappa_{int} \right )^{2}}{ \kappa_{ext}}.
\end{equation}

After solving for $n$ in the above equation, we can calculate the scattering parameter $S_{21}$ from  Eqs. (\ref{Eq: photon_number}) and (\ref{Eq: input-ouput}) in terms of the scale invariant quantities (Eqs.~(\ref{Eq: detla}), (\ref{Eq: xi}), (\ref{Eq: photon_number})) 

\begin{equation}
\label{eq: bare scattering}
    S_{21}=\frac{\left \langle \hat{b}_{out}^{(r)} \right \rangle}{\left \langle \hat{b}_{int}^{(r)}  \right \rangle} = 1-\frac{\kappa_{ext}}{\kappa_{ext}+\kappa_{int}}\frac{1}{1+2j(\delta-\xi n)}.
\end{equation}

Note that, to be consistent with other results in the literature~\cite{four_wave_mixing,Qi_scattering}, we have written the scattering coefficient using the electrical engineering convention for the imaginary unit. In this convention, the imaginary unit is defined as $j=-\sqrt{-1}$, instead of the common physics convention in which the imaginary unit is $i=\sqrt{-1}$~\cite{Qi_scattering,Girvin_Houche}. To perform a direct fit of the experimental data for $S_{21}$, it is necessary to introduce a correction factor that takes into account the net attenuation or gain of the line and the phase shift introduced by the finite speed of the field and the cable length. These corrections are done by multiplying Eq.~(\ref{eq: bare scattering}) by $S_{env}=a e^{j\alpha} e^{-j \omega\tau} $, where $a$ is an additional amplitude, $\alpha$ is a phase shift and $\tau$ is the electronic delay~\cite{Probst_2015}. In addition, following the diameter correction method~\cite{Khalil_2012}, we also introduce a factor of $e^{j\phi}/\cos{\phi}$ to compensate for any impedance mismatch. The corrected scattering coefficient is 

\begin{equation}
\label{Eq: final_fit}
    S_{21}= a e^{j\alpha}e^{-j\omega \tau}\left (1-\frac{\kappa_{ext}}{\kappa_{ext}+\gamma_{int}} \frac{e^{j\phi}}{\cos{\phi}}  \frac{1}{1+2j(\delta-\xi n)}  \right ).
\end{equation}

In the limit of low photon number ($n\to 0$), nonlinear effects are negligible and Eq.~(\ref{Eq: final_fit}) can be fitted directly to extract $\kappa_{ext}$ and $\kappa_{int}$. Note that the dephasing rates $\kappa_{\phi}$ and the losses through the flux line $\kappa_{F}$ are all included in $\kappa_{int}$ by this approximation. At higher input power, one must first solve for $n$ using Eq.~\ref{Eq: photon_number}, and subsequently substitute into Eq.~\ref{Eq: final_fit}. Fitting at higher input power fit allows us to extract $\xi$ from which we can either deduce $U$. However, this requires knowing the incoming photon flux  
\begin{equation}
    \left | \beta_{in}^{(r)} \right |^{2}=\frac{10^{(P_{d}+A)/10}}{\hbar \omega} 10^{-3},
\end{equation}

with $P_{d}$ the power in dbm at room temperature and $A$ is the attenuation of the input line.

\section{Parameter estimation}

All through the work, we use the model in Eq.~\eqref{Eq:lindblad} to perform our numerical simulation.
To study the photon number, we compute the steady state by numerically solving the system $\mathcal{L} \sss =0$.
For the computation of the Liouvillian gaps, instead, we block-diagonalize the Liouvillian \cite{minganti2023dissipative} and perform an Arnoldi-iteration algorithm using shifted-inverted strategy to find the minimal eigenvalue. 
For each simulation, convergence in the cutoff is verified by confirming that, increasing the size of the Hilbert space, data are within $1\%$ difference.

As detailed in the main text, the parameters $\kappa_{\phi}$, $\kappa_{2}$, $G$, and $n_{\rm th}$ cannot be directly argued from straightforward measures, such as those described in \ref{section: supplementary scattering response }.
We thus resort to an optimization strategy, aimed at reproducing the experimental curves for the photon number $n_{\rm ss}$, as well as the Liouvillian gaps $\lambda_{\rm 1st}$ and $\lambda_{\rm SSB}$.

\subsection{Testing the validity of the model}

First, we aim at verifying the validity of the proposed model in \eqref{Eq:lindblad}, in particular in determining the Liouvillian gaps.
To do that, we see if the result of our estimation retrieves a value of $U$ and $\kappa$ that is comparable with the experimentally measured ones.

First, we introduce the cost function.
Let $\vec{p} = [U, \kappa, \kappa_{\phi}, \kappa_2, G, n_{\rm}]$ be the set of parameters.
We compute, for the same set of frequencies experimentally measured $\{\omega_1 \dots \omega_n  \}$, the photon number $\vec{n}^{\rm th} = \{n(\omega_1), \dots  n(\omega_n) \}$.
We then select 6 points around the minimum of the first-order gap, and define $\vec{\lambda}^{\rm exp}_{\rm 1st}$, and numerically obtain the Liouvillian gap $\vec{\lambda}^{\rm th}_{\rm 1st}$ at the corresponding . We do the same for the SSB gap $\vec{\lambda}^{\rm exp}_{\rm SSB}$, and obtain the corresponding numerical data $\vec{\lambda}^{\rm th}_{\rm SSB}$.
We then introduce the cost function
\begin{equation}
    C = d(\vec{n}^{\rm exp}, \vec{n}^{\rm th})
    + 4 d(\vec{\lambda}^{\rm exp}_{\rm 1st}, \vec{\lambda}^{\rm th}_{\rm 1st})
    + 6 d(\vec{\lambda}^{\rm exp}_{\rm SSB}, \vec{\lambda}^{\rm th}_{\rm SSB}),
\end{equation}
where $d$ evaluates the distance between the theoretical result and the corresponding experimental one, once they have been normalized.

We then sort to a simulated annealing algorithm for  probabilistic optimization, as we have to search for the optimal solution in a multi-dimensional space. 
The routine is
\begin{itemize}
    \item Initialization: We set initially the parameters based on the experimental estimation and preliminary simulation of the remainder parameters.
    \item
    Set the temperature that controls the probability of accepting more costly solutions. Heuristically, we find a $50\%$ rejection rate by setting the initial temperature to $T_0 = 0.1$.
    \item Compute the initial cost $C$.
    \item We then enter an interative loop where, for 20 iterations, we repeat
    \begin{itemize}
        \item Update the temperature according to, in our case, an exponential schedule, reading
        $T = T_0 \times 0.93^{\rm iteration}$. 
        \item We extract a new set of random parameters, differing from the old one by a Gaussian function with variance $5\%$ of the parameter.
        \item Compute the cost function $C_{\rm new}$ for this new set of parameters.
        \item If $C_{\rm new}<C$, accept the move. Otherwise, extract a random number $r$, and accept the move if $r < \exp[(C_{\rm new} -C)/T]$.
        In both cases, set $C = C_{\rm new}$. Otherwise, discard the move.
    \end{itemize}
\end{itemize}

\begin{figure}
    \centering    
    \includegraphics[width=\textwidth]{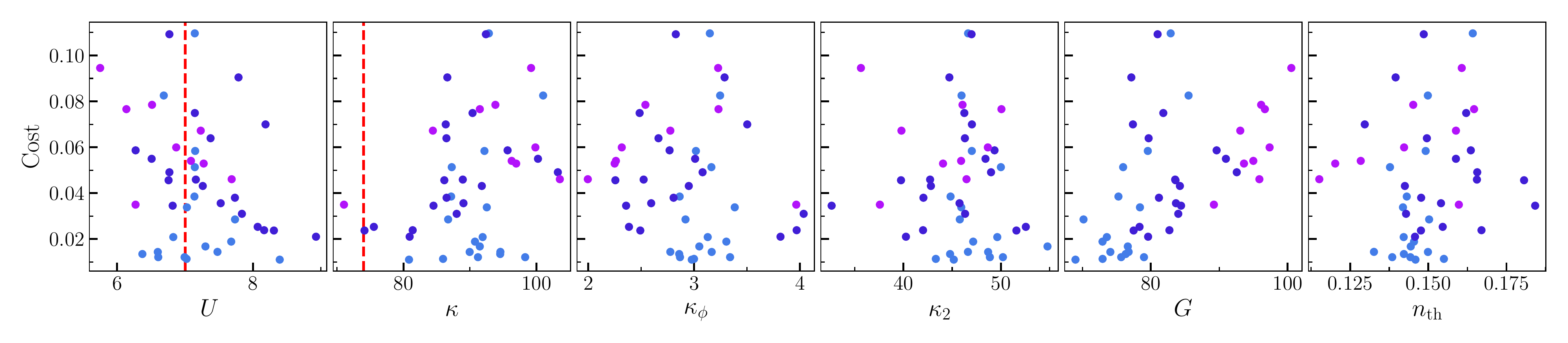}
    \centering
    \begin{tabular}{|c|c||c|c|c|c|c|c||c|}
    \hline
       Pump (A.u.) & $L$ & $U$ [kHz] & $\kappa$ [kHz]& $\kappa_{\phi}$ [KHz]  & $\kappa_2$ [Hz]& $G$ [KHz] & $n_{\rm th}$ & Cost Range \\
       \hline
        0.85 & 1 & 6.3 -- 8.3 & 80.8 -- 98.3 & 2.7 -- 3.3 & 43 -- 55 & 68 -- 79 & 0.13 -- 0.15 & 0.011 -- 0.029\\
         \hline
         0.95 & 1.12    & 6.7 -- 8.9 & 74 -- 103  & 2.2 -- 4.0  & 32 -- 52 & 77 -- 93  & 0.14 -- 0.18 & 0.021--0.046 \\
         \hline
         1.05 & 1.29 & 5.8 -- 7.6  & 71 -- 103 & 2.0 -- 4.0 & 35 --   50  &  89 -- 100 & 0.11 -- 0.16 & 0.034 -- 0.094   \\
         \hline
    \end{tabular}
    \caption{
    Distribution of the obtained parameter vs their cost function, and table summarizing the best estimated parameters obtained through simulated annealing. The red vertical lines indicate the experimentally obtained estimation.}
    \label{fig:annealing_parameter}
\end{figure}

Following this strategy, we find the set of parameters indicated in Fig.~\ref{fig:annealing_parameter}.
Importantly, the data agree with the experimental findings, and allow us to conclude that the model used is indeed predictive of the emergent physics.

\subsection{Determination of the final parameter set}
As we obtain consistent results through the various pump powers, we then assume a ``one parameter-fit-all'' strategy, fixing the values of the the Kerr nonlinearity $U$ and total photon loss rate $\kappa$ to the experimentally obtained value.
We then re-run the simulated annealing algorithm, but this time compering the theoretical data with the experimental ones for all points at $\Delta/2\pi<1$MHz.
The result of this optimization are the parameters reported in the main text.

\section{Theory of dissipative phase transitions}
\subsection{Open-system dynamics and Quantum Trajectories}

\renewcommand{\L}{\mathcal{L}}

The Lindblad master equation in \eqref{Eq:lindblad} describes the state of an open quantum system at a time $t$ via a density matrix $\rhot$.
Within a quantum trajectory approach, instead, the density matrix can be thought as a statistical mixture of  pure states
\begin{equation}
\label{Eq:Correspondence_quantum_traj_density_matrix}
\rho (t) =\lim_{N\to\infty} \frac{1}{N}  \sum_{n=1}^N  \ket{\psi_n(t)}\bra{\psi_n(t)},
\end{equation}
where the evolution of the pure quantum states composing the mixture $\left\{ \ket{\psi_n(t)} \right\}$ evolves according to a given stochastic protocol. The quantum expectation values can be obtained by averaging the over many of those states. When the number of trajectories $N$ is large enough one recover the result of the Lindblad master equation with a statistical error that scales as $N^{-1/2}$.
Below, we will use the following notation:
\begin{itemize}
	\item $\langle{\hat{o}} \rangle(t) = {\rm Tr}\left[\rhot(t) \hat{o}\right]$ indicates the average obtained either by the master equation or by an infinite number of trajectories.
	\item $\langle{\hat{o}} \rangle_{\Psi}(t) = \expval{\Psi(t)|\hat{o}|\Psi(t)}$ indicates the expectation value of a single quantum trajectory $\ket{\Psi(t)}$.
	\item Given a generic function $f$, we define the average over (ideally infinitely many) trajectories as 
	$\overline{f(\langle{\hat{o}} \rangle_{\Psi}(t))} = \lim_{N\to\infty}\frac{1}{N}\sum_{n=1}^N f(\langle{\hat{o}} \rangle_{\Psi_n}(t))$.
\end{itemize}

\subsubsection{Ergodicity}
\label{App:Ergodicity}

In many experimental data reported above, instead of performing a measure over many quantum trajectories, we rather measured an extremely long one, and then averaged over such measures. This property is nothing but an \textit{ergodicity} of the quantum trajectory, that explores the entirety of the probability space of the steady state.

The proof of the ergodicity of a single quantum trajectory (in system with a single zero eigenvalue) goes as follows:\\
 \textbullet \quad  Consider an initial condition $\ket{\Psi (t=0)}$ and evolve it to the long time limit, $\ket{\Psi (t=T)}$ with $T \gg 1/\lambda_{\rm 1st}\gg 1/\lambda_{\rm 2nd}$. At this point in time, the quantum trajectory will have lost any memory of the initial condition.
 Indeed, of we were to average over many quantum trajectories, we would recover $\sss$. 
 In other words, $\ket{\Psi (T)}$ is one of the states of $\sss$.
\\ \textbullet \quad We reinitialise the system, by considering a set of trajectories $\ket{\Phi_{n} (t=0)}=\ket{\Psi (t=T)}$, that is the initial state of our second simulation is the final state of the first one.
\\ \textbullet \quad Since the average over many $\ket{\Phi_n (t=T)}$ must recover $\sss$, we deduce that each $\ket{\Phi_n (t=0)}$ can evolve with a certain probability towards one of the states composing $\sss$.
\\ \textbullet \quad Given that the time evolution of the quantum trajectory is Markovian, every one of the trajectories $\ket{\Phi_n (t=0)}$ is a legitimate evolution for $\ket{\Psi (t=T)}$.
\\ \textbullet \quad Since this line of reasoning can be extended to all the states $\ket{\Psi (t>T)}$, we conclude that a single quantum trajectory must explore all the states of $\sss$, and the average for long times must exactly recover the average over many trajectories.

\subsection{The Liouvillian spectrum and phase transitions}

The steady-state of a system can display a nonanalitical behavior as a function of one parameter.
In this case, we say that a phase transition is taking place.
Here, we briefly recall the spectral properties of the Liouvillian, and how they can signal the emergence of phase transitions \cite{MingantPRA18_Spectral}. 
The interested reader may find a more detailed discussion of the peculiar properties of DPTs also in \cite{minganti2021liouvillian,Minganti2021continuous,KesslerPRA12,CarmichaelPRX15,minganti2023dissipative,PRLLieu20,soriente2021distinctive,Soriente2018,LeePRL13,HuberPRA20,BiondiPRA17}.

Given any Liouvillian $\LL$, we can introduce its eigenvalues $\lambda_i$ and eigenmatrices $\eig{i}$, defined via the relation
\begin{equation}\label{Eq:Spectrum_sup}
\LL \eig{i}=\lambda_i \eig{i}.
\end{equation}
It can be proved \cite{BreuerBookOpen,RivasBOOK_Open} that $\Re{\lambda_i}\leq 0, \forall i$.
For convenience, we sort the eigenvalues in such a way that $\abs{\Re{\lambda_0}}<\abs{\Re{\lambda_1}} < \ldots < \abs{\Re{\lambda_n}}$. 
Usually, there exists a unique steady state density matrix $\sss\propto \eig{0}$ such that $\LL \sss=0$, i.e., the steady state does not evolve anymore under the action of the Liouvillian superoperator.
In this configuration, the real part of the eigenvalues is responsible for the relaxation towards the steady-state, while the complex values of $\lambda_i$ describe oscillation processes in the dynamics.
The eigenmatrix $\eig{1}$ (the one associated to the smallest nonzero eigenvalue $\lambda_1$) describes the slowest relaxing state towards the steady state, and $\sss = \lim\limits_{t \to + \infty} e^{-{\mathcal L} t} \rho(0)$.

Knowing the full Liouvillian spectrum, and aside from points where the Liouvillian is defective \cite{Minganti2019} one can immediately write the dynamics of any density matrix as \cite{MacieszczakPRL16}
\begin{equation}\label{Eq:decomposition_dynamics}
    \rhot(t) = \sss + \sum_j c_j e^{-\lambda_j t} \eig{j}
\end{equation}
where the coefficient $c_j$ can be determined using the left eigenoperators of the Liouvillian.

Consider now a system which, in a certain region of the space parameters, admits a unique steady state.
In the thermodynamic limit $N\to+\infty$, a transition between two different phases is characterized by the nonanalytical behavior of some $\zeta$-independent observable $\hat{o}$ when the parameter $\zeta$ tends to the critical value $\zeta_c$.
Formally, we say that there is a phase transition of order $M$ if
\begin{equation}\label{Eq:DPTDefinition}
\lim_{\zeta \to  \zeta_c}\left| \lim_{N\rightarrow +\infty} \frac{\partial^M}{\partial \zeta^M} \Tr{\sss(\zeta, N) \hat{o}}\right|=+ \infty.
\end{equation}
Roughly speaking, this phase transition takes place in the thermodynamic limit when some eigenvalues pass from being nonzero to become exactly zero, both in its real and imaginary parts, as a function of the parameter $\zeta$.

In finite-size systems, phase transitions cannot be observed.
Nevertheless, the study of the Liouvillian eigenvalues provides much useful information about the scaling and nature of the transition \cite{VicentiniPRA18}.

\subsubsection{Second order phase transition}

A weak symmetry of an open quantum system is described by the presence of a superoperator $\mathcal{U}=\hat{V}\bigcdot \hat{V}^{-1}$ \cite{BaumgartnerJPA08}, such that
\begin{equation}
\mathcal{U}^{-1}\mathcal{L}\,\mathcal{U}=\mathcal{L},
\end{equation}
or, equivalently, $[\mathcal{L},\mathcal{U}]=0$. 
The symbol $\bigcdot$ is a placeholder, and $\mathcal{U} \rho =(\hat{V}\bigcdot \hat{V}^{-1}) \rho = \hat{V}\rho \hat{V}^{-1}$  
The presence of a symmetry in the system fixes many properties of the system.
Indeed, if $\sss$ is the only eigenmatrix with zero eigenvalue of $\L$ (unique steady state) before the transition, one can prove that it must also be an eigenmatrix of $\mathcal{U}$, and $\mathcal{U} \sss= \sss$ ($\sss$ is a symmetric state).

Let us consider now the $Z_2$ symmetry of the model in \eqref{Eq:lindblad}.
In this case, the symmetry superoperator is $\mathcal{U} = \hat{\Pi} \bigcdot  \hat{\Pi}$ where $\hat{\Pi}= \exp{i \pi \hat{a}^\dagger \hat{a}}$ is the parity operator.
In this case, the phase transition is associated to one eigenvalue $\lambda_{\rm SSB}$ becoming and remaining zero in a whole region.
The corresponding states $\eig{\rm SSB}$, allow to construct the symmetry-breaking metastable state.
Indeed, one can construct $\rho_{\rm SSB}^{\pm} = \sss \pm \eig{\rm SSB} $, which are well-defined density matrices that decay at a rate $\lambda_{\rm SSB}$, and $\mathcal{U} \rho_{\rm SSB}^{\pm} = \rho_{\rm SSB}^{\mp} $ (they are not symmetric).

\subsubsection{First-order dissipative phase transition and the Liouvillian spectrum}
\label{App:Liouvillian}

For a first-order phase transition, $\sss$ must be discontinuous and the transition is signalled by $\lambda_{\rm 1st}$
In Ref.~\cite{MingantPRA18_Spectral}, it was proved that $\eig{\rm 1st}\propto \eig{\rm 1st}^+ - \eig{\rm 1st}^-$, $\eig{\rm 1st}^+$ ($\eig{\rm 1st}^-$) being the density matrix just before (after) the phase transition.
Moreover, at the critical point $\sss\propto \eig{\rm 1st}^+ + \eig{\rm 1st}^-$. 
Let us note that this  equation has a clear physical interpretation: at the critical point, for a finite-size system, the steady state is the equiprobable mixture of the two phases, which are encoded in the spectral decomposition of $\eig{\rm 1st}$. 
In a region at the left (right) of the critical point, $\eig{\rm 1st}^+$ ($\eig{\rm 1st}^-$) are metastable \cite{LandaPRL20,LandaPRB20}.
This means that if the system is initialized in one of these two states it will remain stuck, for a time proportional to $1/\lambda_{\rm 1st}$, before reaching the steady-state \cite{MacieszczakPRL16}.
This gives rise to hysterical behaviour, typical of first-order phase transitions \cite{RodriguezPRL17}.

\subsection{Extracting the Liouvillian gap from symmetry breaking trajectories}

We explore here the relation between the dynamics of a single quantum trajectory and the Liouvillian eigenvalues, and demonstrate \eqref{Eq:Correlators}.

Let us consider a two-point correlation function for quantum trajectories, i.e.,
\begin{equation}
\overline{\expval{\hat{o}}_{\Psi}(t)\expval{\hat{p}}_{\Psi}(t')}=  \lim_{N\to \infty}  \sum_{n=1}^{N} \frac{\expval{\hat{o}}_{\Psi_n}(t)
\expval{\hat{p}}_{\Psi_n}(t')}{N}.
\end{equation}

To understand the meaning of these objects, we need to introduce the idea of the probability space of the trajectories \cite{VicentiniPRA18,Vicentini2019}.
In this formalism, the density matrix initial pure state can be formally written as an integral over the space of trajectories $\mathcal{H}$ as
\begin{equation}\label{Eq:Evolution_in_probability_space}
\rhot(t) =e^{-\mathcal{L} t} \ketbra{\Psi(0)} = \int_{\mathcal{H}} \de  \Psi(t)\;  p\big[\Psi(t)|\Psi(0) \big] \hat{\rho}_\Psi (t),
\end{equation}
where $p\big[\Psi(t)|\Psi(0) \big]$ indicates the conditional probability of obtaining $\ket{\Psi(t)}$ given the initial condition $\ket{\Psi(0)}$, and $\hat{\rho}_\Psi (t)= \ket{\Psi(t)}\bra{\Psi(t)}$.
Since the steady state is independent of the initial condition, we have
\begin{equation}
\sss = \int_{\mathcal{H}}\de \Psi \; p_{\rm ss }\big[\Psi \big] \hat{\rho}_\Psi.
\end{equation}

In this notation, we have
\begin{equation}\label{Eq:Two-time_correlator}
\begin{split}
\overline{\expval{\hat{o}}_{\Psi}(t)\expval{\hat{p}}_{\Psi}(t')}= & \iint_{\mathcal{H}} \de  \Psi(t)\; \de  \Psi(t')\;  p\big[\Psi(t)|\Psi(0), \Psi(t')|\Psi(0) \big] \expval{\hat{o}}_{\Psi}(t) \expval{\hat{p}}_{\Psi}(t') \\ &=\int_{\mathcal{H}} \de  \Psi(t)  p\big[\Psi(t)|\Psi(0) \big]  \expval{\hat{o}}_{\Psi}(t) \int_{\mathcal{H}}  p\big[\Psi(t')|\Psi(t) \big] \expval{\hat{p}}_{\Psi}(t').
\end{split}
\end{equation}
where $p\big[\Psi(t), \Psi(t')|\Psi(0) \big]$ is the joint probability of having $\ket{\Psi(t)}$ and $\ket{\Psi(t')}$ given the initial condition $\ket{\Psi(0)}$, and  we used the fact that
\begin{equation}
p\big[\Psi(t)|\Psi(0), \Psi(t')|\Psi(0) \big]= p\big[\Psi(t)|\Psi(0) \big] p\big[\Psi(t')|\Psi(t) \big].
\end{equation}
The latter follows from the fact that the conditional probability of $\Psi(t')$ depends only on the intermediate state $\Psi(t)$, but not on the previously visited state such as $\Psi(0)$, since a quantum trajectory is a Markovian process.

The second term of Eq.~\eqref{Eq:Two-time_correlator} is now identical to Eq.~\eqref{Eq:Evolution_in_probability_space}, and thus we can re-write it as
\begin{equation}
\begin{split}
\int_{\mathcal{H}}  p\big[\Psi(t')|\Psi(t) \big] \expval{\hat{p}}_{\Psi}(t') &= \operatorname{Tr}\left[ \hat{p} \int_{\mathcal{H}}  p\big[\Psi(t')|\Psi(t) \big]  \ket{\Psi(t')}\bra{\Psi(t')} \right]=
 \operatorname{Tr}\left[ \hat{p} \left(e^{-\LL (t'-t)} \hat{\rho}_{\Psi} (t)\right).
 \right]
\end{split}
\end{equation}

Passing back to the definition in terms of quantum trajectories, we have
\begin{equation}
\begin{split}
\overline{\expval{\hat{o}}_{\Psi}(t)\expval{\hat{p}}_{\Psi}(t')}= \lim_{N\to \infty} \sum_{n=1}^{N} \frac{ \operatorname{Tr}\left[ \hat{o}  \hat{\rho}^n_{\Psi} (t) \right] \operatorname{Tr}\left[ \hat{p} \left(e^{\LL (t'-t)} \hat{\rho}^n_{\Psi} (t)\right) \right]}{N}.
\end{split}
\end{equation}
One can now use Eq.~\eqref{Eq:decomposition_dynamics} and write 
\begin{equation}
    \hat{\rho}^n_{\Psi} = \sss + \sum_j c_j^n(t) \, \eig{j}
\end{equation}
Using the spectral decomposition of the Liouvillian we can rewrite as
$\hat{\rho}^n_\Psi(t)= \sss +\sum c_i^n(t) \eig{i}$,
and finally
\begin{equation}
\begin{split}
\overline{\expval{\hat{o}}_{\Psi}(t)\expval{\hat{p}}_{\Psi}(t')}= \lim_{N\to \infty} \frac{1}{N} \sum_n \operatorname{Tr}\left[ \hat{o}  \hat{\rho}^n_\Psi(t) \right] \operatorname{Tr}\left[ \hat{p} \left(\sss +\sum_i  e^{- \lambda_i (t'-t)} c_i^n(t) \eig{i}\right) \right].
\end{split}
\end{equation}

\begin{figure}
    \centering
    \includegraphics[width=\textwidth]{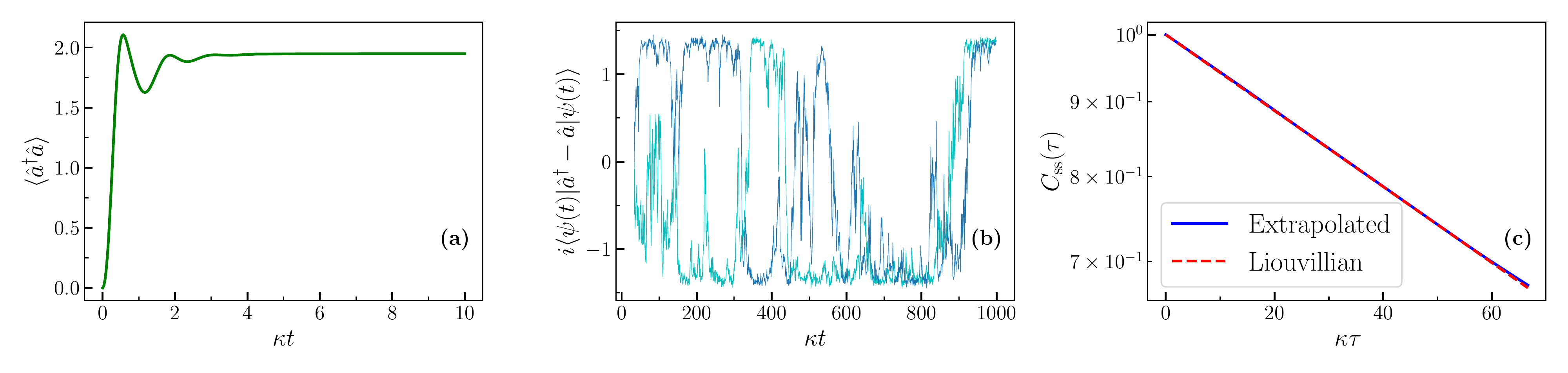}
    \caption{\textbf{Correlation function and Liouvillian gap}. (a) Photon number evolution according to the Lindblad master equation in \eqref{Eq:lindblad}. (b) Single quantum trajectories performing heterodyne measurement \cite{Wiseman_BOOK_Quantum}. (c) As a function of time, the correlation function $C_{\rm ss}(t)$, and $\exp{\lambda_{\rm SSB} t}$, the latter having being obtained by numerical diagonalization of the Liouvillian. Parameters: $\Delta =0$, $U/\kappa = 1$,  $G/\kappa = 2$, $\kappa_\phi/\kappa = 0$, $\kappa_2/\kappa = 0.1$, and $n_{\rm th} =0$.}
    \label{fig:correlation_theory}
\end{figure}

We send now $t \to \infty$, so that the coefficient $c_i^n(t) \to c_{i, {\rm ss}}^n$ are time-independent.
We then call the time difference $\Delta t=t' -t$, and we suppose that $\Delta t$ is large, so we can neglect all but the eigenvalue $\lambda_1$ the closest to zero in real part.
We obtain
\begin{equation}\label{Eq:final_correlation}
\begin{split}
\overline{\expval{\hat{o}}_{\Psi}^{\rm ss}\expval{\hat{p}}_{\Psi}(\Delta t)} &\simeq  \lim_{N\to \infty} \frac{1}{N} \sum_n \operatorname{Tr}\left[ \hat{o}  \hat{\rho}^n_\Psi(t) \right] \operatorname{Tr}\left[ \hat{p} \left(\sss +  e^{-\lambda_1 \Delta t} c_{i, {\rm ss}}^n \eig{i}\right) \right] \\ 
& =\lim_{N\to \infty} \frac{1}{N} \sum_n \operatorname{Tr}\left[ \hat{o}  \hat{\rho}^n_\Psi(t) \right] \left\{\expval{\hat{p}}_{\rm ss} +  \operatorname{Tr}\left[ \hat{p}  e^{-\lambda_1 \Delta t} c_i^n \eig{i} \right]\right\} \\ 
&= \expval{\hat{o}}_{\rm ss} \expval{\hat{p}}_{\rm ss} + \lim_{N\to \infty}  \frac{e^{-\lambda_1 \Delta t}}{N} \sum_n  \operatorname{Tr}\left[ \hat{o}  \hat{\rho}^n_{\Psi} \right] \operatorname{Tr}\left[ \hat{p} c_1^n \eig{1} \right] = \expval{\hat{o}}_{\rm ss} \expval{\hat{p}}_{\rm ss} + e^{-\lambda_1 \Delta t} \mathcal{R},
\end{split}
\end{equation}
where $\mathcal{R} = \sum_n  \operatorname{Tr}\left[ \hat{o}  \hat{\rho}^n_{\Psi} \right] \operatorname{Tr}\left[ \hat{p} c_1^n \eig{1} \right] $.
The correlation function in Eq.~\eqref{Eq: autocorellation}  is a re-normalization of Eq.~\eqref{Eq:final_correlation}.
Hence, we conclude that, by studying the evolution of a single trajectory level, we can access the value of the Liouvillian gap. This is numerically demonstrated in Fig.~\ref{fig:correlation_theory}.

%

\end{document}